\documentclass[fleqn,twoside]{article}
\usepackage{amsmath,amssymb,espcrc2,color}

\usepackage{graphicx}
\usepackage[figuresright]{rotating}
\newcommand{\url}[1]{{\tt #1}}
\newcommand{\lsim}
{\;\raisebox{-.3em}{$\stackrel{\displaystyle <}{\sim}$}\;}
\newcommand{\gsim}
{\;\raisebox{-.3em}{$\stackrel{\displaystyle >}{\sim}$}\;}

\newcommand{\gmt}{$(g-2)_\mu$}
\newcommand{\br}{{\rm BR}}
\newcommand{\bsg}{BR($b \to s \gamma$)}
\newcommand{\btn}{BR($B_u \to \tau \nu_\tau$)}
\newcommand{\bmm}{BR($B_s \to \mu^+\mu^-$)}
\newcommand{\ssi}{\sigma^{\rm SI}_p}
\newcommand{\Och}{\ensuremath{\Omega_\chi h^2}}

\newcommand{\MW}{M_W}
\newcommand{\MZ}{M_Z}
\newcommand{\Mh}{M_h}
\newcommand{\MA}{M_A}
\newcommand{\mt}{m_t}
\newcommand{\mgl}{m_{\tilde g}}

\newcommand{\neu}[1]{\tilde \chi^0_{#1}}
\newcommand{\mneu}[1]{m_{\tilde \chi^0_{#1}}}
\newcommand{\mste}{m_{\tilde t_1}}
\newcommand{\mstaue}{m_{\staue}}
\newcommand{\staue}{\tilde \tau_1}
\newcommand{\tb}{\tan\beta}
\newcommand{\ecm}{\sqrt{s}}
\newcommand{\tev}{\,\, \mathrm{TeV}}
\newcommand{\gev}{\,\, \mathrm{GeV}}

\definecolor{Orange}{named}{Orange}
\definecolor{Purple}{named}{Purple}

\graphicspath{{figs/}}

\graphicspath{{figs/}}

\title{Likelihood Functions for Supersymmetric Observables 
in Frequentist Analyses of the CMSSM and NUHM1} 

\author{
O.~Buchmueller\address[Imperial]
   {High Energy Physics Group, Imperial College, Blackett Laboratory, 
    Prince Consort Road, London SW7  2AZ, UK},
R.~Cavanaugh\address[FNAL]
   {Fermi National Accelerator Laboratory, P.O. Box 500, 
    Batavia, Illinois 60510, USA}\hbox{$^{\rm ,}$}\address[UIC]
   {Physics Department, University of Illinois at Chicago, Chicago, 
    Illinois 60607-7059, USA},
A.~De Roeck\address[CERN]
   {CERN, CH-1211 Gen\`eve 23, Switzerland}\hbox{$^{\rm ,}$}\address[Antwerpen]
   {Antwerp University, B-2610 Wilrijk, Belgium},
J.R.~Ellis\addressmark[CERN],
H.~Fl\"acher\address[Rochester]
   {Department of Physics and Astronomy, University of Rochester, 
    Rochester, New York 14627, USA},
S.~Heinemeyer\address[Santander]
   {Instituto de F\'{\i}sica de Cantabria (CSIC-UC), 
    E--39005 Santander, Spain},
G.~Isidori\address[Isidori]
   {INFN, Laboratori Nazionali di Frascati, Via E. Fermi 40, 
    I--00044 Frascati, Italy}\hbox{$^{\rm ,}$},
K.A.~Olive\address[Minnesota] 
   {William I. Fine Theoretical Physics Institute, University of Minnesota, Minneapolis,
    Minnesota 55455, USA}, 
F.J.~Ronga\address[ETHZ]
   {Institute for Particle Physics, ETH Z\"urich, CH-8093 Z\"urich, 
   Switzerland},
G.~Weiglein\address[Durham]
   {IPPP, University of Durham, Durham DH1 3LE, U.K.}
}
       
\begin{document}

\begin{abstract}
On the basis of frequentist analyses of experimental constraints
from electroweak precision data, \gmt, $B$ physics and cosmological data, we
investigate the parameters of the constrained MSSM (CMSSM) 
with universal soft supersymmetry-breaking mass parameters, and a model
with common non-universal Higgs masses (NUHM1). We present $\chi^2$ likelihood 
functions for the masses of supersymmetric particles and Higgs bosons,
as well as \bsg, \bmm\ and the spin-independent dark matter
scattering cross section, $\ssi$. In the CMSSM we find preferences for
sparticle masses that are relatively light.
In the NUHM1 the best-fit values for many sparticle masses are
even slightly
smaller, but with greater uncertainties. The likelihood
functions for most sparticle masses are cut off sharply at small masses,
in particular by the LEP Higgs mass constraint. Both in the CMSSM
and the NUHM1, the
coannihilation region is favoured over the focus-point region at
about the
3-$\sigma$ level, largely but not exclusively because of \gmt. 
Many sparticle masses are highly correlated in both the CMSSM and NUHM1, 
and most of the regions preferred at the 95\% C.L.\ are accessible to
early LHC running, though high-luminosity running would be needed to cover the
regions allowed at the 3-$\sigma$ levels. Some slepton and
  chargino/neutralino masses should be in reach at the ILC.
The masses of the heavier Higgs
bosons should be accessible at the LHC and the ILC in 
portions of the preferred regions in the $(\MA, \tb)$ plane.
In the CMSSM,
the likelihood function for \bmm\ is peaked close to
the Standard Model value, but much larger values are possible in the NUHM1.
We find that values of $\ssi > 10^{-10}$~pb are preferred in both the
CMSSM and the NUHM1. We study the effects of dropping the \gmt, \bsg,
\Och\ and $\Mh$ constraints, demonstrating that they are not in tension
with the other constraints.

% Preprint numbers
\bigskip
\begin{center}
CERN-PH-TH/2009-124, DCPT/09/112, FTPI-MINN-09/24, 
IPPP/09/56, UMN-TH-2804/09
\end{center}
%\vspace{-0.5cm}
\end{abstract}

% typeset front matter (including abstract)
\maketitle

%---------------------------------------------------------------------
\section{Introduction}
\label{sec:intro}
%---------------------------------------------------------------------

Supersymmetry (SUSY)~\cite{Nilles:1983ge,Haber:1984rc,Barbieri:1982eh}
is one of the favoured ideas for physics beyond
the Standard Model (SM) that may soon be explored at the Large Hadron
Collider (LHC). In a recent paper~\cite{Master2}, we presented some
results from frequentist analyses of 
the parameter spaces of the constrained minimal supersymmetric
extension of the Standard Model (CMSSM) --- in which the
soft supersymmetry-breaking scalar and gaugino masses are each constrained
to universal values $m_0$ and $m_{1/2}$, respectively \cite{Drees:1992am,Baer:1995nc,bb2,Barger:1997kb,Ellis:2001msa,bk,Roszkowski:2001sb,Lahanas:2001yr,Djouadi:2001yk,deBoer:2001nu,Chattopadhyay:2001va,eos2,baer,arndut,eoss,baer2,lahnan,nath,munoz,adh}
 --- and the NUHM1 --- in which the soft supersymmetry-breaking
contributions to the Higgs masses are allowed a different but common value
\cite{bmpbt1,bmpbt2,EOS08}.
Other statistical analyses in these models can be found in 
~\cite{deBoer:2003xm,Belanger:2004ag,Ellis:2003si,Ellis:2004tc,Ellis:2005tu,Ellis:2006ix,Ellis:2007aa,Ellis:2007ka,Ellis:2007ss,Heinemeyer:2008fb,Bechtle:2004pc,Lafaye:2007vs} and Markov Chain Monte Carlo (MCMC) analyses in 
\cite{Baltz:2004aw,Allanach:2005kz,Allanach:2006jc,Allanach:2006cc,Allanach:2007qj,Allanach:2007qk,Allanach:2008iq,Feroz:2008wr,deAustri:2006pe,Roszkowski:2006mi,Roszkowski:2007fd,Roszkowski:2007va,Trotta:2008bp,Master1,othermod,nonunivgaugino,fitmssm19,fittinoMC}.
For comparison, see also~\cite{nmssm1,nmssm2} for recent analyses in the
next-to-minimal extension of the SM, as well
as~\cite{fut,flat1,flat2} for other analyses in supersymmetric 
models without a dedicated fit.

The results presented in~\cite{Master2} included the parameters of the
best-fit points in the CMSSM and the NUHM1, as well as the 68 and
95\%~C.L.\ regions  
found with default implementations of the phenomenological, experimental
and cosmological constraints. These include precision electroweak data,
the anomalous magnetic moment of the muon, \gmt, 
$B$-physics observables (the rates for \bsg\ and \btn,
$B_s$ mixing, and the upper limit on \bmm), the bound on the lightest MSSM Higgs
  boson mass, $\Mh$, and the cold dark matter (CDM) density
inferred from astrophysical and cosmological data~\footnote{We did not 
include the constraint imposed by the experimental
upper limit on the spin-independent DM scattering
cross section $\ssi$, which is subject to astrophysical and hadronic uncertainties,
as discussed below.}, assuming that this is
dominated by the relic density of the lightest neutralino, $\Och$.
We also discussed in~\cite{Master2} the sensitivities of the areas of the preferred
regions to changes in the ways in which the major constraints are
implemented. 
We found that the smallest sensitivity was to the CDM density,
and the greatest sensitivity was that to \gmt.

In this paper we adopt the frequentist approach 
from~\cite{Master2}, which is different from 
the Bayesian approach adopted in \cite{Allanach:2005kz,Allanach:2006jc,Allanach:2006cc,Allanach:2007qj,Allanach:2007qk,Allanach:2008iq,Feroz:2008wr,deAustri:2006pe,Roszkowski:2006mi,Roszkowski:2007fd,Roszkowski:2007va,Trotta:2008bp}. 
A key issue in a Bayesian approach is the appropriate choice of priors. 
As discussed in some recent Bayesian analyses of the 
CMSSM~\cite{Allanach:2007qk,Allanach:2008iq,Trotta:2008bp},
conclusions for preferred regions of parameter space can depend 
the choice of priors. 
In our view, the results of a Bayesian approach should not be
considered definitive unless they are shown to be sufficiently
independent of 
plausible variations in the choice of priors. In our frequentist
analysis, we use the MCMC technique to sample
efficiently the CMSSM and NUHM1 parameter spaces, and we generate
sufficiently many chains to sample adequately these parameter
spaces, as discussed in more detail in Section~2 of this paper.

Our treatments of the experimental constraints from electroweak
precision observables, $B$-physics observables and cosmological data
are, in general, very similar to those
in~\cite{Master2}. Accordingly, we do not discuss details in this paper,
contenting ourselves with a brief recapitulation and update.

In Section~3 we extend the presentation of results 
from our MCMC frequentist analysis to include the global $\chi^2$ likelihood
functions for various observables, including $\Mh$, \bsg, \bmm\ and the 
spin-independent DM scattering cross section, $\ssi$,
as well as sparticle masses.
We also discuss the correlations between pairs of these observables,
and compare the results in the CMSSM and NUHM1. We pay particular attention
to the prospects for detecting SUSY in forthcoming experiments,
including searches at the LHC and the ILC as well as 
$B$ physics and direct searches for CDM.

We present an update on the prediction of $\Mh$ in the CMSSM~\cite{Master1} 
and the first prediction for $\Mh$ in the NUHM1. For these analyses the
experimental constraints 
on $\Mh$ itself have been left out of the fit. The result in the
CMSSM of \cite{Master1} is confirmed with a best-fit value slightly
below the LEP bound. Within the NUHM1, however, a value 
{\em above} the LEP bounds arises naturally.
For other observables, however, the $\Mh$ information {\it is}
included in calculating the 
$\chi^2$ likelihood functions. 
The likelihood functions for generic sparticle masses are skewed, being
cut off at low masses by the LEP lower limit on $\Mh$, in particular. On
the other hand the likelihood functions rise more gradually for large
masses, with the largest contribution arising from \gmt. We see that the
role of the $\Mh$ constraint is smaller in the NUHM1 than in the CMSSM,
reflecting the fact that the other constraints suggest, in the NUHM1, a
value of $\Mh$ somewhat larger than the LEP lower limit. 

As remarked in~\cite{Master2},
the preferred values of the sparticle masses are generally somewhat
lower in the NUHM1 than in the CMSSM. This is because the extra degree of
freedom in the Higgs sector allows lower values of $m_{1/2}$ to be reconciled
with upper limits on deviations from the SM and the LEP lower
limit on $\Mh$. Recall that in the CMSSM, the Higgs mass mixing parameter,
$\mu$, and the Higgs pseudoscalar mass, $\MA$, are fixed by the minimization
of the Higgs potential ensuring electroweak symmetry breaking when
$\tb$ is chosen as an input parameter.  In contrast, in the NUHM1,
either $\mu$ {\em or} $\MA$ can be chosen as an additional input parameter
\footnote{The choice of either $\mu$ or $\MA$ as an input is equivalent
  to a choice of the soft Higgs mass $m_{h_1} = m_{h_2} \ne m_0$ 
at the GUT scale.},
thus allowing substantial additional freedom in the light Higgs scalar
mass for a given set of CMSSM parameters ($m_{1/2}, m_0, A_0, \tb$).
The greater freedom in the Higgs sector also results in
different mass ranges being favoured for the heavier Higgs bosons $H, A,
H^\pm$ and for the heavier neutralinos, as observed in~\cite{Master2}.

We find here that
sparticle masses are mostly highly correlated. This could be expected
for $\mneu{1}$ and $\mgl$, which are both determined essentially
uniquely by $m_{1/2}$. However, the correlation is only slightly
weakened for the slepton and squark masses, including 
$\mste$. This is partly because the largest contributions to the
preferred values of most of these particles
are due to $m_{1/2}$, rather than to $m_0$. This tendency
is reinforced by the fact that our likelihood analysis finds that the
coannihilation regions are favoured in both the CMSSM and the
NUHM1. However, this preference is slightly weakened in the NUHM1, where
direct-channel annihilation through the heavy Higgs ($A, H$) poles may
also play a subsidiary role, and larger values of $m_0$ become
possible. In particular, the correlation between $\mstaue$
and $\mgl$ is particularly weak in the NUHM1, reflecting the
appearance of preferred regions of the parameters away from the
coannihilation strip.
In general, there are good prospects for discovering SUSY
in early LHC running, in both the CMSSM and the NUHM1. 

We find that \bmm\ is expected to be  close
to its SM value in the CMSSM, because of the strong
preference for relatively low $\tb$ where the supersymmetric
contributions to this channel are small. They may be
much larger in the NUHM1 
because of the freedom to choose $\MA$ below its nominal CMSSM value.
Spin-independent scattering of supersymmetric dark matter may well be
observable in planned experiments in both the CMSSM and the NUHM1, 
where a somewhat larger range for $\ssi$
is preferred in the NUHM1 \cite{bmpbt2,Ellis:2009ai}.

However, these optimistic conclusions rely critically on the
implementation of the \gmt\ constraint using low-energy $e^+ e^-$ data,
as used in our analysis, and we discuss in Section \ref{sec:dropgmt} the
implications of removing the \gmt\ constraint. We also discuss the
predictions of our fits for \bsg, \Och\ and $\Mh$, presenting the
likelihood functions for each of these observables without
their own contributions. None of these observables exhibits any
significant tension with the others.

%---------------------------------------------------------------------
\section{Description of the Frequentist Statistical Method Employed}
\label{sec:mpfit}
%---------------------------------------------------------------------

We define a global $\chi^2$ likelihood function, which combines all
theoretical predictions with experimental constraints:
\begin{align}
\chi^2 &= \sum^N_i \frac{(C_i - P_i)^2}{\sigma(C_i)^2 + \sigma(P_i)^2}
\nonumber \\[.2em]
&+ {\chi^2(\Mh) + \chi^2(\br(B_s \to \mu\mu))}
\nonumber \\[.2em]
&+ {\chi^2(\mbox{SUSY search limits})}
\nonumber \\[.2em]
&+ \sum^M_i \frac{(f^{\rm obs}_{{\rm SM}_i}
              - f^{\rm fit}_{{\rm SM}_i})^2}{\sigma(f_{{\rm SM}_i})^2}
\label{eqn:chi2}
\end{align} 
Here $N$ is the number of observables studied, $C_i$ represents an
experimentally measured value (constraint) and each $P_i$ defines a
prediction for the corresponding constraint that depends on the
supersymmetric parameters.
The experimental uncertainty, $\sigma(C_i)$, of each measurement is
taken to be both statistically and systematically independent of the
corresponding theoretical uncertainty, $\sigma(P_i)$, in its
prediction. We denote by
$\chi^2(\Mh)$ and $\chi^2(\br(B_s \to \mu\mu))$ the $\chi^2$
contributions from the two measurements for which only one-sided
bounds are available so far, as discussed below.
Furthermore we include the lower limits from the direct searches
for SUSY particles at LEP~\cite{LEPSUSY} as one-sided limits, denoted by 
``$\chi^2(\mbox{SUSY search limits})$'' in eq.~(\ref{eqn:chi2}).

We stress that, as in~\cite{Master2,Master1},
the three standard model parameters
$f_{\rm SM} = \{\Delta\alpha_{\rm had}, \mt, \MZ \}$ are included as fit
parameters and allowed to vary with their current experimental
resolutions $\sigma(f_{\rm SM})$. We do not
include $\alpha_s$ as a fit parameter, 
which would have only a minor impact on the analysis.

Formulating the fit in this fashion has the advantage that the
$\chi^2$ probability, $P(\chi^2, N_{\rm dof})$,
properly accounts for the number of degrees of freedom, $N_{\rm dof}$,
in the fit and thus represents a quantitative and meaningful measure for
the ``goodness-of-fit.'' In previous studies \cite{Master1},
$P(\chi^2, N_{\rm dof})$ has been verified to have a flat distribution,
thus yielding a reliable estimate of the confidence level for any particular
point in parameter space.
Further, an important aspect of the formulation is that all model parameters are
varied simultaneously in the MCMC sampling, and care is exercised to
fully explore the
multi-dimensional space, including possible interdependencies between
parameters.  All confidence levels for selected model parameters are performed 
by scanning over the desired parameters while 
minimizing the $\chi^2$ function with respect to all other model parameters. 
That is, in order to determine the function $\chi^2(x)$ for some model parameter $x$, 
all the remaining free parameters are set to values 
corresponding to a new $\chi^2$ minimum determined for fixed $x$. 
The function values where $\chi^2(x)$ is found to be equal to $\chi^2_{min}+ \Delta \chi^2$ 
determine the confidence level contour. For two-dimensional parameter scans we 
use $\Delta \chi^2 =2.28 (5.99)$ to determine the 68\%(95\%) confidence level contours.

Only experimental constraints are imposed when deriving confidence level
contours, without any arbitrary or direct constraints placed on model
parameters themselves.%
\footnote{For reasons of stability of higher-order contributions, 
we limit the range of $\tb$ to values below $\tb = 60$.
As explained in Section~\ref{sec:summary} below, we furthermore impose a
cut on parameter regions where the higher-order corrections relating the
running mass to the on-shell mass of the pseudo-scalar Higgs boson get
unacceptably large.}
This leads to robust and statistically meaningful
estimates of the total 68\% and 95\% confidence levels,
which may be composed of multiple separated contours.
Finally, the sensitivity of the global fit to different constraint
scenarios can be studied by removing one of the
experimental constraints or by rescaling one of the experimental
uncertainties, as discussed in Sect.~3 in \cite{Master2}.
Studies of such scenarios are particularly helpful in
identifying which experimental data are most useful in constraining the
theoretical model and hence in precisely studying how hyper-volumes in
parameter space become more tightly constrained (either now or in
the future).

Since each new scenario in which a parameter is removed or an
uncertainty re-scaled represents, fundamentally, a new $\chi^2$ function
which must be minimized, multiple re-samplings of the full
multi-dimensional parameter space are, in principle, required to
determine the most probable fit regions for each scenario. However,
these would be
computationally too expensive. To avoid this difficulty, we exploit
the fact that independent $\chi^2$ functions are additive and result in
a well defined $\chi^2$ probability.  Hence, we define ``loose'' $\chi^2$
functions, $\chi^2_{\rm loose}$, in which the term
representing some constraint, e.g., $\Omega_{\rm CDM}$, is removed from the
global $\chi^2$ function.   The $\chi^2_{\rm loose}$ function represents the
likelihood that a particular set of model parameter values is compatible
with a sub-set of the experimental data constraints, without any
experimental knowledge of the removed constraint.

An exhaustive, and computationally expensive, 25 million point
pre-sampling of the $\chi^2_{\rm loose}$ function 
is then performed in the full multi-dimensional model parameter space 
using a MCMC.
Constraint terms representing the various experimental scenarios are
then re-instated or removed to form different $\chi^2$ functions, one for each scenario studied.
If the scenario requires an additional constraint to be removed from
the $\chi^2_{\rm loose}$ function, the density of points pre-sampled for 
the $\chi^2_{\rm loose}$ function was carefully tested and verified to also 
be an unbiased and sufficiently complete sampling of the studied model parameter 
space for the full $\chi^2$ function by using dedicated MCMC samples of 
approximately one million sampling points each, where the particular constraint 
in question was removed. Specifically,
we use this technique to study the effects of removing individually
the \gmt, \bsg, \Och\ and $\Mh$ constraints.
The precise values of the most probable fit parameters are determined via a
full MINUIT minimization of the $\chi^2$ for each different scenario,
but are performed only within the general parameter space regions
not already excluded from the pre-sampling of the $\chi^2_{\rm loose}$
function.  An MCMC final sampling is subsequently
used to determine the 68\% and 95\% confidence-level contours for each
constraint scenario studied~%
\footnote{We note that for parameter space regions having low
probability density, statistical fluctuations can appear in the form of an
``archipelago of islands'' near the 95\% confidence levels.
Such statistical fluctuations simply reflect the lower
MCMC sampling density in regions of low probability.}.

For example, in~\cite{Master2} we showed that the effect of dropping
the $\Omega_{\rm CDM}$ experimental data from the fit is not very
important in constraining the allowed regions in the $(m_{1/2}, m_0)$
and $(m_0, \tb)$~planes.  The reason for this can be understood by
recalling that the WMAP strips in the CMSSM $(m_{1/2},
m_0)$ planes found for different, but {\em fixed}, values of $\tan
\beta$ move around as this and other CMSSM parameters are
varied.  Indeed, for fixed $A_0$, the strips can
be shown to nearly foliate the $(m_{1/2}, m_0)$
plane~\cite{eoss,Battaglia:2003ab}.
Since $\tb$ is only weakly constrained by the experimental data
but gets correlated through the fit to the other parameters
($m_{1/2}$, $m_0$, $A_0$),
the effect of the $\Omega_{\rm CDM}$ constraint is to reduce the
dimensionality of the allowed parameter-space to a certain
``hyper-sheet'' which, when viewed by fixing $\tb$ to a
particular value (i.e.\ slicing the sheet along the $\tb$-axis),
reduces to the observed strips in the $(m_{1/2}, m_0)$ planes. However,
since this sheet is generally embedded in the full parameter space
hyper-volume and is not diagonalized along some particular parameter
axis, a large range of values for $(m_0, m_{1/2}, A_0, \tb)$
remain statistically probable when considering the global fit and, from
a strict statistical consideration, there are no strips of preferred
regions.

When we apply here a similar analysis to the \gmt\ constraint, we find
a very different picture. We exhibited already in~\cite{Master2}
the effect of relaxing this constraint by some fraction, showing
that the preferred areas of the $(m_{1/2}, m_0)$
and $(m_0, \tb)$~planes changed substantially. Here we
illustrate the effect of removing the \gmt\ constraint entirely,
which relaxes very considerably the upper limits on sparticle masses.
However, the other observables still disfavour
very large values of $m_0$ and $m_{1/2}$ by $\Delta \chi^2 \sim 2$,
as we discuss below.

%---------------------------------------------------------------------
\section{Summaries of the CMSSM and NUHM1 Analyses}
\label{sec:summary}

The experimental constraints used in our analyses are listed in
Table~\ref{tab:constraints}. The notations for the observables are
standard, and were defined in~\cite{Master1,Master2}. Their
values generally have only minor updates from the values quoted there, but
one important comment concerns our implementation of the LEP
constraint on $\Mh$. The value quoted in the Table was derived within
the SM, and is applicable to the CMSSM, in which
the relevant Higgs couplings are very similar to those in the 
SM~\cite{Ellis:2001qv,Ambrosanio:2001xb}, so that the SM
exclusion results can be used, supplemented with an additional theoretical
uncertainty whose implementation we now describe.

%%%%%%%%%%%%%%%%%%%%%% T A B L E %%%%%%%%%%%%%%%%%%%%%%%%%%%%%%%%%%%%%%%%%
\begin{table*}[htb!]
\renewcommand{\arraystretch}{1.25}
\begin{center}
\begin{tabular}{|c|c|c|c|c|} \hline
Observable & Th.\ Source & Ex.\ Source & Constraint & Add.\ Th.\ Unc. \\
\hline \hline
$\mt$ [GeV] & \cite{Heinemeyer:2006px,Heinemeyer:2007bw}
            & \cite{mt1731} & $173.1 \pm 1.3$ & -- \\
\hline
$\Delta\alpha_{\rm had}^{(5)}(m_{\rm Z})$ 
     &\cite{Heinemeyer:2006px,Heinemeyer:2007bw} 
     &\cite{ADLOS:2005em} &$0.02758\pm0.00035$ & -- \\
\hline
$\MZ$ [GeV]         
     &\cite{Heinemeyer:2006px,Heinemeyer:2007bw}   
     &\cite{ADLOS:2005em} &$91.1875\pm0.0021$  & -- \\ 
\hline\hline
$ \Gamma_{Z}$ [GeV]    
     &\cite{Heinemeyer:2006px,Heinemeyer:2007bw}   
     &\cite{ADLOS:2005em} &$2.4952\pm0.0023$   & 0.001  \\ 
\hline
$\sigma_{\rm had}^{0}$ [nb] 
     &\cite{Heinemeyer:2006px,Heinemeyer:2007bw} 
     &\cite{ADLOS:2005em} &$41.540\pm0.037$    & -- \\
\hline
$R_l$ &\cite{Heinemeyer:2006px,Heinemeyer:2007bw}   
      &\cite{ADLOS:2005em} &$20.767\pm0.025$    & -- \\ 
\hline
$ A_{\rm fb}(\ell)$ &\cite{Heinemeyer:2006px,Heinemeyer:2007bw}   
                   &\cite{ADLOS:2005em} &$0.01714\pm0.00095$ & -- \\ 
\hline
$ A_{\ell}(P_\tau)$ &\cite{Heinemeyer:2006px,Heinemeyer:2007bw}   
                  &\cite{ADLOS:2005em} & 0.1465 $\pm$ 0.0032 & -- \\ 
\hline
$ R_{\rm b}$ &\cite{Heinemeyer:2006px,Heinemeyer:2007bw}   
            &\cite{ADLOS:2005em} & 0.21629 $\pm$ 0.00066 & -- \\ 
\hline
$ R_{\rm c}$ &\cite{Heinemeyer:2006px,Heinemeyer:2007bw}   
            &\cite{ADLOS:2005em} & 0.1721 $\pm$ 0.003 & -- \\ 
\hline
$ A_{\rm fb}({b})$ &\cite{Heinemeyer:2006px,Heinemeyer:2007bw}   
                  &\cite{ADLOS:2005em} & 0.0992 $\pm$ 0.0016 & -- \\ 
\hline
$ A_{\rm fb}({c})$ &\cite{Heinemeyer:2006px,Heinemeyer:2007bw}   
                  &\cite{ADLOS:2005em} & 0.0707 $\pm$ 0.0035 & -- \\ 
\hline
$ A_{b}$  &\cite{Heinemeyer:2006px,Heinemeyer:2007bw}   
         &\cite{ADLOS:2005em} & 0.923 $\pm$ 0.020 & -- \\ 
\hline
$ A_{c}$ &\cite{Heinemeyer:2006px,Heinemeyer:2007bw}   
         &\cite{ADLOS:2005em} & 0.670 $\pm$ 0.027 & -- \\ 
\hline
$ A_\ell({\rm SLD})$ &\cite{Heinemeyer:2006px,Heinemeyer:2007bw}   
                     &\cite{ADLOS:2005em} & 0.1513 $\pm$ 0.0021 & -- \\ 
\hline
$ \sin^2 \theta_{\rm w}^{\ell}(Q_{\rm fb})$ 
        &\cite{Heinemeyer:2006px,Heinemeyer:2007bw}   
        &\cite{ADLOS:2005em} & 0.2324 $\pm$ 0.0012 & -- \\ 
\hline
$\MW$ [GeV]
     &\cite{Heinemeyer:2006px,Heinemeyer:2007bw}
     &\cite{verzocchi,lepewwg} & $80.399 \pm 0.025$  & 0.010 \\
\hline\hline
BR$_{\rm b \to s \gamma}^{\rm exp}/ {\rm BR}_{\rm b \to s \gamma}^{\rm SM} $
     &\cite{Misiak:2006zs,Ciuchini:1998xy,Degrassi:2000qf,Carena:2000uj,D'Ambrosio:2002ex}
     &\cite{hfag}
     &$1.117 \pm 0.076_{\rm exp} \pm 0.082_{\rm th(SM)}$  & 0.050 \\
\hline
BR$(B_{s} \to \mu^{+} \mu^{-})$
     &\cite{Isidori:2001fv,BCRS,Isidori:2006pk,Isidori:2007jw} &\cite{hfag}
     &$ < 4.7 \times 10^{-8}$   & $0.02\times10^{-8}$ \\
\hline
\hline
BR$_{\rm B \to \tau\nu}^{\rm exp}/ {\rm BR}_{\rm B \to \tau\nu}^{\rm SM} $
     &\cite{Isidori:2006pk,Isidori:2007jw,Akeroyd:2003zr} 
     &\cite{CKMbook,Latticefb,notebtaun}
     &  $1.25 \pm 0.40_{\rm [exp+th]}$  & -- \\
\hline
${\rm BR}({B_d \to \mu^+ \mu^-})$
     & \cite{Isidori:2001fv,BCRS,Isidori:2006pk,Isidori:2007jw} &\cite{hfag}
     & $ <2.3 \times 10^{-8}$ & $0.01 \times 10^{-9}$\\
\hline
${\rm BR}_{B \to X_s \ell \ell}^{\rm exp}/{\rm BR}_{B \to X_s \ell  
\ell}^{\rm SM}$
     & \cite{Bobeth:2004jz}&\cite{hfag,Huber:2005ig}
     & $0.99 \pm 0.32$ & -- \\
\hline
BR$_{K \to \mu \nu}^{\rm exp}/{\rm BR}_{K \to \mu \nu}^{\rm SM}$
     & \cite{Isidori:2006pk,Akeroyd:2003zr} &\cite{Antonelli:2008jg}
     & $1.008 \pm 0.014_{\rm [exp+th]}$   & -- \\
\hline
BR$_{K \to \pi \nu \bar{\nu}}^{\rm exp}/{\rm BR}_{K \to \pi \nu \bar 
{\nu}}^{\rm SM}$
     & \cite{Buras:2000qz}&\cite{Artamonov:2008qb}
     & $ < 4.5 $ & -- \\
\hline
$\Delta M_{B_s}^{\rm exp}/\Delta M_{B_s}^{\rm SM}$
     & \cite{Buras:2000qz} &\cite{Bona:2007vi,Lubicz:2008am}
     & $0.97 \pm 0.01_{\rm exp} \pm 0.27_{\rm th(SM)}$ & -- \\
\hline
$\frac{(\Delta M_{B_s}^{\rm exp}/\Delta M_{B_s}^{\rm SM})}{
(\Delta M_{B_d}^{\rm exp}/\Delta M_{B_d}^{\rm SM})}$
     & \cite{Isidori:2001fv,BCRS,Isidori:2006pk,Isidori:2007jw}
     &\cite{hfag,Bona:2007vi,Lubicz:2008am}
     & $1.00 \pm 0.01_{\rm exp} \pm 0.13_{\rm th(SM)} $  & -- \\
\hline
$\Delta \epsilon_K^{\rm exp}/\Delta \epsilon_K^{\rm SM}$
     & \cite{Buras:2000qz} &\cite{Bona:2007vi,Lubicz:2008am}
     & $1.08 \pm 0.14_{\rm [exp+th]}$ & -- \\
\hline\hline
$ a_{\mu}^{\rm exp} - a_{\mu}^{\rm SM}$
     &\cite{Moroi:1995yh,Degrassi:1998es,Heinemeyer:2003dq,Heinemeyer:2004yq}
     &\cite{Bennett:2006fi,Davier:2007ua,Hertzog:2007hz}
                                         &$(30.2 \pm 8.8)\times10^{-10}$
                                         &$2.0\times10^{-10}$ \\
\hline
$\Mh$ [GeV]
     & \cite{Degrassi:2002fi,Heinemeyer:1998np,Heinemeyer:1998yj,Frank:2006yh}
     & \cite{Barate:2003sz,Schael:2006cr} & $> 114.4$ (see text) &  1.5  \\
\hline
$\Omega_{\rm CDM} h^2$
     &\cite{Belanger:2006is,Belanger:2001fz,Belanger:2004yn} 
     &\cite{Dunkley:2008ie}
         &$0.1099 \pm 0.0062$ & 0.012 \\
\hline\hline
\end{tabular}
\caption{\it List of experimental constraints used in this work.
  The values and errors shown are the current best understanding of these
  constraints. The rightmost column displays additional theoretical
  uncertainties taken into account when implementing these constraints
  in the MSSM. We have furthermore taken into account the direct
    searches for SUSY particles at LEP~\cite{LEPSUSY}.
  \label{tab:constraints}}  
\end{center}
\end{table*}
%%%%%%%%%%%%%%%%%%%%%% T A B L E %%%%%%%%%%%%%%%%%%%%%%%%%%%%%%%%%%%%%%%%%

We evaluate the $\chi^2(\Mh)$ contribution within the CMSSM using the
formula
\begin{align}
\chi^2(\Mh) = \frac{(\Mh - \Mh^{\rm limit})^2}{(1.1 \gev)^2 + (1.5 \gev)^2}~,
\label{chi2Mh}
\end{align}
with $\Mh^{\rm limit} = 115.0 \gev$ 
for $\Mh < 115.0 \gev$~\footnote{We use $115.0 \gev$ so as to incorporate
a conservative consideration of experimental systematic effects.}. 
Larger masses do not receive a $\chi^2(\Mh)$ contribution. 
The $1.5 \gev$ in the denominator corresponds to a convolution of
 the likelihood function with a Gaussian function, $\tilde\Phi_{1.5}(x)$,
normalized to unity and centered around $\Mh$, whose width is $1.5 \gev$,
representing the theory uncertainty on $\Mh$~\cite{Degrassi:2002fi}.
In this way, a theoretical uncertainty of up to $3 \gev$ is assigned for 
$\sim 95\%$ of all $\Mh$ values corresponding to one CMSSM parameter point. 
The $1.1 \gev$ term in the denominator corresponds to a parametrization
of the $CL_s$ curve given in the final SM LEP Higgs
result~\cite{Barate:2003sz}. 

Within the NUHM1 the situation is somewhat more involved, since, for
instance, a strong suppression of the $ZZh$ coupling can occur,
invalidating the SM exclusion bounds. 
In order to find a more reliable 95\% C.L.\ exclusion limit for $\Mh$ in the
case that the SM limit cannot be applied, we use the following procedure.
The main exclusion bound from LEP searches comes from the channel
$e^+e^- \to ZH, H \to b \bar b$. The Higgs boson mass limit in this
channel is given as a function of the $ZZH$ coupling in~\cite{Schael:2006cr}. 
A reduction in the $ZZh$ coupling in the
NUHM1 relative to its SM value can be translated into a lower
limit on the lightest NUHM1 Higgs mass, $\Mh^{{\rm limit},0}$, shifted
to lower values with respect to the SM limit of $114.4 \gev$. (The
actual number is obtained using the code {\tt HiggsBounds}~\cite{higgsbounds}
that incorporates the LEP (and Tevatron) limits on neutral Higgs boson
searches.) 
For values of $\Mh \lsim 86 \gev$ the reduction of the $ZZh$
  couplings required to evade the LEP bounds becomes very strong, and we
add a brick-wall contribution to the $\chi^2$ function below this value
(which has no influence on our results).
Finally, eq.~(\ref{chi2Mh}) is used with 
$\Mh^{\rm limit} = \Mh^{{\rm limit},0} + 0.6 \gev$ to ensure a smooth
transition to the SM case, where we use $\Mh^{\rm limit} = 115.0 \gev$
to allow for experimental systematics, as discussed above.
This is a conservative approach in the sense that the 
$1.1 \gev$ term used in eq.~(\ref{chi2Mh}) can be regarded as a lower
limit on the spread of the $CL_s$ curve in the vicinity of 
$\Mh^{{\rm limit},0}$.

The numerical evaluation of the frequentist likelihood function
using these constraints has been performed with the 
{\tt MasterCode}~\cite{Master1,Master2},
which includes the following theoretical codes. For the RGE running of
the soft SUSY-breaking parameters, it uses
{\tt SoftSUSY}~\cite{Allanach:2001kg}, which is combined consistently
with the codes used for the various low-energy observables. 
At the electroweak scale we have included various codes:
{\tt FeynHiggs}~\cite{Degrassi:2002fi,Heinemeyer:1998np,Heinemeyer:1998yj,Frank:2006yh}  
is used for the evaluation of the Higgs masses and  (optionally)
$a_\mu^{\rm SUSY}$  (see also
\cite{Moroi:1995yh,Degrassi:1998es,Heinemeyer:2003dq,Heinemeyer:2004yq})%
\footnote{We recall that the experimental value appears to differ by over
three standard deviations from the best SM calculation based on
low-energy $e^+ e^-$ 
data~\cite{Moroi:1995yh,Bennett:2006fi,Davier:2007ua,Czarnecki:2001pv,Miller:2007kk,Jegerlehner:2007xe,Passera:2008jk},
but that the discrepancy is significantly reduced if $\tau$ decay 
data are used to evaluate the SM prediction. We note that recently a new 
$\tau$~based analysis has appeared~\cite{g-2taunew}, which yields 
a $\sim 1.9\,\sigma$ deviation from the SM prediction. A new SM
prediction based on radiative-return data from 
{\it B{\tiny A}B{\tiny AR}} is in the offing.}.
For flavour-related observables we use 
{\tt SuFla}~\cite{Isidori:2006pk,Isidori:2007jw} as well as 
{\tt SuperIso}~\cite{Mahmoudi:2008tp,Eriksson:2008cx}, and
for the electroweak precision data we have included 
a code based on~\cite{Heinemeyer:2006px,Heinemeyer:2007bw}.
Finally, for dark-matter-related observables, 
{\tt MicrOMEGAs}~\cite{Belanger:2006is,Belanger:2001fz,Belanger:2004yn} and
{\tt DarkSUSY}~\cite{Gondolo:2005we,Gondolo:2004sc} 
have been used.
We made extensive use of the SUSY Les Houches
Accord~\cite{Skands:2003cj,Allanach:2008qq} 
in the combination of the various codes within the {\tt MasterCode}.

It is well known from previous
comparisons that different RGE codes for the running of the soft
SUSY-breaking parameters give quite different results 
in parameter regions where higher-order corrections get very 
large~\cite{CodeComp}.
This happens in general for very large values of $\tb$, but
instabilities can also occur in, e.g., the evaluation of $\MA$ in
the CMSSM. In such a case the evaluation of the impacts of constraints
that are affected by the heavy Higgs bosons can become unreliable.
Motivated by these observations, we made two cuts on the
pre-sampled CMSSM and NUHM1 points in deriving the results
presented below: we do not consider points with 
$\tb > 60$, and we have discarded parameter points where the 
difference between the running ($\overline{{\rm DR}}$)
mass of the pseudoscalar Higgs, $\MA(Q)$, and the physical (on-shell)
mass, $\MA$, gets unacceptably large. For the latter, we have applied 
the condition $|\sqrt{\MA^2(Q)} - \MA|/\MA > 0.6$.
Imposing these cuts has no effect on the best-fit point,
nor on the 68\% C.L.\ range of any parameter of the fit~\footnote{However, 
discarding these points does reduce the 95\% C.L.\
upper limit on $m_{1/2}$ in the NUHM1 by about 10\%, from $\sim 1000 \gev$ to
$\sim 900 \gev$. This difference may be regarded as a 
theoretical systematic uncertainty in the results.}. Motivated by \gmt\ and 
(to a lesser extent) \bsg, we restrict our study to $\mu > 0$.

For the parameters of the best-fit CMSSM point we find
$m_0 = 60 \gev$,  $m_{1/2} = 310 \gev$,  $A_0 = 130 \gev$, $\tb = 11$
and $\mu = 400 \gev$, 
yielding the overall $\chi^2/{\rm N_{\rm dof}} = 20.6/19$ (36\% probability) 
and nominally $\Mh = 114.2 \gev$%
\footnote{This is acceptable,
taking into account the theoretical uncertainty in the {\tt FeynHiggs}
calculation of $\Mh$~\cite{Degrassi:2002fi}, see the discussion above.}%
. These values are very close to the ones reported in \cite{Master2}.
The corresponding parameters of
the best-fit NUHM1 point are $m_0 = 150 \gev$, $m_{1/2} = 270 \gev$,
$A_0 = -1300 \gev$, $\tb = 11$ and
$m_{h_1}^2  = m_{h_2}^2 = - 1.2 \times 10^6 \gev^2$ or, equivalently,
$\mu = 1140 \gev$, yielding
$\chi^2 = 18.4$ (corresponding to a similar fit probability to the CMSSM)
and $\Mh = 120.7 \gev$. 
The similarity between the
best-fit values of $m_0$, $m_{1/2}$ and $\tb$ in the CMSSM and the NUHM1
suggest that the model frameworks used are reasonably stable: if they
had been very different, one might well have wondered what would be
the effect of introducing additional parameters, as in the
NUHM2~\cite{nuhm2,nuhm2-2} with two non-universality parameters in the
Higgs sector%
\footnote{Computationally, 
exploring adequately the NUHM2 parameter space using the frequentist approach
would be very expensive, but we hope to return to it in the future.}.

\begin{figure*}[htb!]
%%%%%%%%%%%%%%%%%%%%%%%%%%%%%%%
{\resizebox{8cm}{!}{\includegraphics{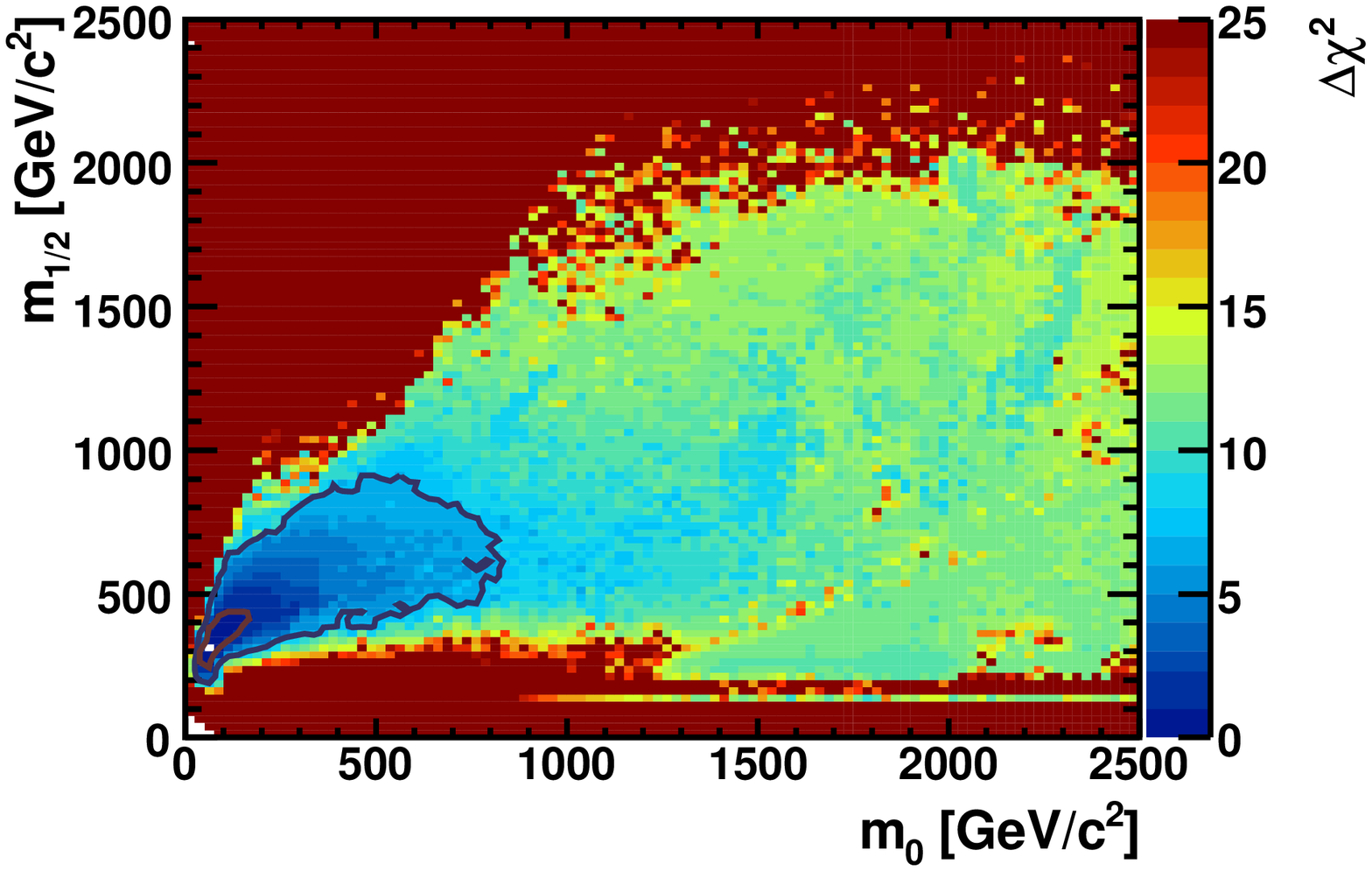}}}  
{\resizebox{8cm}{!}{\includegraphics{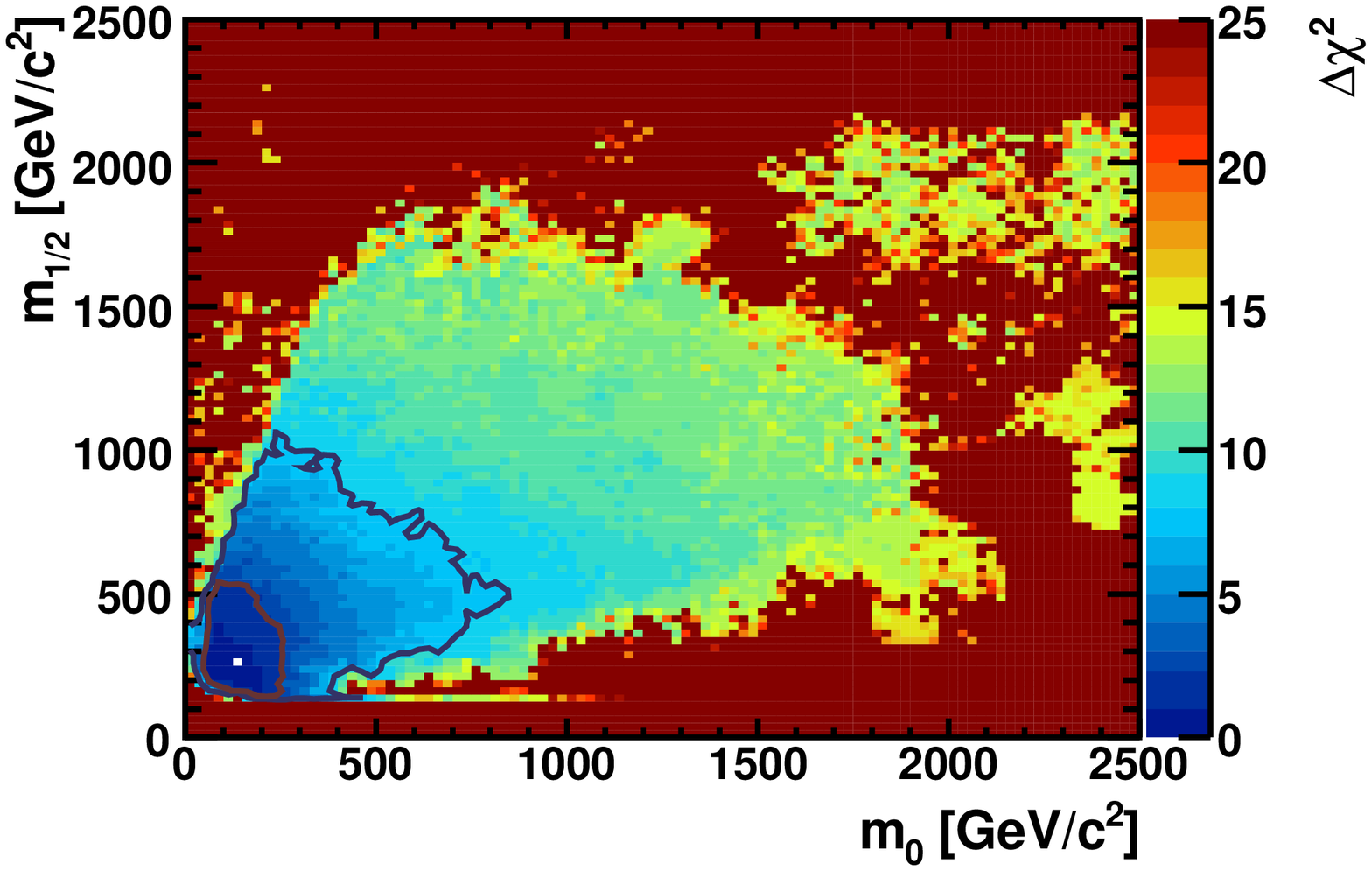}}}
%%%%%%%%%%%%%%%%%%%%%%%%%%%%%%%
\caption{\it The $\Delta\chi^2$ functions in the $(m_0, m_{1/2})$ planes for
  the CMSSM (left plot) and 
  for the NUHM1 (right plot). We see that the coannihilation regions at
  low $m_0$ and $m_{1/2}$ are favoured in both cases.
}
\label{fig:m0m12}
\end{figure*}

These best-fit points are both in the coannihilation region of the
$(m_0, m_{1/2})$ plane, as can be seen in Fig.~\ref{fig:m0m12},
which displays contours of the $\Delta\chi^2$ function in the CMSSM (left)
and the NUHM1 (right). The C.L.\ contours extend to slightly larger
values of $m_0$ in the CMSSM, while they extend to slightly larger
values of $m_{1/2}$ in the NUHM1, as was
already shown in~\cite{Master2} for
the 68\% and 95\% C.L.\ contours.
However, the qualitative features of the $\Delta\chi^2$ contours are quite
similar in the two models, indicating that the preference for small
$m_0$ and $m_{1/2}$ are quite stable and do not depend on details
of the Higgs sector. We recall that it was found in~\cite{Master2} that
the focus-point region was disfavoured at beyond the 95\% C.L. in both
the CMSSM  and the NUHM1. We see in Fig.~\ref{fig:m0m12} that this
region is disfavoured at the level $\Delta\chi^2 \sim 8$ in the CMSSM
and $> 9$ in the NUHM1. 

%%%%%%%%%%%%%%%%%%%%%% F I G U R E %%%%%%%%%%%%%%%%%%%%%%%%%%%%%%%%%%%
%%%%%%%%%%%%%%%%%%%%%% F I G U R E %%%%%%%%%%%%%%%%%%%%%%%%%%%%%%%%%%%

\begin{figure*}[htb!]
%%%%%%%%%%%%%%%%%%%%%%%%%%%%%%%
{\resizebox{8cm}{!}{\includegraphics{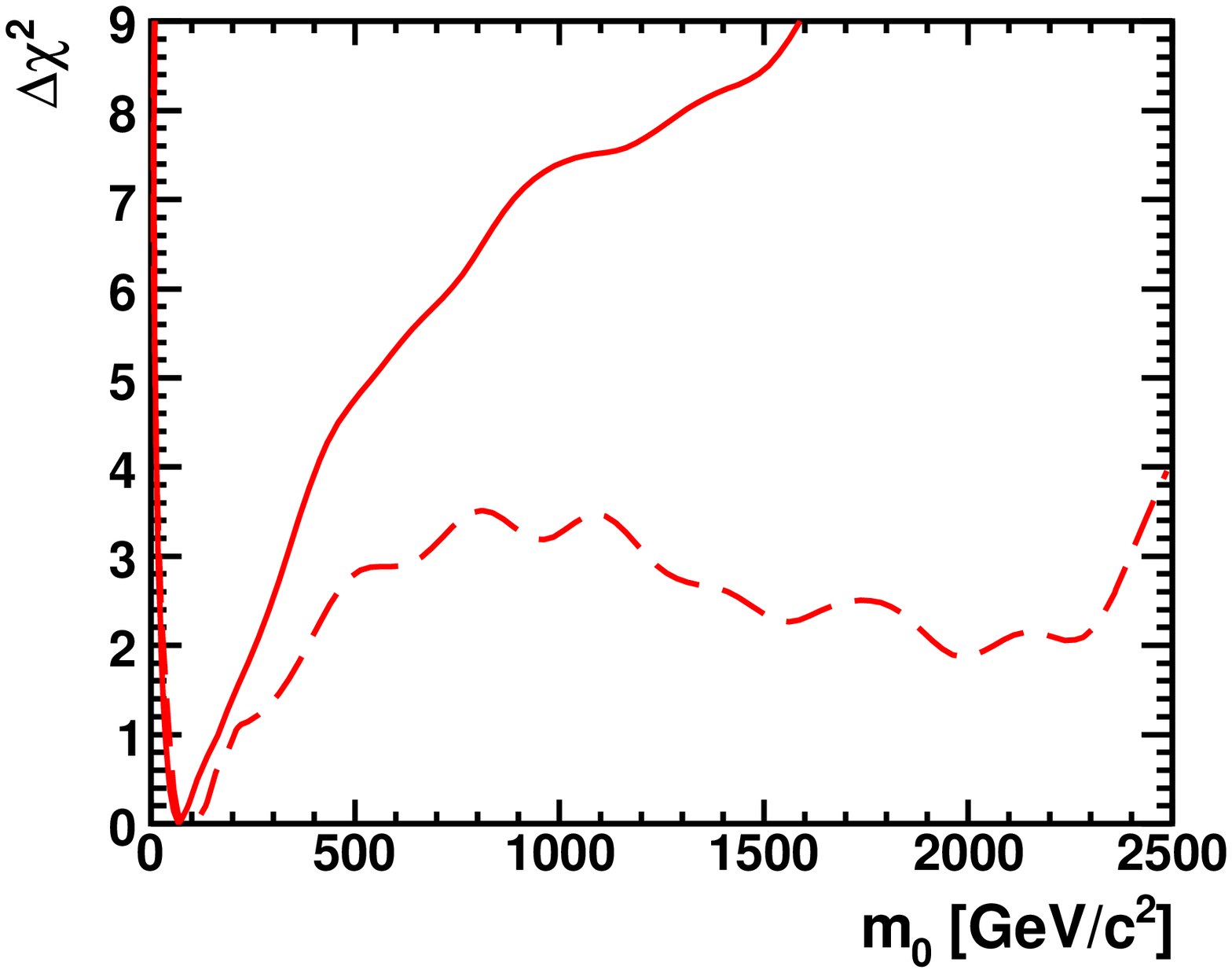}}}  
{\resizebox{8cm}{!}{\includegraphics{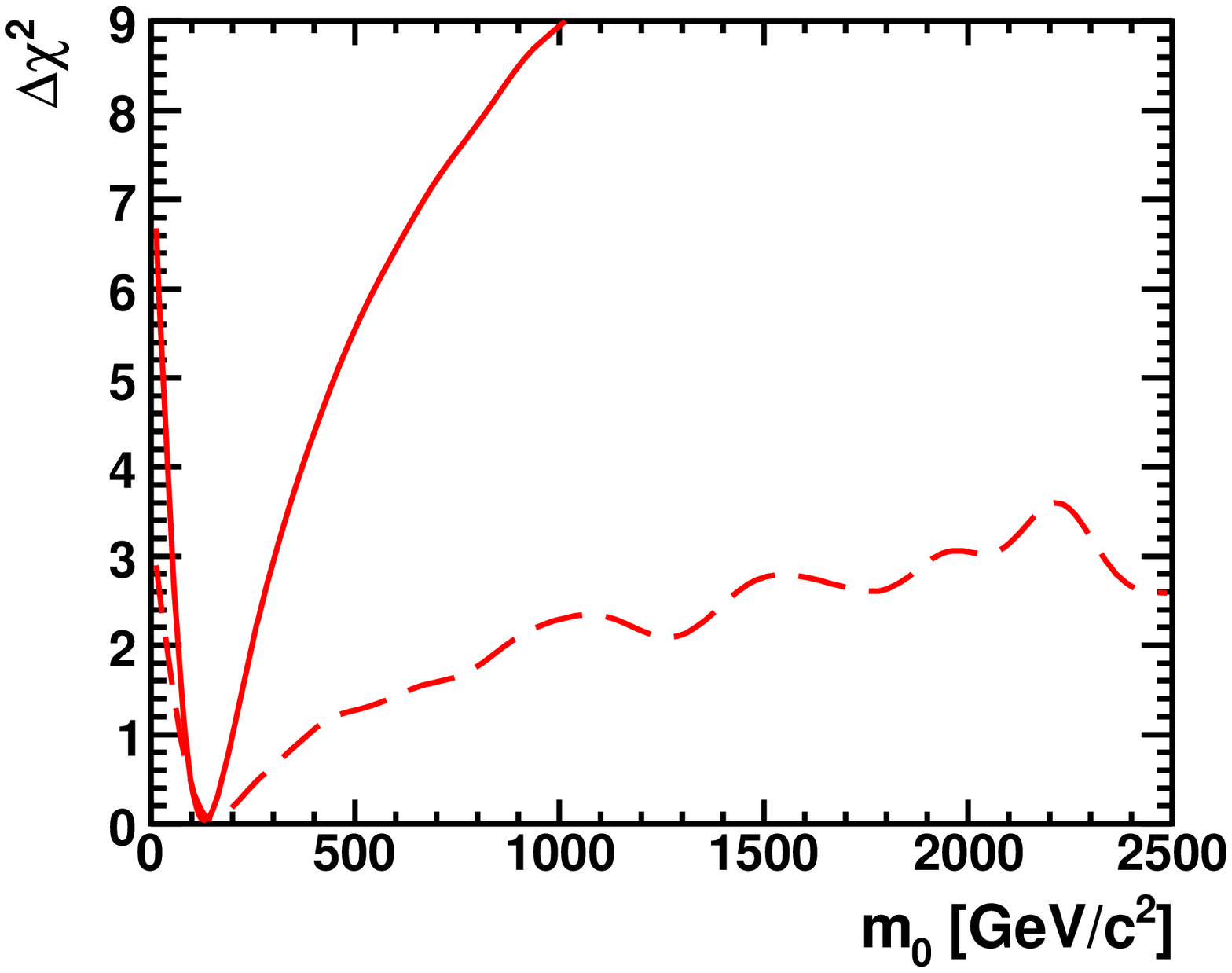}}}
%%%%%%%%%%%%%%%%%%%%%%%%%%%%%%%
\caption{\it The likelihood functions for $m_0$ in the CMSSM (left plot) and
  in the NUHM1 (right plot). The $\chi^2$ values are shown including
  (excluding) the \gmt\ constraint as the solid (dashed) curves.
}
\label{fig:m0}
\end{figure*}

This feature is seen explicitly in the left and right panels of
Fig~\ref{fig:m0}, 
which display the likelihood functions for $m_0$ in the CMSSM and NUHM1,
respectively. (We recall that the focus-point region would be found at 
$m_0 \gsim 1500 \gev$.) Looking first at the solid lines corresponding
to the full 
global fit, we also see explicitly that low values of $m_0 \sim 100 \gev$ are
favoured in both cases, reflecting the fact that coannihilation points
are generally favoured.
The favoured regions also have relatively low values of 
$m_{1/2}$, as seen in Fig.~\ref{fig:m0m12}. As we discuss in more detail
later, the minimum in both cases is found at low $\tb \sim 11$.

%%%%%%%%%%%%%%%%%%%%%% F I G U R E %%%%%%%%%%%%%%%%%%%%%%%%%%%%%%%%%%%
%%%%%%%%%%%%%%%%%%%%%% F I G U R E %%%%%%%%%%%%%%%%%%%%%%%%%%%%%%%%%%%

The large values of
$\Delta \chi^2$ in the focus-point region are largely, but not entirely,
due to the \gmt\ constraint, as can be seen in the dashed
lines in Fig.~\ref{fig:m0}, where this constraint has been removed. In the
CMSSM case without \gmt, the global minimum at $m_0 \sim 100 \gev$ is
followed by a
local maximum around $m_0 \sim 1000 \gev$ with $\Delta \chi^2 \sim 3$. 
This is in turn followed by a secondary local minimum around 
$m_0 \sim 2000 \gev$ with $\Delta \chi^2 \sim 2$. The absolute minimum
occurs in the coannihilation region, and the secondary minimum occurs in
the focus-point region. The local maximum at intermediate $m_0$ reflects
the fact that such values of $m_0$ are compatible with the dark matter
constraint only at relatively large values of $\tb$ and $m_{1/2}$
that are disfavoured by other constraints. This is not the case
in the NUHM1, where intermediate values of $m_0$ with relatively low
values of $\tb$ are compatible with the \Och\ constraint (thanks to the
possible appearance of direct-channel Higgs poles), as well as the other
constraints. 
See Section~\ref{sec:drop} below for a more detailed discussion of
the impact of dropping the \gmt\ constraint.

We summarize in
Table~\ref{tab:chi2} the contributions to the global $\chi^2$ likelihood
function at the best-fit points in the CMSSM and NUHM1 due to the most
important observables as well as their total $\chi^2$. 
We also list the contributions to $\chi^2$ for
the best fit we find in the focus-point (FP) region for the CMSSM
(considered to be that with $m_0 > 1000 \gev$). This point has
$m_0 = 2550 \gev$, $m_{1/2} = 370 \gev$, $A_0 = 1730 \gev$ 
and $\tb = 51$.
It is apparent from Table~\ref{tab:chi2}
that the focus-point region is disfavoured by \gmt, but also by
$\MW$,  
and that the contributions of the other observables 
fail to overcome this disadvantage. 
Indeed, many of the other observables favour independently
the coannihilation region, e.g., \btn, $A_\ell({\rm SLD})$~%
\footnote{We note, however, that within the SM there
  is significant tension between the experimental value of 
  $A_\ell({\rm SLD})$ and $A_{\rm fb}(b)$(LEP), 
and that this tension is not reduced significantly in the CMSSM or
NUHM1, see also~\cite{Heinemeyer:2007bw}.}
and $R_\ell$ --- though the difference here is relatively small. Only 
$\Mh$ and \bsg\ and $A_{\rm fb}$(b)(LEP) favour the FP region, but not with
high significance.

%%%%%%%%%%%%%%%%%%%%%% T A B L E %%%%%%%%%%%%%%%%%%%%%%%%%%%%%%%%%%%%%%%%%
\begin{table*}[htb!]
\renewcommand{\arraystretch}{1.2}
\begin{center}
\begin{tabular}{|c||c|c|c|} \hline
Observable & Best CMSSM fit & Best NUHM1 fit &
            Best CMSSM FP fit \\ \hline \hline
\gmt       &  0.44 & 0.002 & 8.4 \\
\hline
\btn  &  0.20 & 0.41  & 0.85 \\
\hline
$\MW$      &  0.53 & 0.08  & 1.5  \\
\hline
$A_\ell({\rm SLD})$
          &  2.84 & 3.22  & 3.56 \\
\hline
$A_{\rm fb}(b)$(LEP)
          &  7.61  &  7.08 & 6.74     \\
\hline
$R_\ell $  &  0.96 & 1.01  & 1.05 \\
\hline
$ {\rm BR}_{\rm b \to s \gamma}^{\rm SUSY}/ 
  {\rm BR}_{\rm b \to s \gamma}^{\rm SM} $
          &  1.16 & 0.001 & 0.95 \\
\hline
$\Mh$      &  0.17 & 0    & 0  \\
\hline\hline
$\chi^2_{\rm tot}$
          & 20.6 & 18.5   & 29.8    \\
\hline
\end{tabular}
\caption{\it The principal contributions to the global $\chi^2$ likelihood
function from the experimental and phenomenological constraints used in
this work, as well as the total $\chi^2$, for the best-fit points in the
CMSSM and NUHM1, which both 
lie in the coannihilation region. For comparison, we also show the analogous
numbers for the best CMSSM fit we find in the focus-point (FP) region
with $m_0 > 1000 \gev$. Only those observables yielding the main
contributions to the total $\chi^2$ are listed in the table.
\label{tab:chi2}} 
\end{center}
\end{table*}
%%%%%%%%%%%%%%%%%%%%%% T A B L E %%%%%%%%%%%%%%%%%%%%%%%%%%%%%%%%%%%%%%%%%

%---------------------------------------------------------------------
\section{Likelihood Distributions for Sparticle Masses and Other Observables}
\label{sec:onedim}
%---------------------------------------------------------------------

In our previous paper~\cite{Master2} we discussed, in addition to the
spectra at the best-fit 
points in the CMSSM and NUHM1, the regions of the $(m_0, m_{1/2})$ planes
preferred in these scenarios at the 68 and 95\% C.L. Here we complement
those discussions by providing directly the likelihood functions for certain
sparticle masses, noting in particular the impacts of the most relevant
constraints%
\footnote{In each case, we show $\Delta\chi^2$, 
the difference between the total $\chi^2$ function and its value at the
minimum for the relevant model.}.

\begin{figure*}[htb!]
%%%%%%%%%%%%%%%%%%%%%%%%%%%%%%%
\resizebox{8.0cm}{!}{\includegraphics{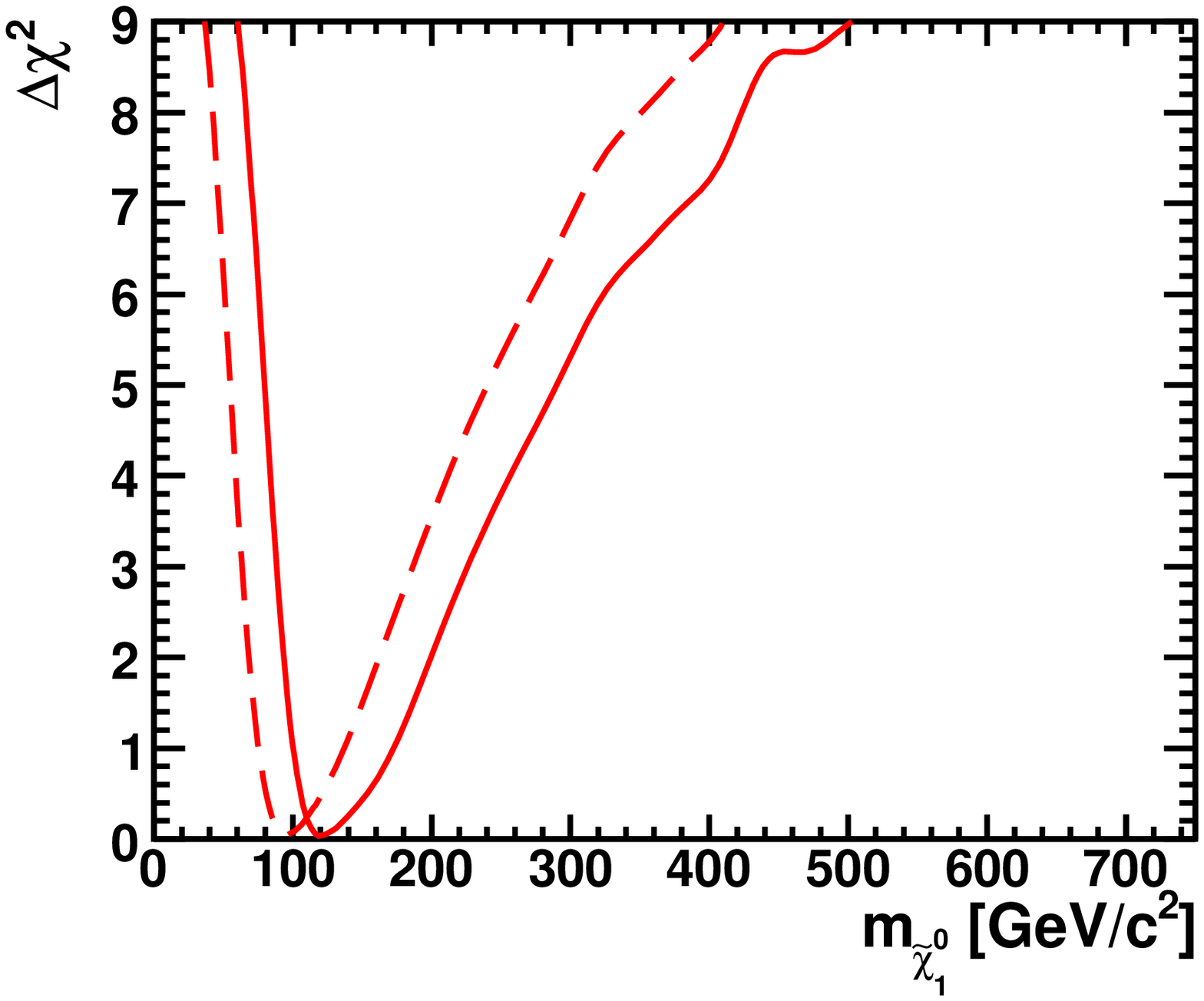}}
\resizebox{8.0cm}{!}{\includegraphics{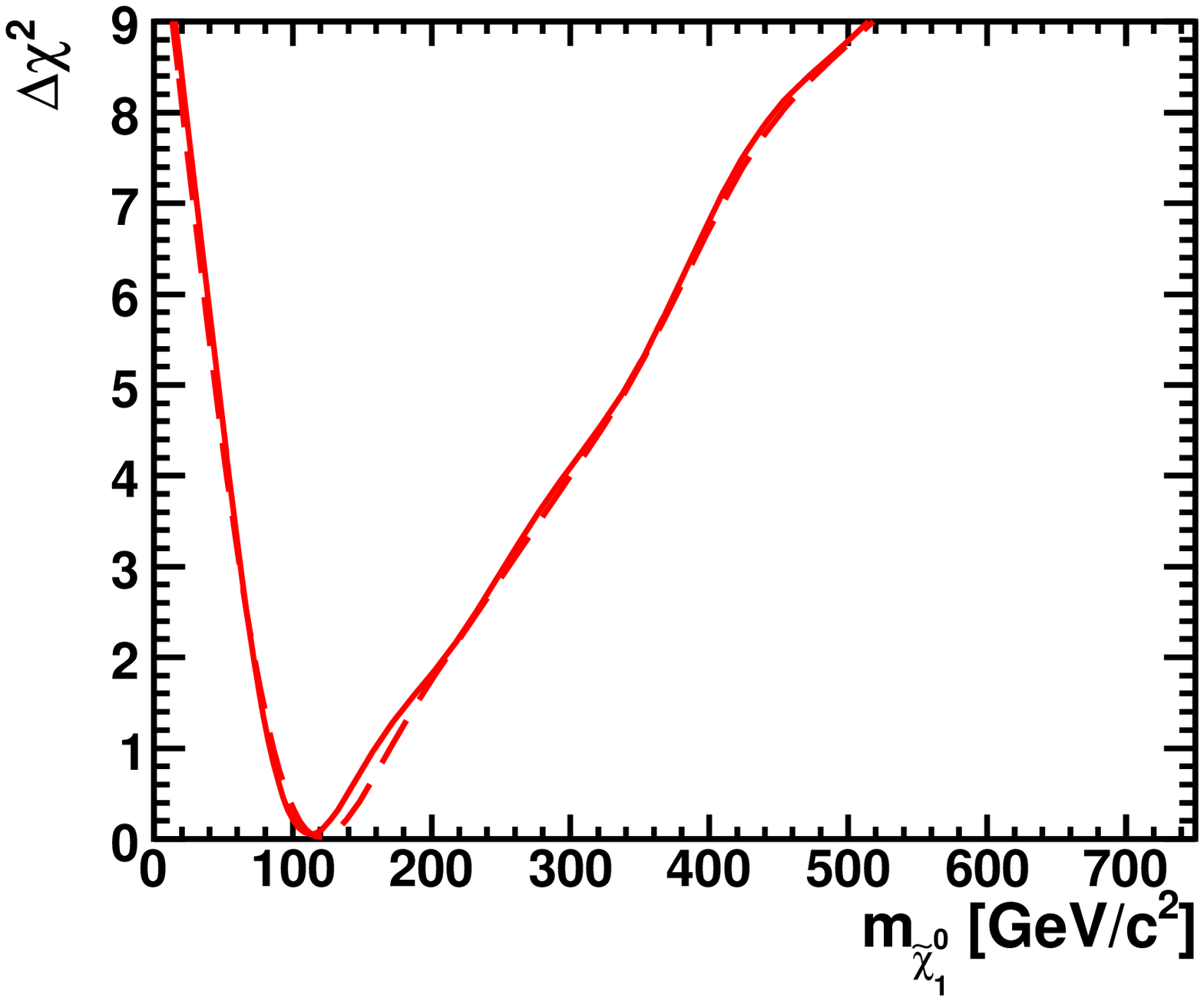}}
%%%%%%%%%%%%%%%%%%%%%%%%%%%%%%%
\vspace{-1em}
\caption{\it The likelihood functions for $\mneu{1}$ in the CMSSM (left)
and in the NUHM1 (right), both with (solid lines) and without
(dashed lines) the LEP constraint on $\Mh$.
}
\label{fig:mchi}
\end{figure*}

We start by discussing the likelihood functions for the
mass of the neutralino LSP, $\mneu{1}$, in the CMSSM and NUHM1.
The left panel of Fig.~\ref{fig:mchi} displays the likelihood function
in the CMSSM. 
The solid line shows the result obtained when
incorporating the LEP Higgs limit, while the dashed line corresponds to
the case where the LEP Higgs constraint is removed.
There is a sharp rise in the likelihood function at low values of
$\mneu{1}$, which is caused by the limits from the direct searches for
SUSY particles, but receives also contributions from
\bsg\ and other constraints. This sharp rise in the likelihood function 
persists when the LEP Higgs constraint is removed,
but is shifted towards slightly lower values of $\mneu{1}$ in that
case.
The right panel of Fig.~\ref{fig:mchi}
shows the likelihood function for $\mneu{1}$ in the NUHM1, again with
and without the LEP $\Mh$ constraint imposed.
Including the LEP $\Mh$ constraint we see that the optimal value of
$\mneu{1}$ is somewhat smaller than in the CMSSM case, reflecting the
lower value of $m_{1/2}$ at 
the corresponding best-fit point discussed in~\cite{Master2}~%
\footnote{We recall that, to a very good approximation, 
$\mneu{1} \sim 0.42\, m_{1/2}$ in most of the relevant regions of the
CMSSM and NUHM1 parameter spaces discussed here.}%
. Finally, the dashed line in the right panel of
Fig.~\ref{fig:mchi} displays  
the likelihood function in the NUHM1 with the LEP Higgs constraint removed.
Here we see very little difference from the result
for the NUHM1 with the LEP constraint imposed. This reflects the fact
that in the NUHM1 (unlike the CMSSM) the other constraints do not push
$\Mh$ down to 
quite low values, a point made explicit in Fig.~\ref{fig:mh} below.

%%%%%%%%%%%%%%%%%%%%%% F I G U R E %%%%%%%%%%%%%%%%%%%%%%%%%%%%%%%%%%%
%%%%%%%%%%%%%%%%%%%%%% F I G U R E %%%%%%%%%%%%%%%%%%%%%%%%%%%%%%%%%%%

The gradual rises in the likelihood functions at large $\mneu{1}$ in
both the CMSSM and the NUHM1 are dominated by the contribution of \gmt,
discussed already above, which disfavours large $m_{1/2}$ (and $m_0$). 
We comment later on the impacts on $\mneu{1}$ and other
observables if the \gmt\ constraint is removed.

In order to see explicitly the importance of the $\Mh$ constraint, we
display in Fig.~\ref{fig:mh} the likelihood functions for $\Mh$ in the
CMSSM (left) and the NUHM1 (right), both with (solid lines)
and without (dashed lines) the LEP constraint on $\Mh$. 
Comparing first the two CMSSM results, we see that the other
constraints would prefer a value of $\Mh$ somewhat
below the SM Higgs limit from
LEP~\cite{Barate:2003sz} 
(this was already observed in \cite{Master1}).
The best fit value for $\Mh$ is still acceptable in that case, 
in particular in view of the 
theoretical uncertainties in the CMSSM evaluation of $\Mh$, see the
discussion above.
However, in the case of the NUHM1, shown in the right plot of 
Fig.~\ref{fig:mh}, the best-fit value of
$\Mh$ indicated by the other constraints is significantly higher than
the SM LEP lower limit. As a consequence, incorporating the LEP
constraint (see above for details), shown as the solid line,
does not alter significantly the best-fit 
value of $\Mh$. As a corollary, the differences between the likelihood
functions of the NUHM1 for other masses and observables between the fits
with and without the LEP $\Mh$ constraint are less significant than for
the CMSSM. In the rest of this paper (except in
Section~\ref{sec:dropMh})
we show results with the LEP $\Mh$ constraint imposed. 

%%%%%%%%%%%%%%%%%%%%%% F I G U R E %%%%%%%%%%%%%%%%%%%%%%%%%%%%%%%%%%%
\begin{figure*}[htb!]
%%%%%%%%%%%%%%%%%%%%%%%%%%%%%%%
\resizebox{8.0cm}{!}{\includegraphics{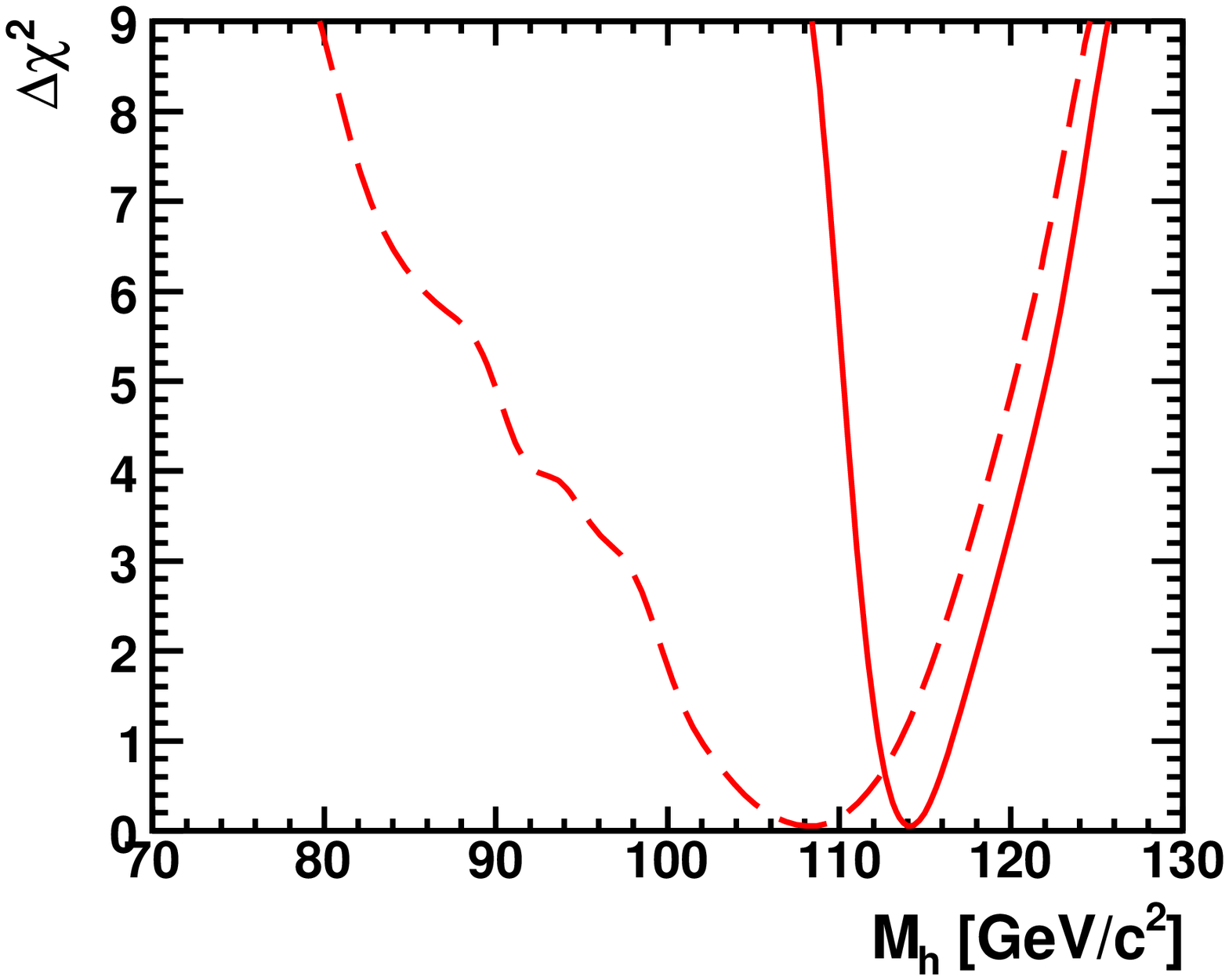}}
\resizebox{8.0cm}{!}{\includegraphics{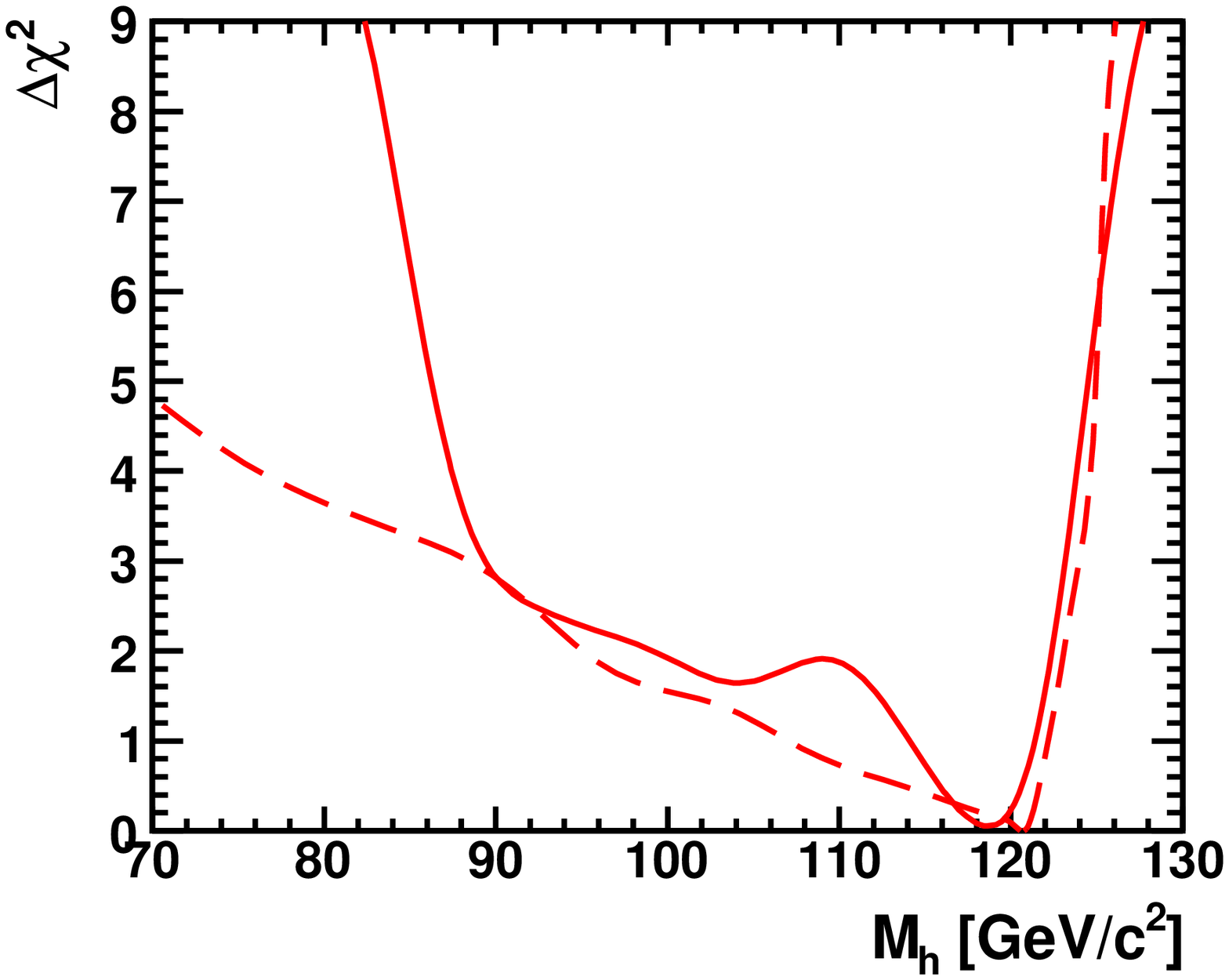}}\\
%%%%%%%%%%%%%%%%%%%%%%%%%%%%%%%
\vspace{-1em}
\caption{\it The likelihood functions for $\Mh$ in the CMSSM (left)
and in the NUHM1 (right), both with (solid lines) and without
(dashed lines) the LEP constraint on $\Mh$.
}
\label{fig:mh}
\end{figure*}
%%%%%%%%%%%%%%%%%%%%%% F I G U R E %%%%%%%%%%%%%%%%%%%%%%%%%%%%%%%%%%%

\begin{figure*}[htb!]
%%%%%%%%%%%%%%%%%%%%%%%%%%%%%%%
\resizebox{8cm}{!}{\includegraphics{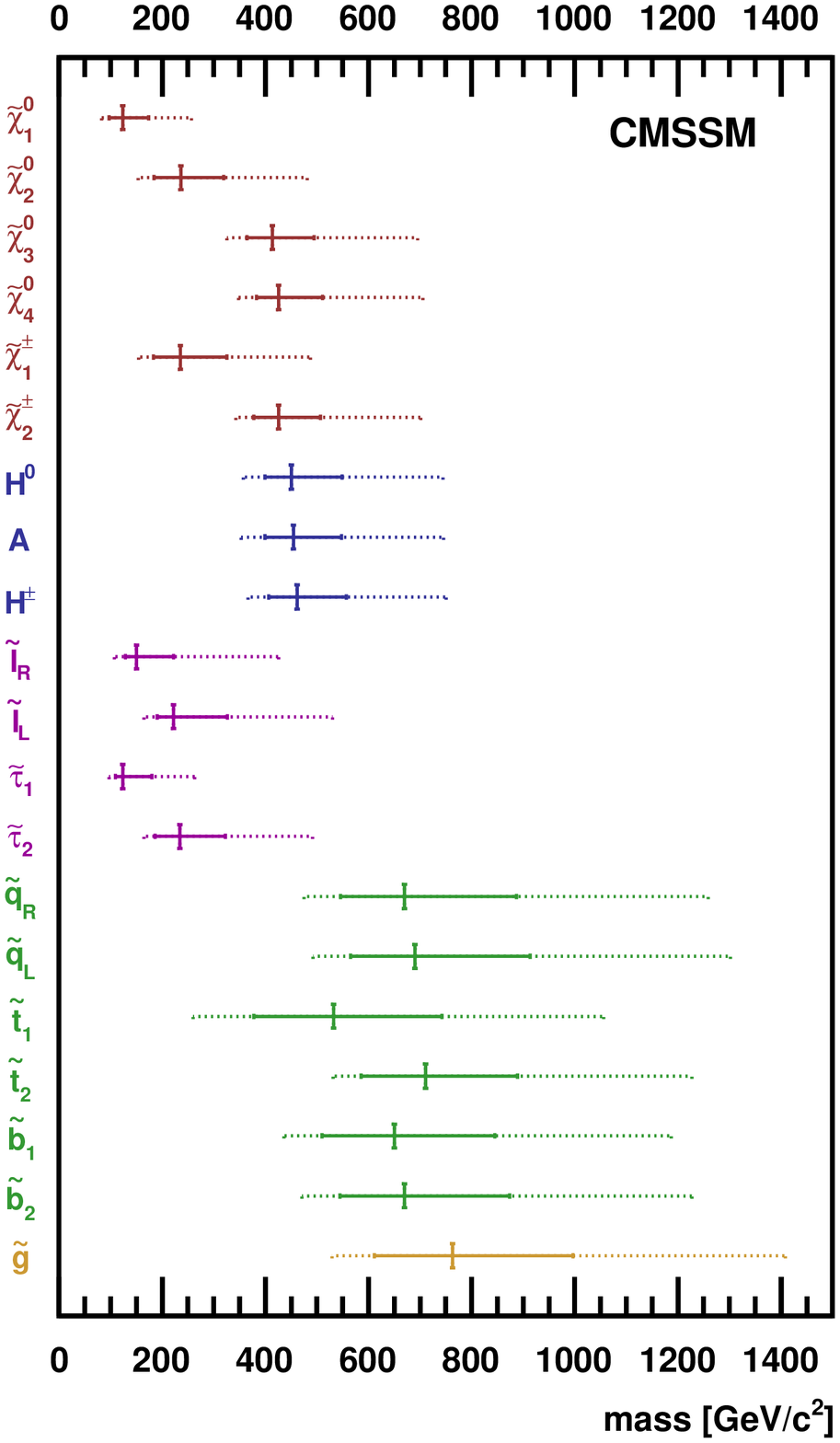}}
\resizebox{8cm}{!}{\includegraphics{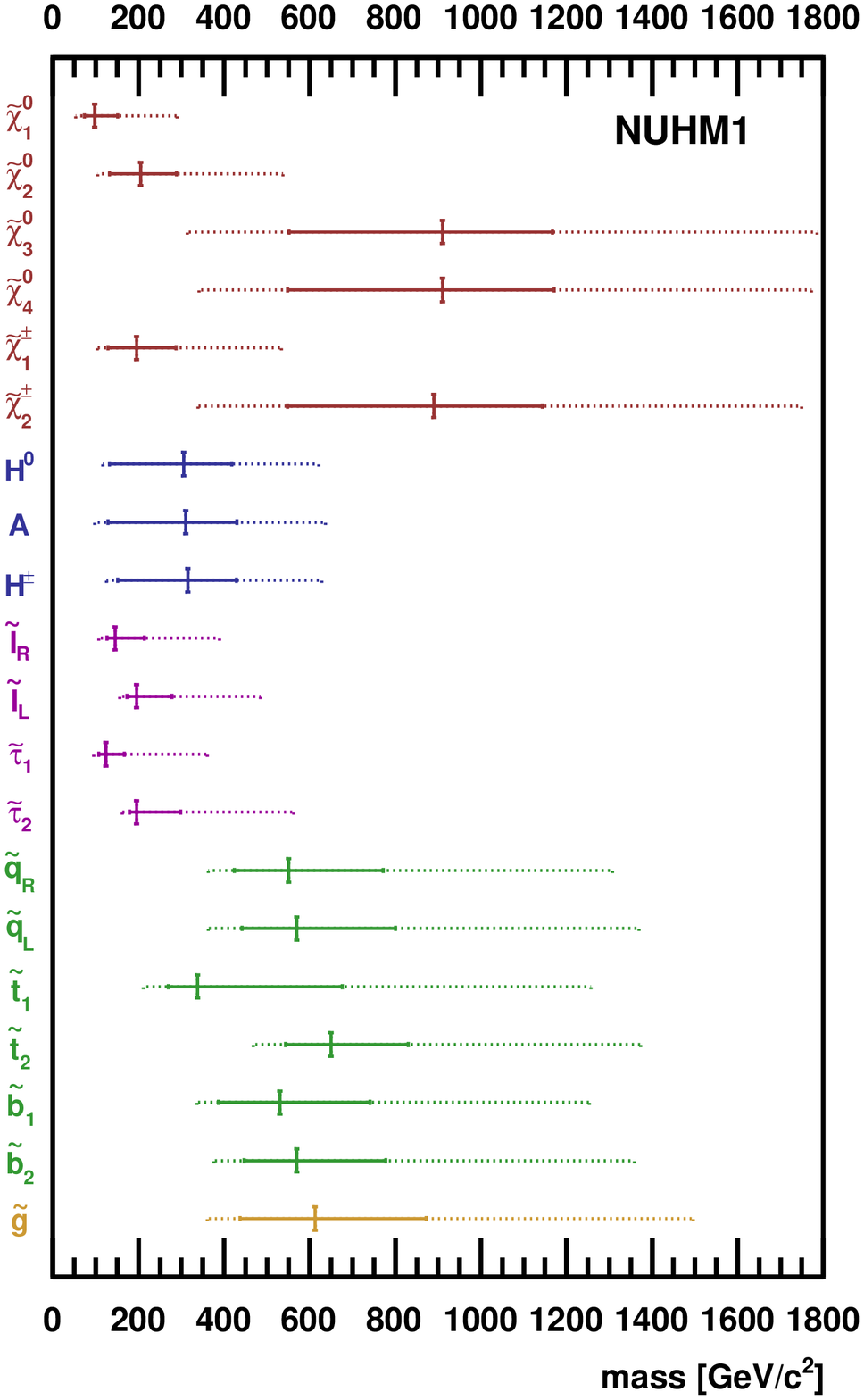}}
%%%%%%%%%%%%%%%%%%%%%%%%%%%%%%%
\caption{\it Spectra in the CMSSM (left) and the NUHM1 (right). The vertical
solid lines indicate the best-fit values, the horizontal solid lines
are the 68\% C.L.\
ranges, and the horizontal dashed lines are the 95\% C.L.\ ranges for the
indicated mass parameters. 
}
\label{fig:spectrum}
\end{figure*}

We discuss next the likelihood functions for various sparticle masses,
which are summarized in Fig.~\ref{fig:spectrum}. The results for the CMSSM
spectrum are shown in the left plot, and for the NUHM1 in the right plot. 
We start our discussion with the gluino mass, $\mgl$.
In both the CMSSM and the NUHM1, the
best-fit points have relatively low values of $\mgl \sim 750$ and $\sim
600 \gev$, respectively. These favoured values are well within the range
even of the early operations of the LHC with reduced centre-of-mass
energy and limited luminosity. However, the effect of the gradual
increase in $\chi^2$ as $m_{1/2}$ increases, due essentially to \gmt\ as
commented before, means that even quite large values of 
$\mgl \lsim 2.5 \tev$
are allowed at the 3-$\sigma$ ($\Delta \chi^2 = 9$) level 
(not shown in Fig.~\ref{fig:spectrum}). The LHC
should be  able to discover a gluino with $\mgl \sim 2.5 \tev$ with
100/fb of integrated luminosity at 
$\ecm = 14 \tev$~\cite{atlastdr,cmstdr}, and the proposed 
SLHC luminosity upgrade to 1000/fb of integrated luminosity at 
$\ecm = 14 \tev$ should permit the discovery of a gluino with 
$\mgl \sim 3 \tev$~ \cite{Gianotti:2002xx}. 
However, Fig.~\ref{fig:spectrum} does demonstrate that, whilst there are good
prospects for discovering SUSY in early LHC running~\cite{Master2}, 
this cannot be `guaranteed', even if one accepts the \gmt\ constraint.

%%%%%%%%%%%%%%%%%%%%%% F I G U R E %%%%%%%%%%%%%%%%%%%%%%%%%%%%%%%%%%%
%%%%%%%%%%%%%%%%%%%%%% F I G U R E %%%%%%%%%%%%%%%%%%%%%%%%%%%%%%%%%%%

The central values of the masses of the supersymmetric
partners of the $u, d, s, c, b$ quarks are slightly lighter than the
gluino, as seen in 
Fig.~\ref{fig:spectrum}. The difference between the gluino and the
squark masses is sensitive primarily to $m_0$.
Since the preferred regions of the
parameter space in both the CMSSM
and the NUHM1 are in the $\neu{1}$-slepton coannihilation
region where $m_0 < m_{1/2}$, $m_0$ makes only
small contributions to the central values of the
squark masses~\footnote{However, this is not true in general, as we discuss
in more detail later.}. The SUSY partners of the left-handed components of
the four lightest quarks, the ${\tilde q_L}$, 
are predicted to be slightly heavier than the corresponding right-handed
squarks, ${\tilde q_R}$, as seen by comparing the mass ranges in
Fig.~\ref{fig:spectrum}. As in the case of the gluino,
squark masses up to $\sim 2.5 \tev$ are allowed at the 3-$\sigma$
level. Comparing 
the left and right panels, we see that the squarks are predicted to be
somewhat lighter in the NUHM1 than in the CMSSM, 
but this difference is small compared
with the widths of the corresponding likelihood functions.

Turning now to the likelihood functions for the mass of the lighter stop, 
$\mste$, we find that it is shifted to values somewhat lower than for
the other squark flavours. It can also be seen
that the 2-$\sigma$ range of its likelihood function differ from those
of the gluino and the other squarks, reflecting the importance of scalar
top mixing. We recall that this depends strongly on the trilinear
soft SUSY-breaking parameter $A_t$ and the
Higgs mixing parameter $\mu$, as well as on the precise value
of $\mt$. As we discuss
below, the favoured range of values of $\mu$ is quite
circumscribed in the CMSSM, whereas a larger variation in $\mu$
is possible in the NUHM1. This has the effect of somewhat
broadening the likelihood function for $\mste$ in the NUHM1.

\begin{figure*}[htb!]
%%%%%%%%%%%%%%%%%%%%%%%%%%%%%%%
\resizebox{8cm}{!}{\includegraphics{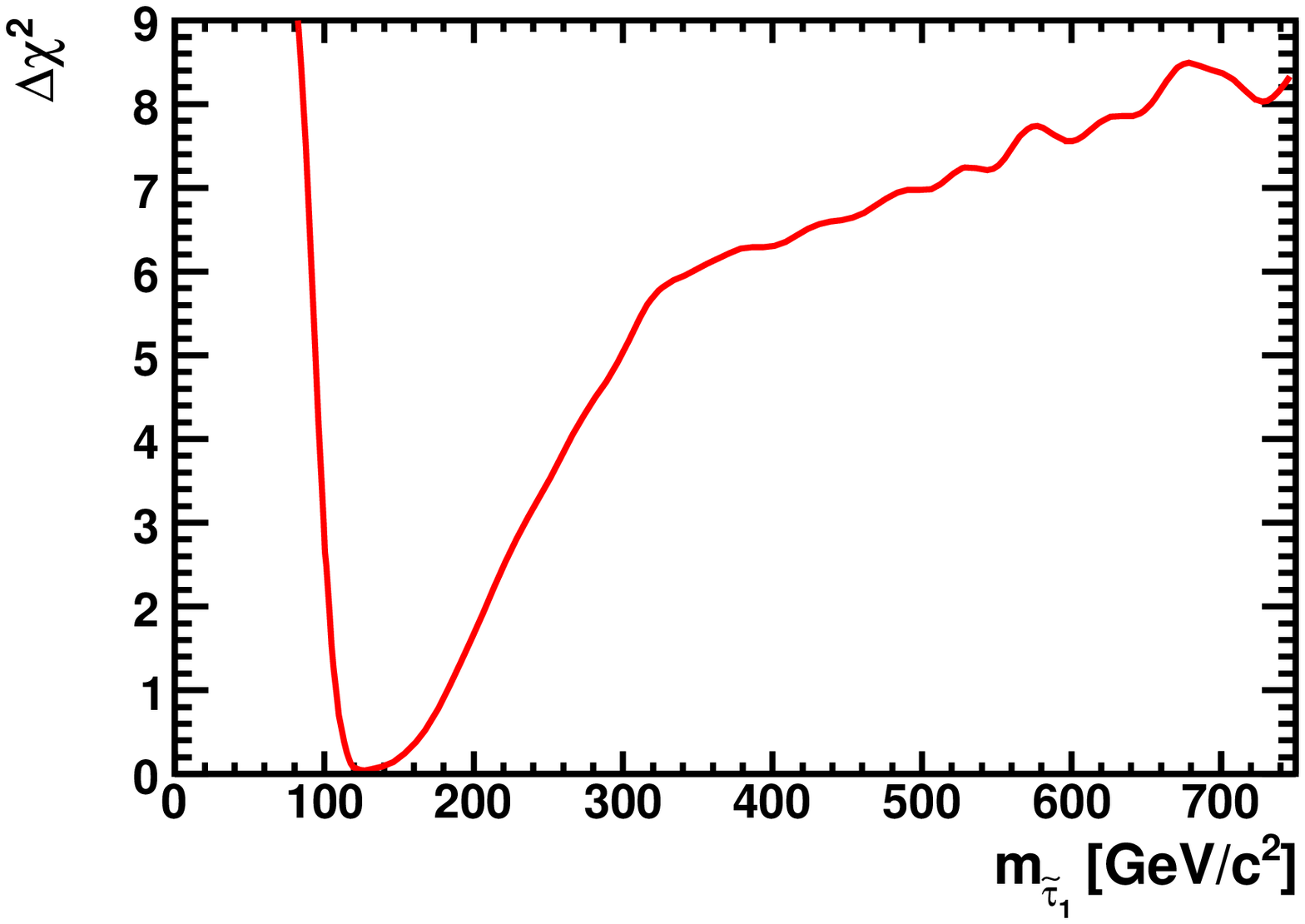}}
\resizebox{8cm}{!}{\includegraphics{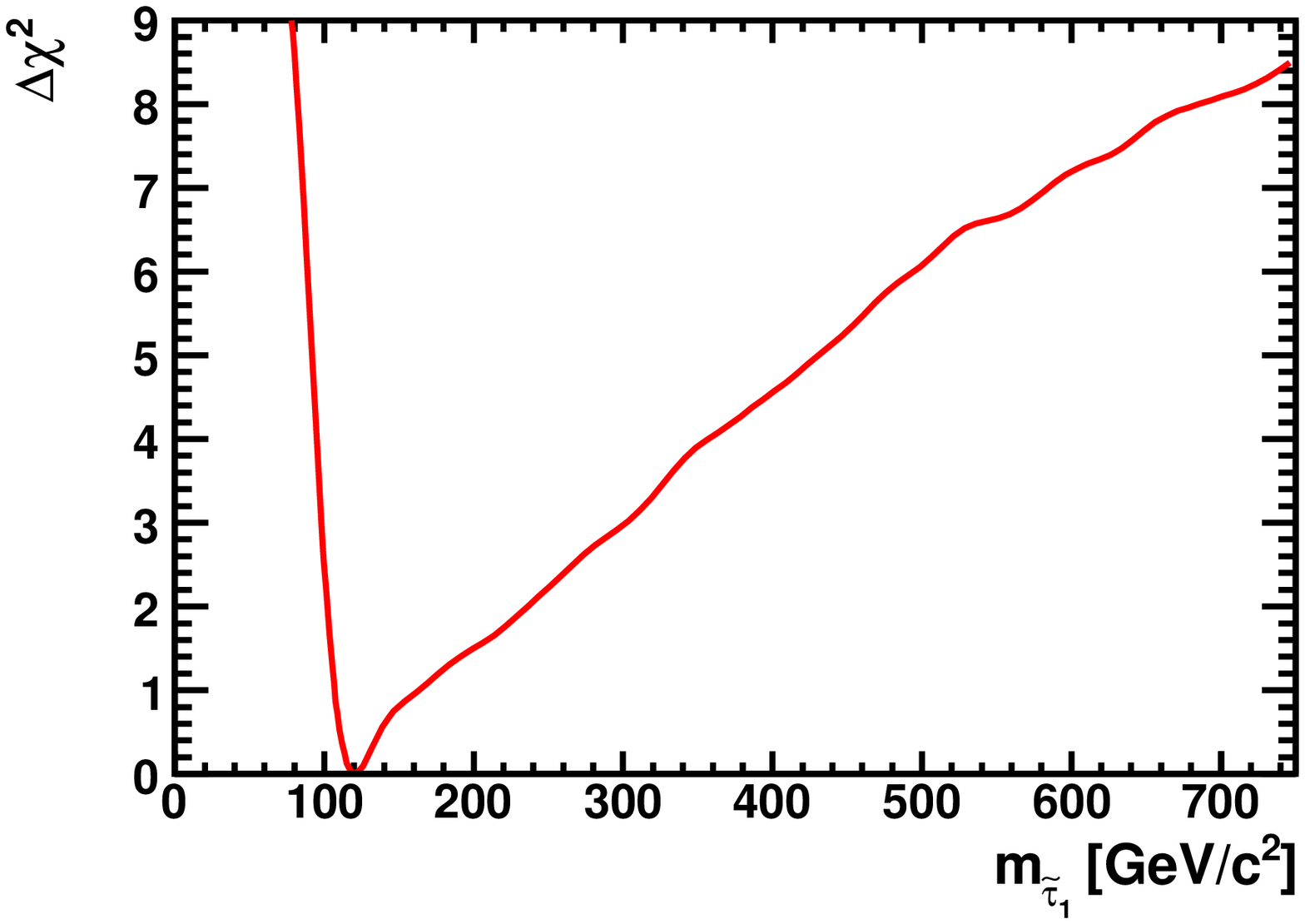}}
%%%%%%%%%%%%%%%%%%%%%%%%%%%%%%%
\vspace{-1cm}
\caption{\it The likelihood functions for $\mstaue$ in the CMSSM
  (left panel) and in the NUHM1 (right panel).
}
\label{fig:mstau}
\end{figure*}

In the case of the lighter stau $\staue$, see its range in
Fig.~\ref{fig:spectrum} and its likelihood function in 
Fig.~\ref{fig:mstau}, the mass is very similar to
that of the LSP $\neu{1}$ in the coannihilation region, but this is not
the case in the rapid-annihilation $H, A$ funnel region. The differences
in the likelihood functions for the $\staue$ and the LSP $\neu{1}$, shown
in Fig.~\ref{fig:mchi}, reflect the importance of this funnel region. In the
case of the CMSSM (left panel of Fig.~\ref{fig:mstau}), the funnel region
appears only at large values of $\tb$ that are relatively disfavoured.
This is why the shape of the $\staue$ likelihood function differs
significantly from that of the $\neu{1}$ only at relatively large
masses. In the case of the NUHM1 (shown in the right panel of
Fig.~\ref{fig:mstau}), rapid annihilation is possible also for low
$\tb$, leading to larger values of $m_0$ than in the CMSSM also for
relatively small values of $\mstaue$.

The scalar taus as well as the other scalar leptons are expected to be
relatively light, as can be seen in Fig.~\ref{fig:spectrum}. They would
partially be in the reach of the ILC(500) (i.e.\ with $\sqrt{s} = 500 \gev$)
and at the 95\% C.L.\ nearly all be in the reach of the
ILC(1000)~\cite{teslatdr,Brau:2007zza}. 
This also holds for the two lighter neutralinos and the light chargino
(In the NUHM1, small parts of the the 95\% C.L.\
regions for the masses of the heavier stau and 
the light chargino are above $500 \gev$.)

%%%%%%%%%%%%%%%%%%%%%% F I G U R E %%%%%%%%%%%%%%%%%%%%%%%%%%%%%%%%%%%
%%%%%%%%%%%%%%%%%%%%%% F I G U R E %%%%%%%%%%%%%%%%%%%%%%%%%%%%%%%%%%%

\begin{figure*}[htb!]
%%%%%%%%%%%%%%%%%%%%%%%%%%%%%%%
\resizebox{8cm}{!}{\includegraphics{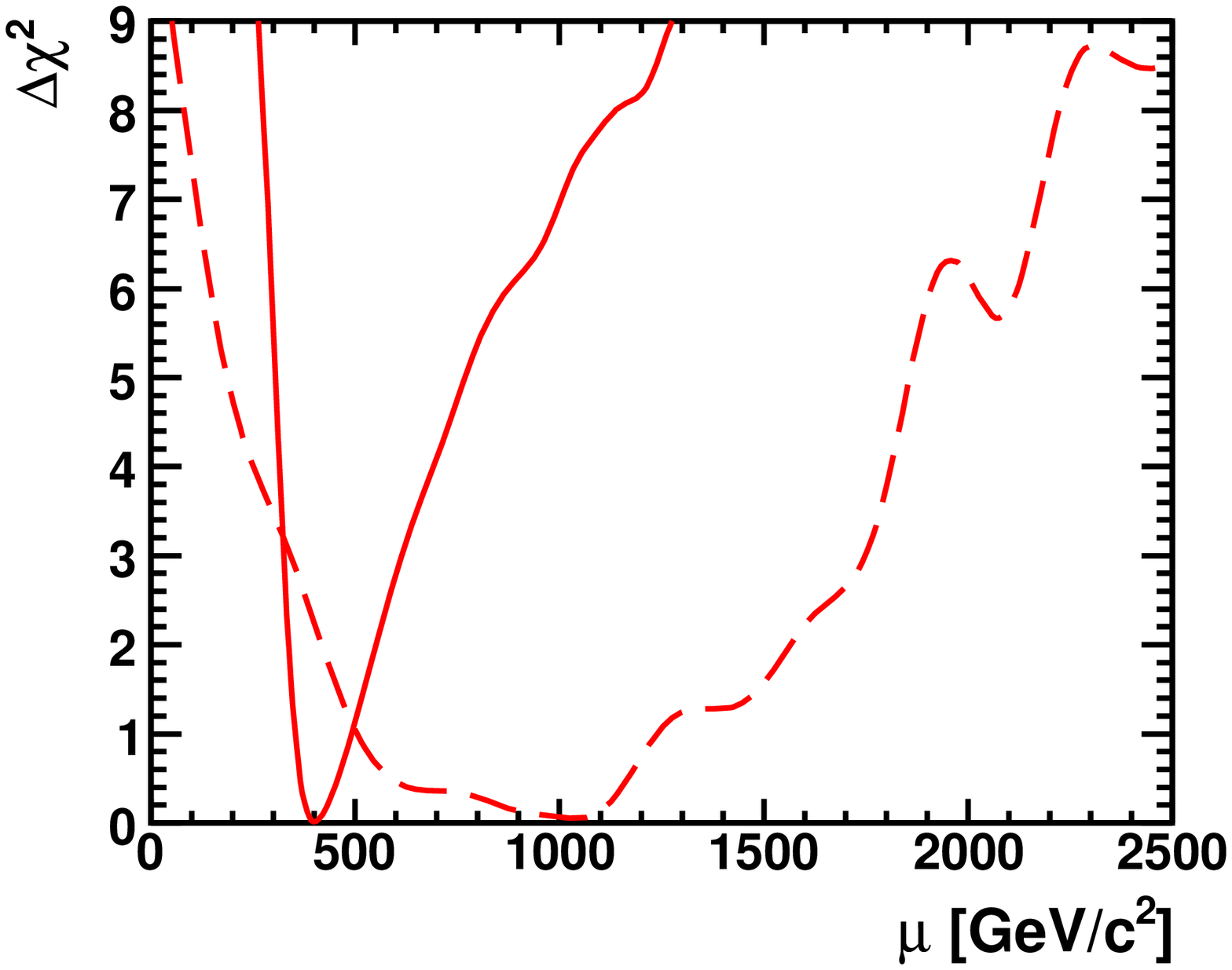}}
\resizebox{8cm}{!}{\includegraphics{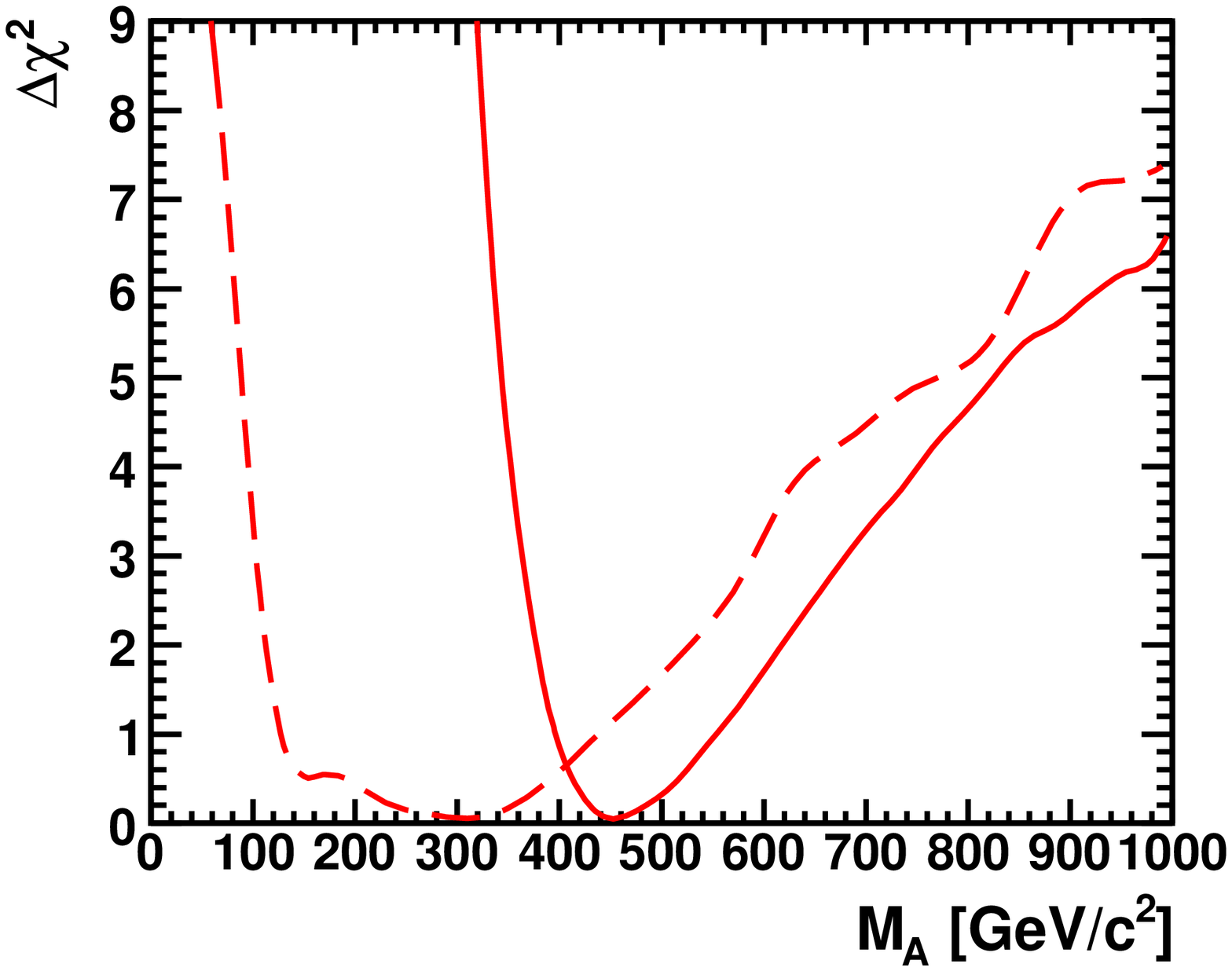}}
%%%%%%%%%%%%%%%%%%%%%%%%%%%%%%%
\vspace{-1em}
\caption{\it The likelihood functions for $\mu$ (left panel) and
$\MA$ (right panel) in the CMSSM (solid lines)
and in the NUHM1 (dashed lines). 
}
\label{fig:muMA}
\end{figure*}

The left plot of Fig.~\ref{fig:muMA} displays the likelihood functions
for $\mu$ in the CMSSM (solid lines) and the NUHM1 (dashed
lines)~\footnote{We recall that, motivated by 
\gmt\ and \bsg, we study only $\mu > 0$.}. 
In the CMSSM, the values
of $|\mu|$ and $\MA$ are fixed in terms of the other model
parameters by the electroweak boundary conditions. Consequently, the
range of values for $\mu$ is quite small in the CMSSM, and the magnitude of
$\mu$ turns out to be relatively small. In the NUHM1, the much larger
range of $\mu$ 
reflects the greater freedom in the Higgs sector. Solving the electroweak
vacuum conditions for models with non-universal Higgs masses broadens
the $\mu$ distribution\footnote{Very large values of 
$|\mu| \gtrsim 1\tev$ are disfavoured by the presence of deep charge- and
colour-breaking minima \cite{eoss3,tachy}, but this constraint is not applied here.}, 
with the implications discussed above for the
likelihood function for the ${\tilde t_1}$. The right panel of
Fig.~\ref{fig:muMA} displays the likelihood functions for $\MA$
(see also the range in Fig.~\ref{fig:spectrum}). The
likelihood function in the CMSSM is again somewhat narrower than
in the NUHM1, reflecting the influence of the electroweak boundary
conditions. The best-fit value in the CMSSM is significantly higher than
in the NUHM1. 
Values up to $\MA \lsim 500 \gev$ could be tested at
the ILC(1000), i.e.\ the preferred regions of both models could be
probed.

%%%%%%%%%%%%%%%%%%%%%% F I G U R E %%%%%%%%%%%%%%%%%%%%%%%%%%%%%%%%%%%
%%%%%%%%%%%%%%%%%%%%%% F I G U R E %%%%%%%%%%%%%%%%%%%%%%%%%%%%%%%%%%%

\begin{figure*}[htb!]
%%%%%%%%%%%%%%%%%%%%%%%%%%%%%%%
\resizebox{8cm}{!}{\includegraphics{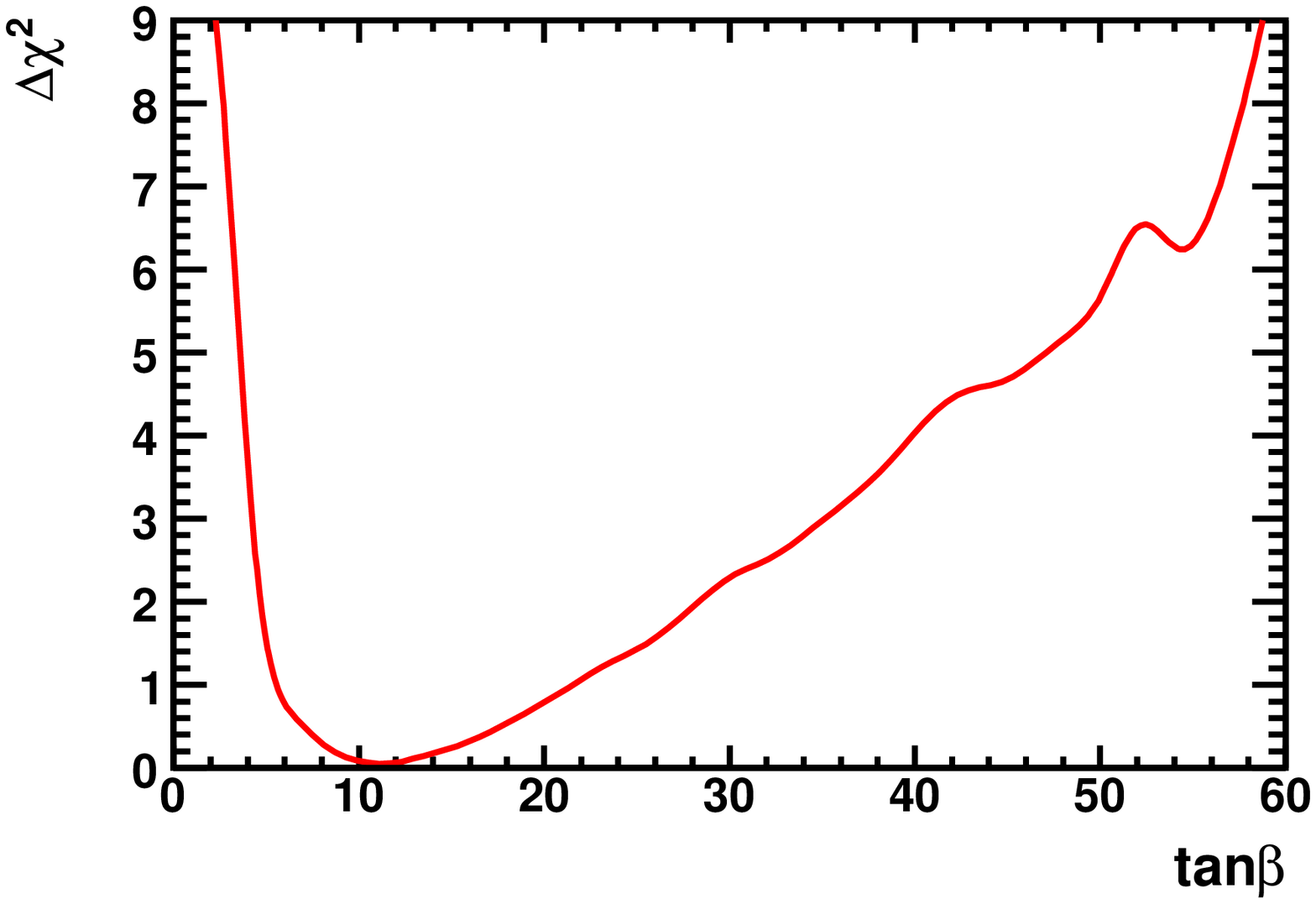}}
\resizebox{8cm}{!}{\includegraphics{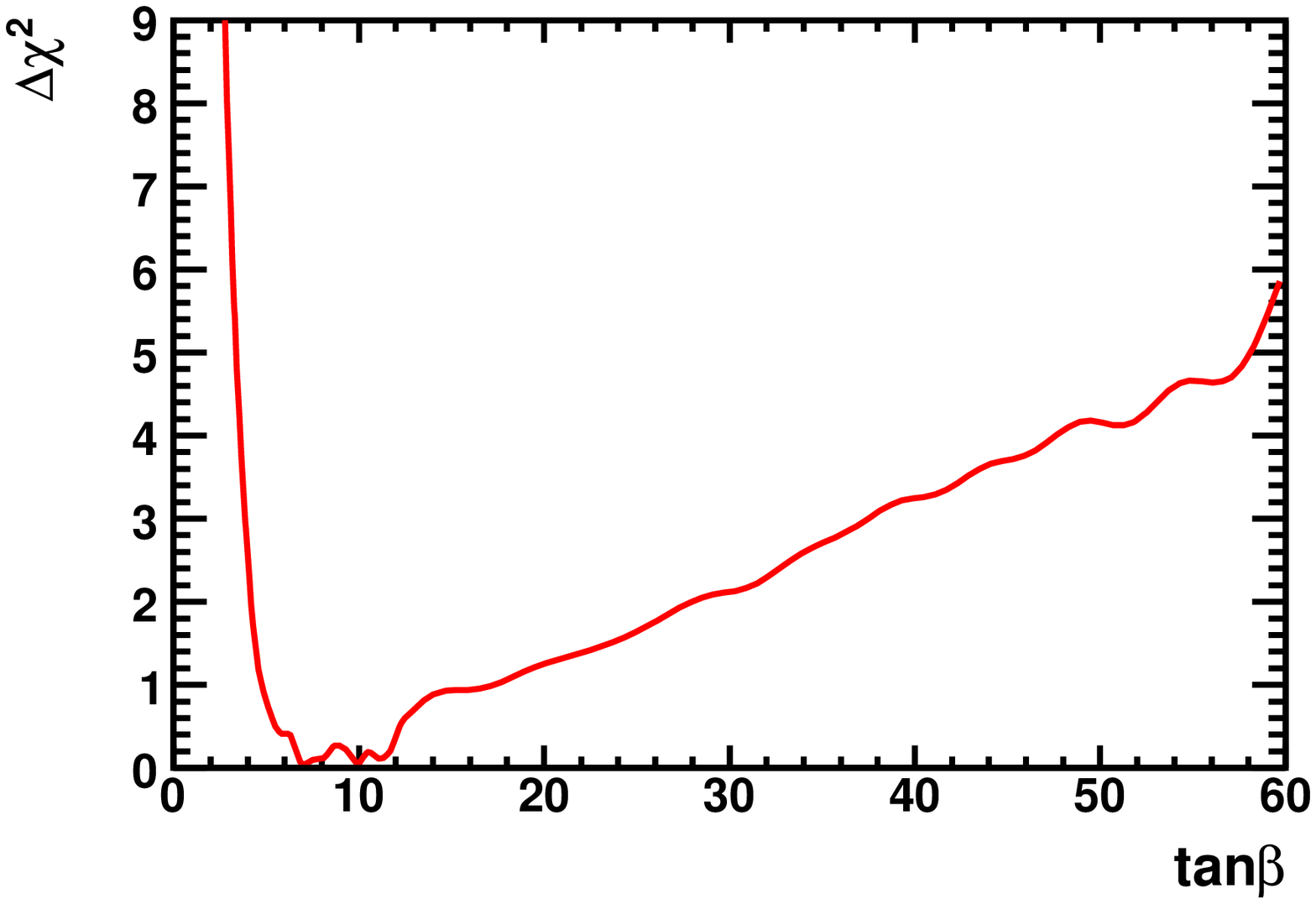}}
%%%%%%%%%%%%%%%%%%%%%%%%%%%%%%%
\vspace{-1cm}
\caption{\it The likelihood functions for $\tb$ in the CMSSM (left panel)
and in the NUHM1 (right panel).
}
\label{fig:tanb}
\end{figure*}

Fig.~\ref{fig:tanb} displays the likelihood
functions for $\tb$. These
are largely similar in the CMSSM and the NUHM1, with $\tb \sim 11$
being favoured in both models.

%%%%%%%%%%%%%%%%%%%%%% F I G U R E %%%%%%%%%%%%%%%%%%%%%%%%%%%%%%%%%%%
%%%%%%%%%%%%%%%%%%%%%% F I G U R E %%%%%%%%%%%%%%%%%%%%%%%%%%%%%%%%%%%

\begin{figure*}[htb!]
%%%%%%%%%%%%%%%%%%%%%%%%%%%%%%%
\resizebox{8cm}{!}{\includegraphics{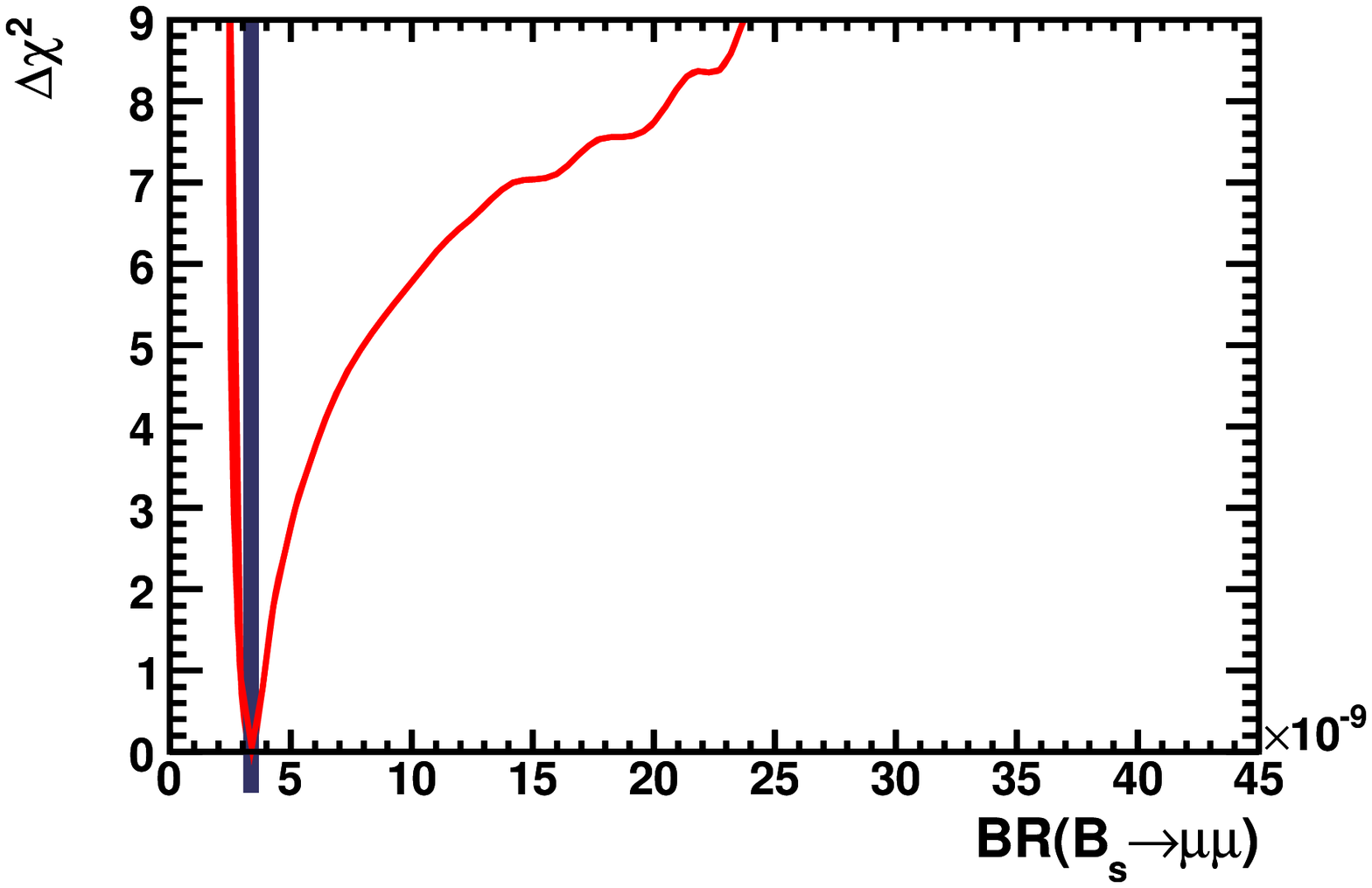}}
\resizebox{8cm}{!}{\includegraphics{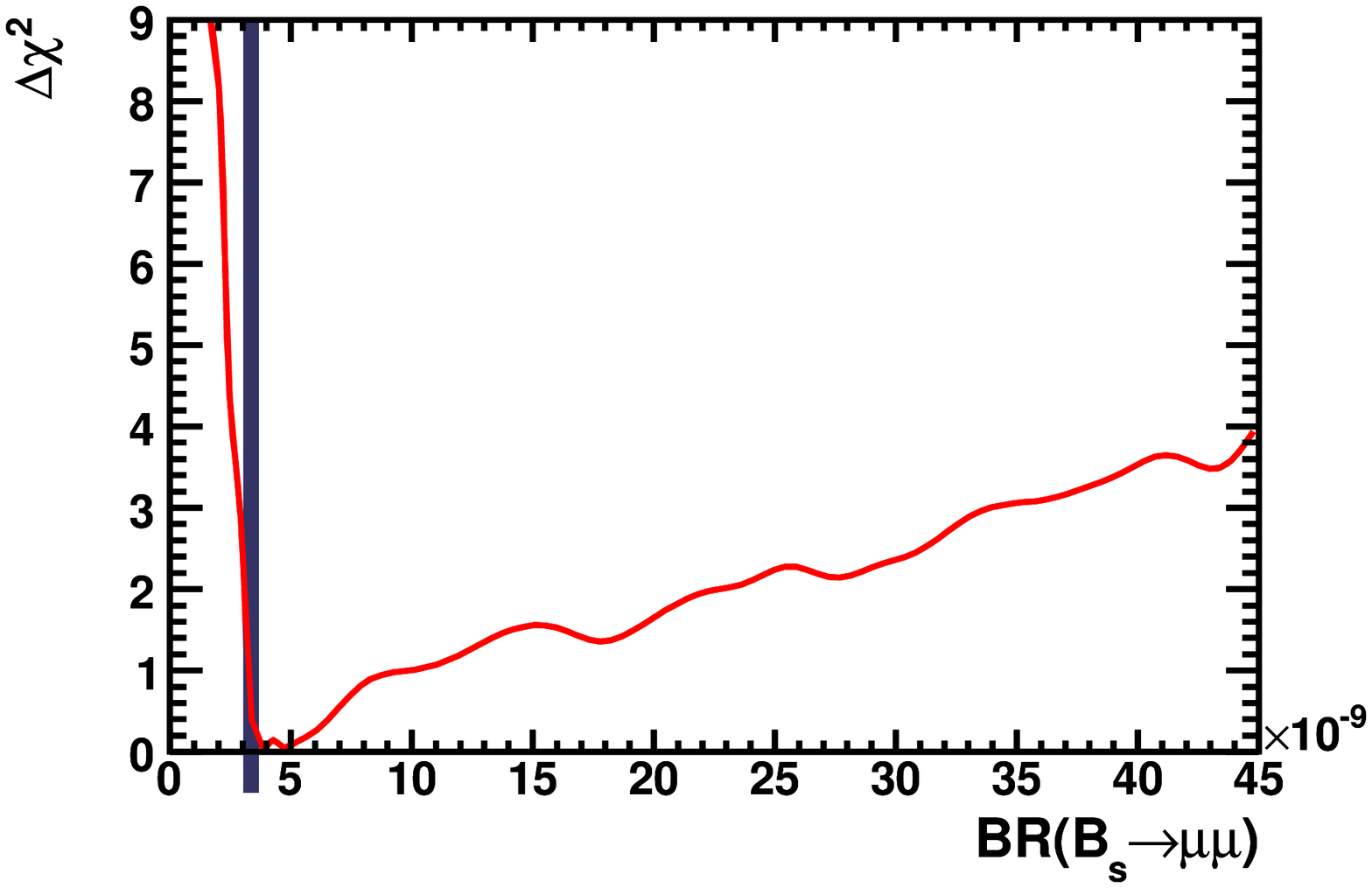}}
%%%%%%%%%%%%%%%%%%%%%%%%%%%%%%%
\vspace{-1cm}
\caption{\it The likelihood functions for the branching ratio 
\bmm\ in the CMSSM (left panel) and in the NUHM1 (right panel).
The vertical lines indicate the SM value with its theoretical error.
}
\label{fig:bsmumu}
\end{figure*}

\begin{figure*}[htb!]
%%%%%%%%%%%%%%%%%%%%%%%%%%%%%%%
\resizebox{8cm}{!}{\includegraphics{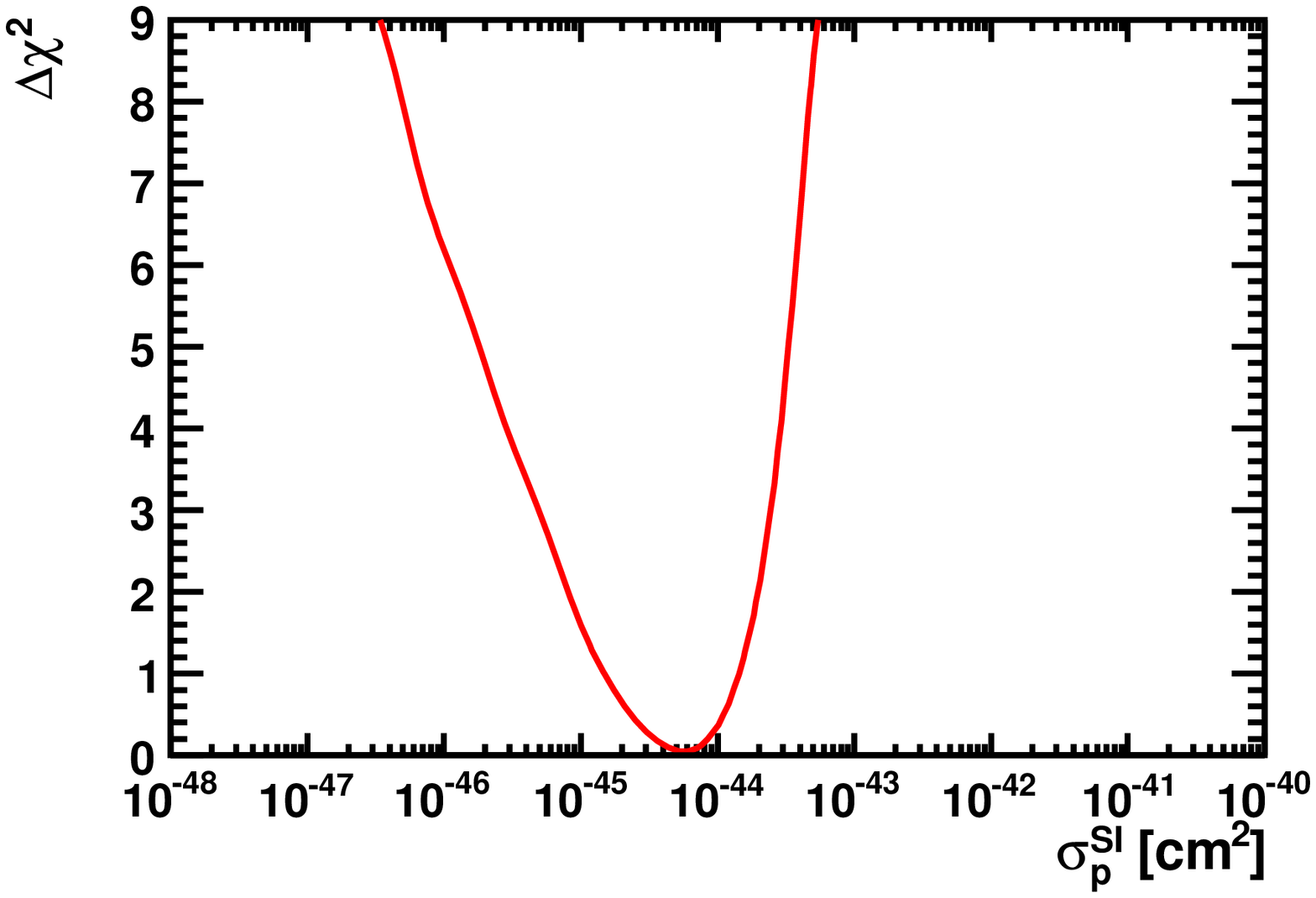}}
\resizebox{8cm}{!}{\includegraphics{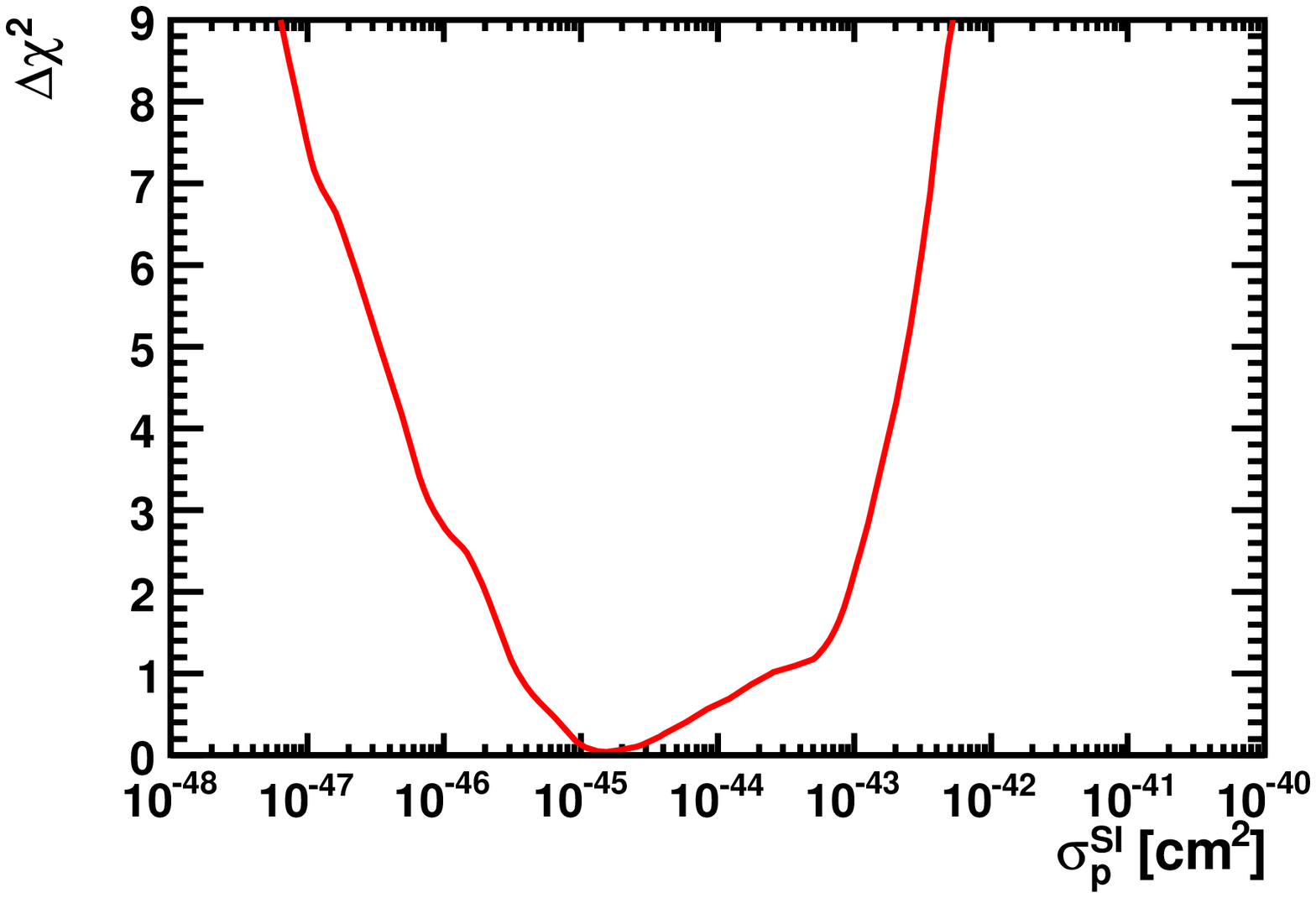}}
%%%%%%%%%%%%%%%%%%%%%%%%%%%%%%%
\vspace{-1cm}
\caption{\it The likelihood functions for the
  spin-independent $\neu{1}$-proton scattering cross section $\ssi$ (in cm$^2$)
in the CMSSM (left panel) and in the NUHM1 (right panel).
}
\label{fig:sig}
\vspace{3em}
\end{figure*}

We turn now to the predictions for two other observables, namely
\bmm\ shown in Fig.~\ref{fig:bsmumu} and the
spin-independent $\neu{1}$-proton scattering cross section $\ssi$
shown in 
Fig.~\ref{fig:sig}~
\footnote{ The spin-independent $\neu{1}$-proton 
and -neutron scattering cross sections are very similar, and
the spin-dependent scattering cross sections (not shown) are much further away
from the prospective experimental sensitivity.}.
We see in the left panel of Fig.~\ref{fig:bsmumu}
that values of the \bmm\ similar
to that in the SM are favoured, particularly for the
preferred lower values of $\tb$. However, large deviations
from the SM prediction (indicated by the vertical lines, which include the
theoretical uncertainty) are still possible at the 3-$\sigma$ level.
The picture in the NUHM1 is completely different, since the $\chi^2$
function is quite flat, with no significant penalty for substantial deviations
from the SM prediction, and very large values of the branching ratio
being allowed at the 2-$\sigma$ level. The difference is largely due
to the fact that smaller masses of the heavier Higgs bosons are
permitted in the NUHM1. A large value of \bmm\
would be a promising harbinger of SUSY
at the LHC, and would favour {\it a priori} the NUHM1 over the CMSSM.
Assuming the SM value, i.e.\
$\br(B_s \to \mu^+ \mu^-) \approx 3.4 \times 10^{-9}$, it has been
estimated~\cite{BDKflavor} that LHCb could observe this process
at the 5$\sigma$ level within a few years of
running. This makes this process a very interesting probe of SUSY that
could help to distinguish between different models.

%%%%%%%%%%%%%%%%%%%%%% F I G U R E %%%%%%%%%%%%%%%%%%%%%%%%%%%%%%%%%%%
%%%%%%%%%%%%%%%%%%%%%% F I G U R E %%%%%%%%%%%%%%%%%%%%%%%%%%%%%%%%%%%

The value of $\ssi$ shown in Fig.~\ref{fig:sig} is calculated assuming a
  $\pi$-N scattering $\sigma$ term 
$\Sigma_N = 64$~MeV: plausible values range between about 45 and 80~MeV,
and $\ssi$ increases quite rapidly with $\Sigma_N$~\cite{eoss8,Ellis:2008hf,giedt}.
We see in Fig.~\ref{fig:sig} that values of the $\neu{1}$-proton cross
section $\ssi \sim 10^{-8}$~pb are expected in the CMSSM, and that much
larger values seem quite unlikely. On the other hand, in the NUHM1,
though the best-fit value of the cross section is somewhat lower, a much
larger range is possible~\footnote{No scaling of the cross-section was
done here to account for regions where \Och\ falls below the WMAP
range, but such points pay a $\chi^2$ penalty.}.
Hence, detection of dark matter with a cross section
much larger than $\sim 10^{-8}$~pb $=10^{-44}$ cm$^{2}$
would also be a good diagnostic for
discriminating between the NUHM1 and the CMSSM. 
The present best upper limits on $\ssi$ from
the CDMS~\cite{CDMS} and Xenon10~\cite{Xe10} experiments are at the
$\sim 10^{-7}$~pb level~\footnote{Assuming a local LSP density of 0.3~GeV/cm$^3$,
which is subject to astrophysical uncertainties.}, and the planned experiments should be sensitive
down to below the $\sim 10^{-10}$~pb level~\cite{XENON100,superCDMS}. 

%%%%%%%%%%%%%%%%%%%%%% F I G U R E %%%%%%%%%%%%%%%%%%%%%%%%%%%%%%%%%%%
%%%%%%%%%%%%%%%%%%%%%% F I G U R E %%%%%%%%%%%%%%%%%%%%%%%%%%%%%%%%%%%

%---------------------------------------------------------------------
\section{Correlations between Sparticle Masses and with other Observables}
\label{sec:correlations}
%---------------------------------------------------------------------

\begin{figure*}[htb!]
%%%%%%%%%%%%%%%%%%%%%%%%%%%%%%%
\resizebox{8cm}{!}{\includegraphics{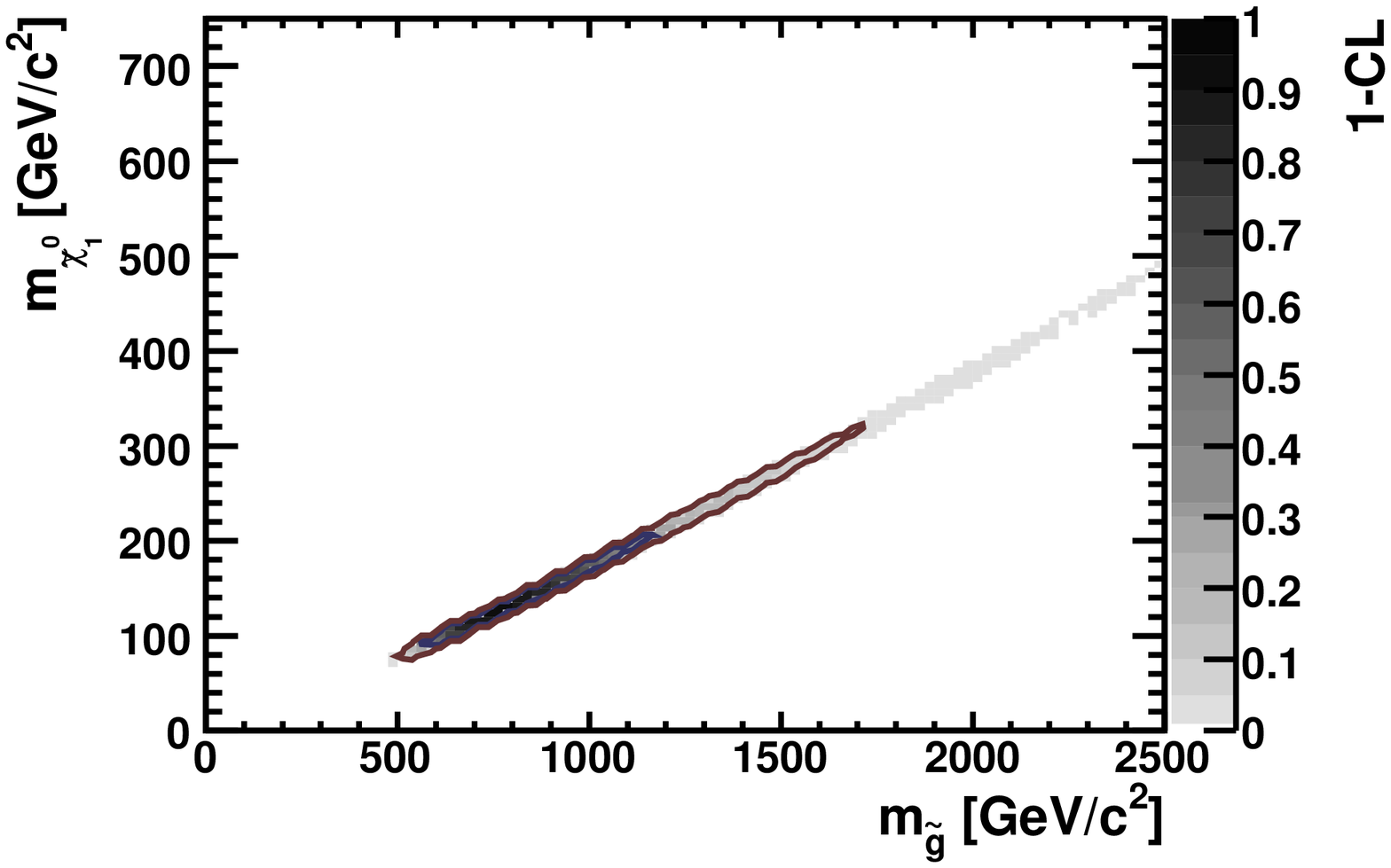}}
\resizebox{8cm}{!}{\includegraphics{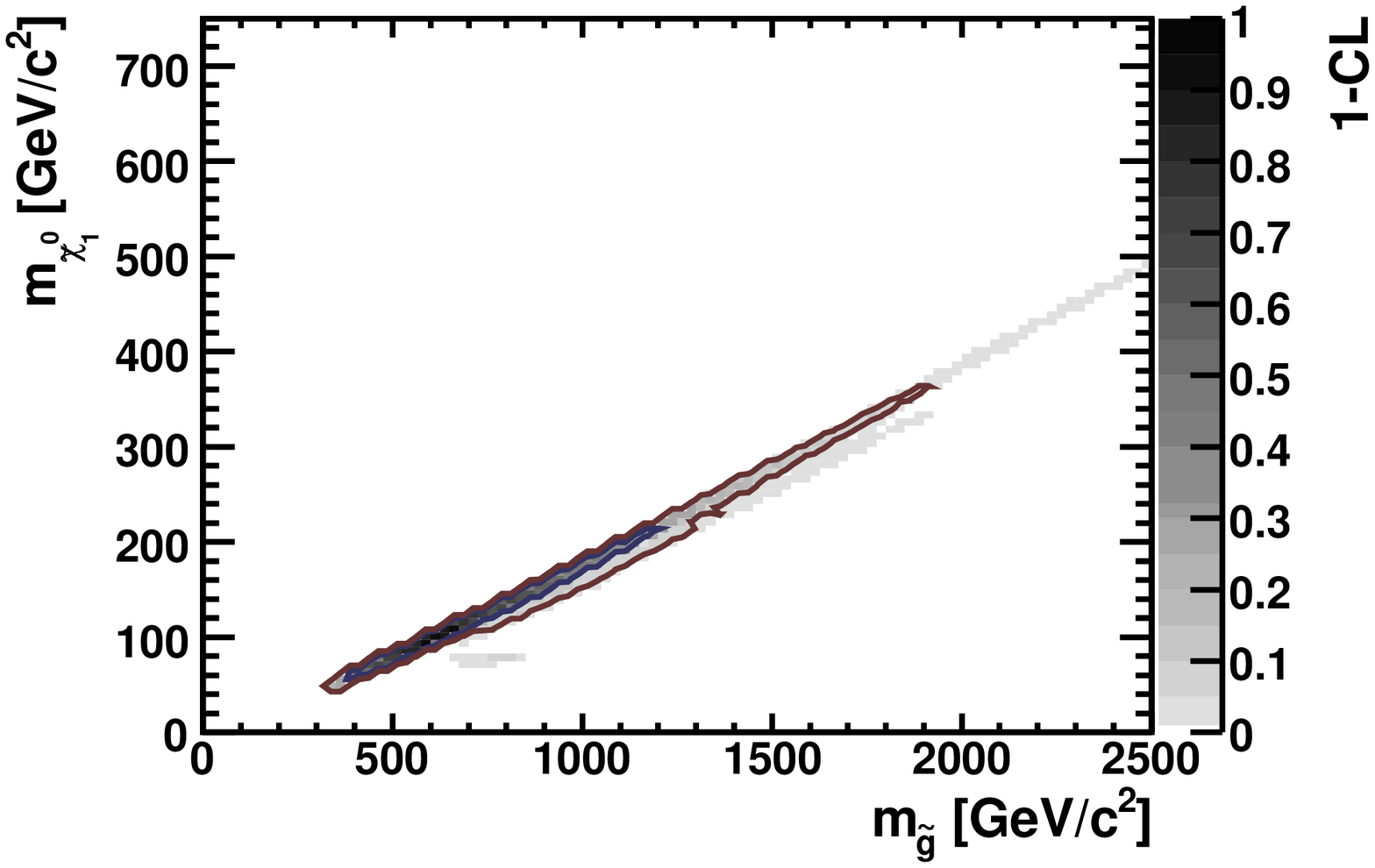}}
%%%%%%%%%%%%%%%%%%%%%%%%%%%%%%%
\vspace{-1cm}
\caption{\it The correlation between the LSP mass, $\mneu{1}$,
and the gluino mass, $\mgl$, in the CMSSM (left panel) and in the NUHM1
(right panel). 
}
\label{fig:mchimg}
\end{figure*}

We now discuss in more detail some of the correlations between
sparticle masses and observables, starting with the LSP mass, $\mneu{1}$,
and the gluino mass, $\mgl$, shown in Fig.~\ref{fig:mchimg}~%
\footnote{For one-dimensional scans, we continue to quote up to 9 units in
$\Delta\chi^2$, which corresponds to $3\sigma$.
For two-dimensional confidence level contour plots, 
we quote $1 - CL$ instead of the $\Delta\chi^2$; the blue (red) lines in the plots
correspond to $1 - CL = 32 (5)$\%, and the white regions
correspond to $1 - CL \le 1\%$, or $\Delta\chi^2 \ge 9.21$ units.}.  
%(or $3\sigma$).}. 
We expect a very strong correlation, since the value of $m_{1/2}$
largely controls both masses. However, in both cases there are
radiative corrections that enter when making the transition
from the SUSY-breaking parameters defined using the
$\overline{{\rm DR}}$ 
prescription to the on-shell masses, that depend on the other MSSM parameters.
Moreover, the LSP is not a pure Bino, and the mixing with other
neutralino states depends on the value of $\mu$, in particular.
Indeed, we see in Fig.~\ref{fig:mchimg} a very strong $\mneu{1} - \mgl$
correlation in the CMSSM (left panel), which is not quite so
strong for the NUHM1 (right 
panel). Moreover, in the latter case we notice a small (grey) island of parameters
where $\mneu{1}$ is substantially lower
than one would have expected for the
corresponding value of $\mgl$. These few examples have a Higgsino-like LSP, 
and have relatively small likelihoods.

%%%%%%%%%%%%%%%%%%%%%% F I G U R E %%%%%%%%%%%%%%%%%%%%%%%%%%%%%%%%%%%
%%%%%%%%%%%%%%%%%%%%%% F I G U R E %%%%%%%%%%%%%%%%%%%%%%%%%%%%%%%%%%%

A corollary of the correlation between $\mgl$ and $\mneu{1}$ seen in
Fig.~\ref{fig:mchimg} is the relation between the mass scale of the heavy 
supersymmetric particles~\cite{Ellis:2004bx} that might be discovered at
the LHC with the 
threshold for producing the lighter sparticles that might be measured at
a future linear $e^+ e^-$ collider. If one observes
a gluino at a certain mass scale (or establishes a lower limit on its
mass), according 
to Fig.~\ref{fig:mchimg} one will have, within the CMSSM or the NUHM1, a
lower bound on the threshold for pair-producing observable sparticles at
a linear collider, $\ecm > 2 \mneu{1}$~%
\footnote{The lightest neutralino might
then be visible in the channel 
$e^+e^- \to \neu{1}\neu{1}\gamma$~\cite{nnga1,nnga2,nnga3}.}%
. The relevant $\mneu{1}$ may
be read directly off the vertical scale of Fig.~\ref{fig:mchimg}. This
is in general related to $\mgl$ by a simple, universal numerical factor,
the only exception being the small island (which has a rather low
likelihood) of models with
unusually low $\mneu{1}$ in the NUHM1, mentioned earlier and seen in the
right panel of Fig.~\ref{fig:mchimg}. 

\begin{figure*}[htb!]
%%%%%%%%%%%%%%%%%%%%%%%%%%%%%%%
\resizebox{8cm}{!}{\includegraphics{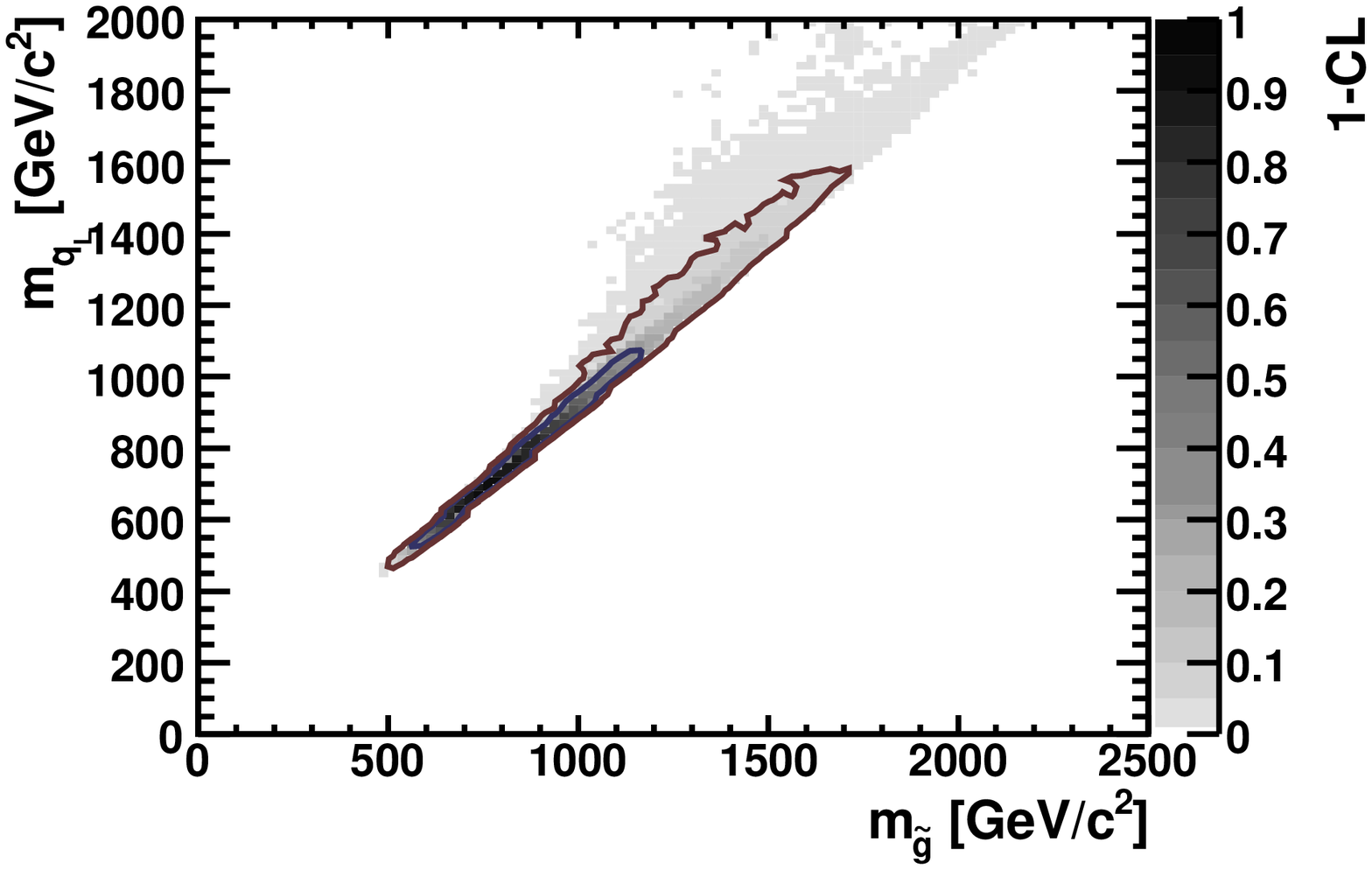}}
\resizebox{8cm}{!}{\includegraphics{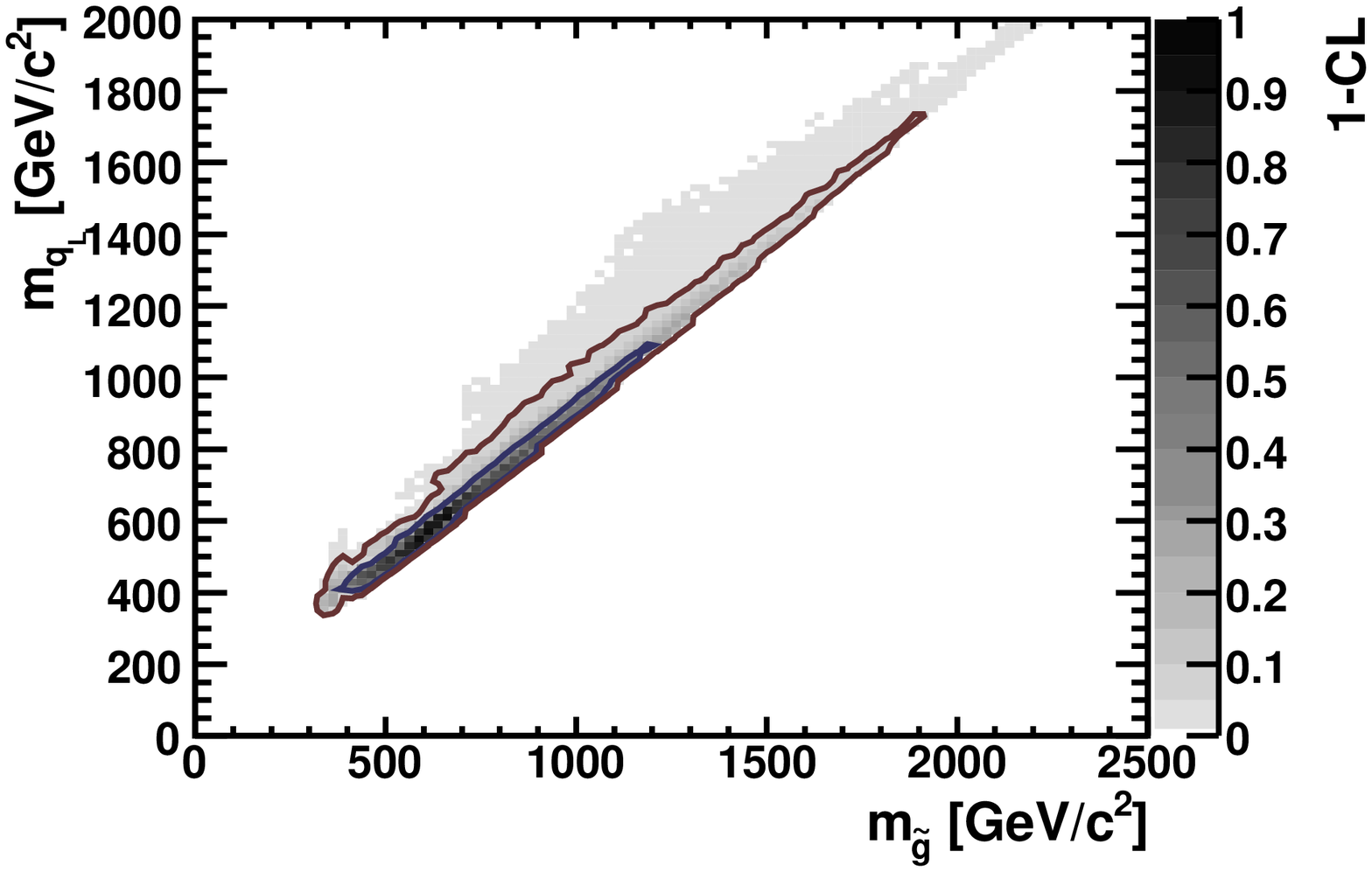}}
\resizebox{8cm}{!}{\includegraphics{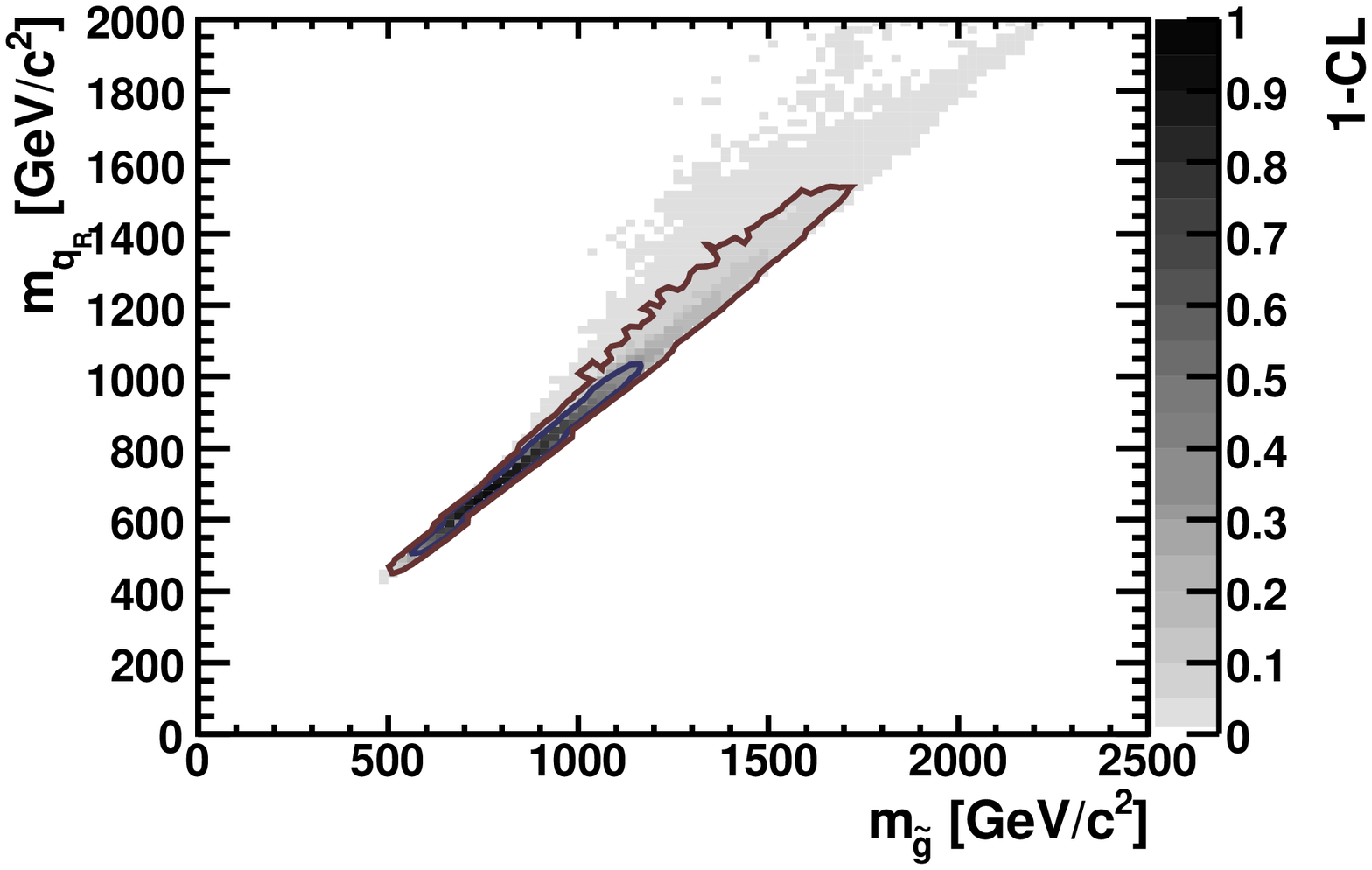}}
\resizebox{8cm}{!}{\includegraphics{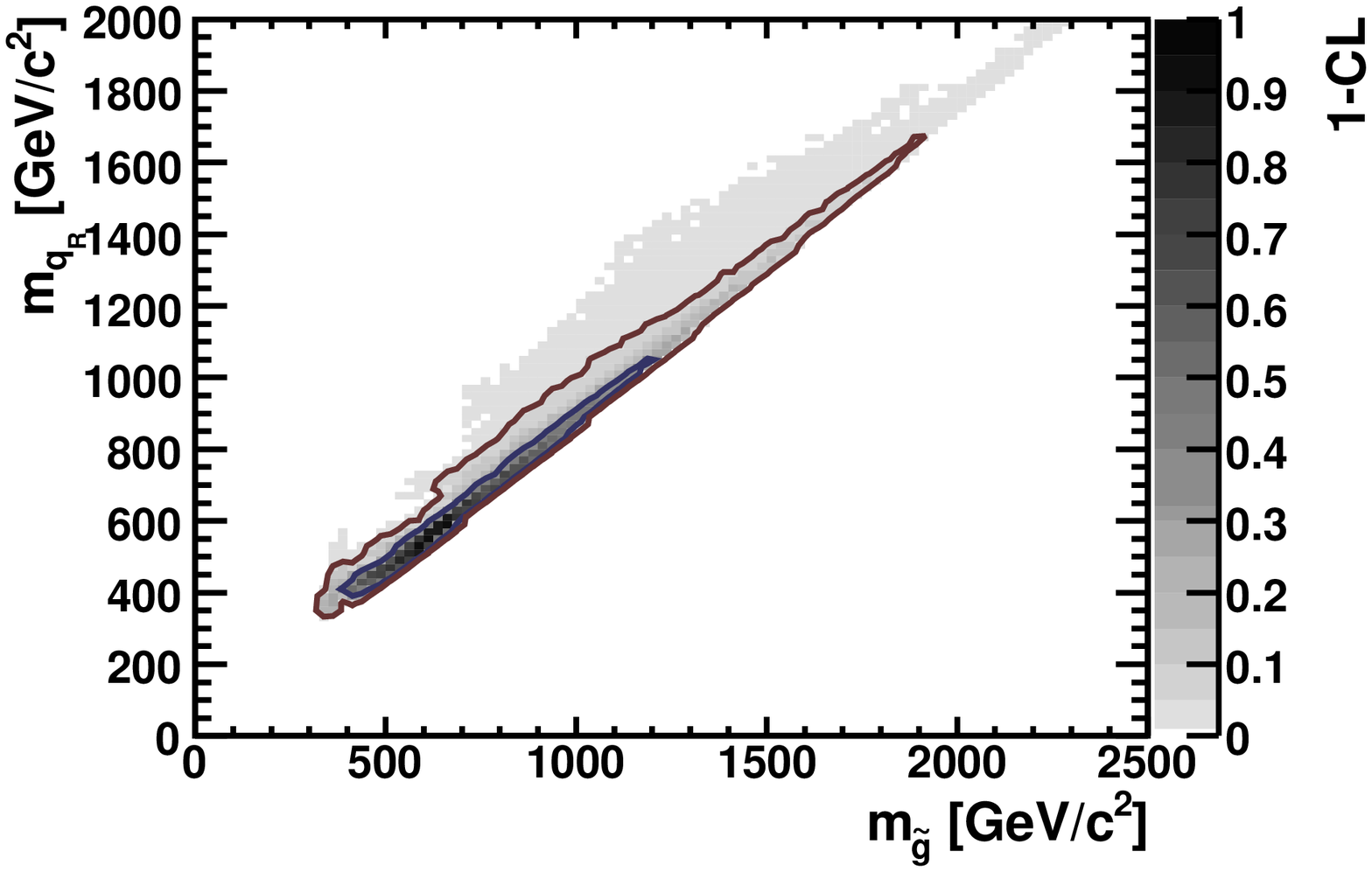}}
%%%%%%%%%%%%%%%%%%%%%%%%%%%%%%%
\vspace{-1em}
\caption{\it The correlations between the gluino mass, $\mgl$,
and the masses of the the left- and right-handed partners
of the five light squark flavours, $m_{\tilde q_{L,R}}$ (upper and
lower panels, respectively) are shown  
in the CMSSM (left panels) and in the NUHM1
(right panels). 
}
\label{fig:mgmsq}
\end{figure*}

In principle, the masses of the squark partners of the five lightest quarks
depend on $m_0$ as well as $m_{1/2}$.
However, as seen in Fig.~\ref{fig:mgmsq}, they are also very highly
correlated with $\mgl$, reflecting the fact that $m_0 < m_{1/2}$ in
the favoured regions of the CMSSM and the NUHM1, and also the fact that
the sensitivities of $m_{\tilde q_{L,R}}$ to $m_0$ are intrinsically smaller
than that to $m_{1/2}$. That said, we see that the correlations of
$m_{\tilde q_{L,R}}$ with $\mgl$ are slightly weakened in the CMSSM
(left panels) 
at large $\mgl$, reflecting the appearance of the rapid-annihilation
funnel with relatively large $m_0$ at large $m_{1/2}$ and $\tb$. The
greater width of the correlations in the NUHM1 (right panels) at small $\mgl$,
compared to the CMSSM, reflects the possibility of greater $m_0$ due to
the appearance of a rapid-annihilation funnel at smaller values of
$m_{1/2}$ and $\tb$ than in the CMSSM.

%%%%%%%%%%%%%%%%%%%%%% F I G U R E %%%%%%%%%%%%%%%%%%%%%%%%%%%%%%%%%%%
%%%%%%%%%%%%%%%%%%%%%% F I G U R E %%%%%%%%%%%%%%%%%%%%%%%%%%%%%%%%%%%

\begin{figure*}[htb!]
%%%%%%%%%%%%%%%%%%%%%%%%%%%%%%%
\resizebox{8cm}{!}{\includegraphics{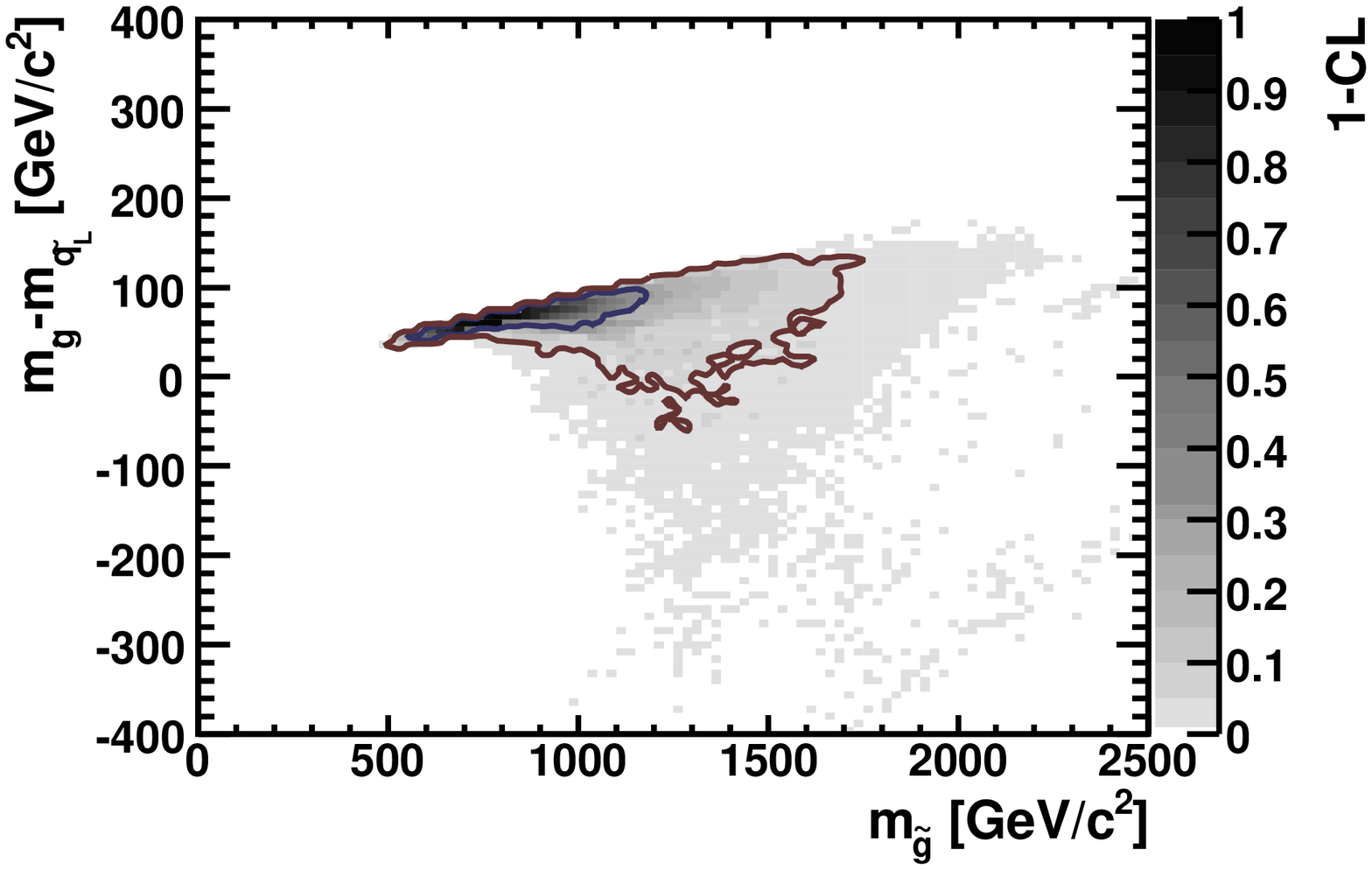}}
\resizebox{8cm}{!}{\includegraphics{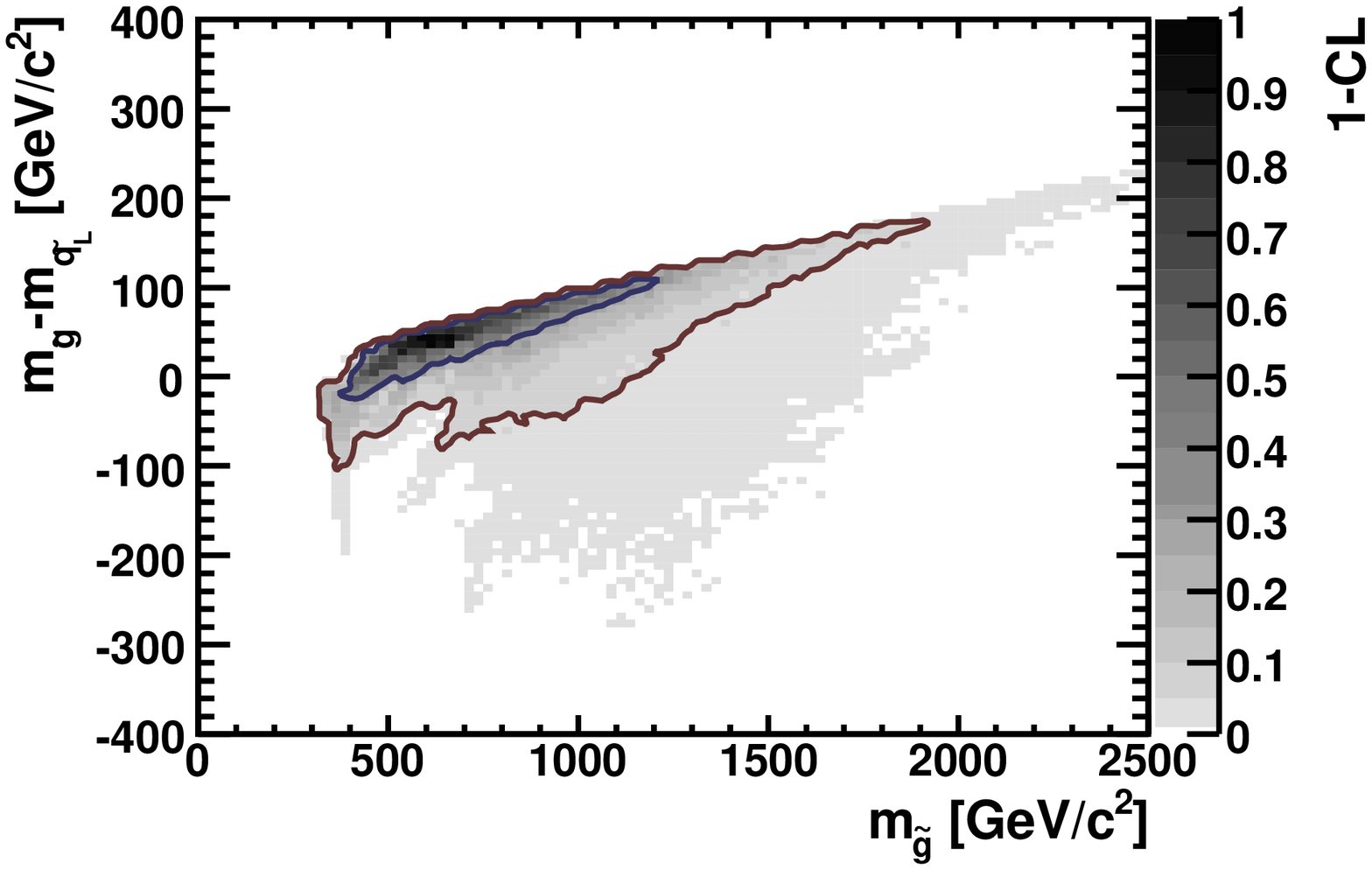}}
\resizebox{8cm}{!}{\includegraphics{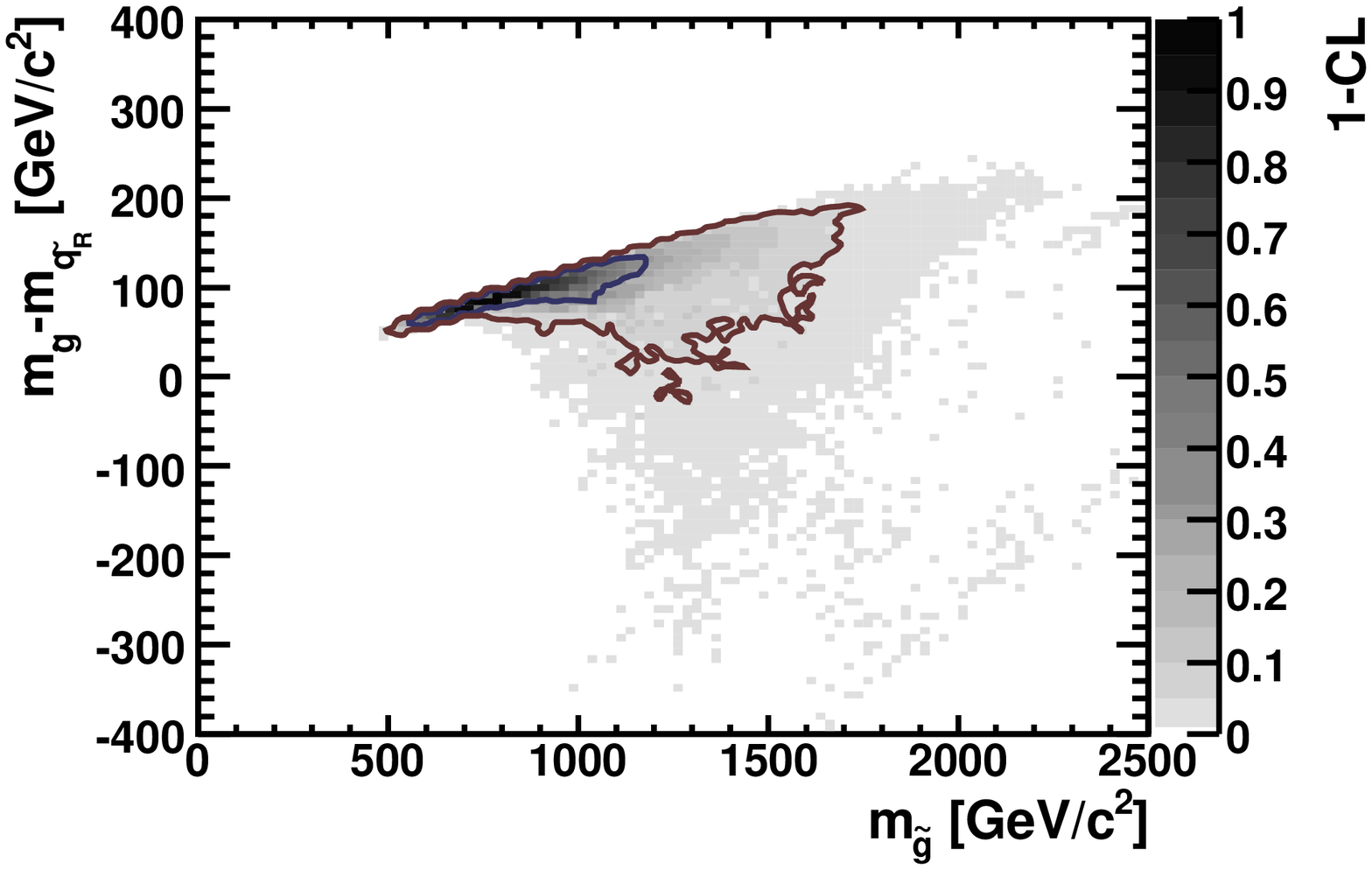}}
\resizebox{8cm}{!}{\includegraphics{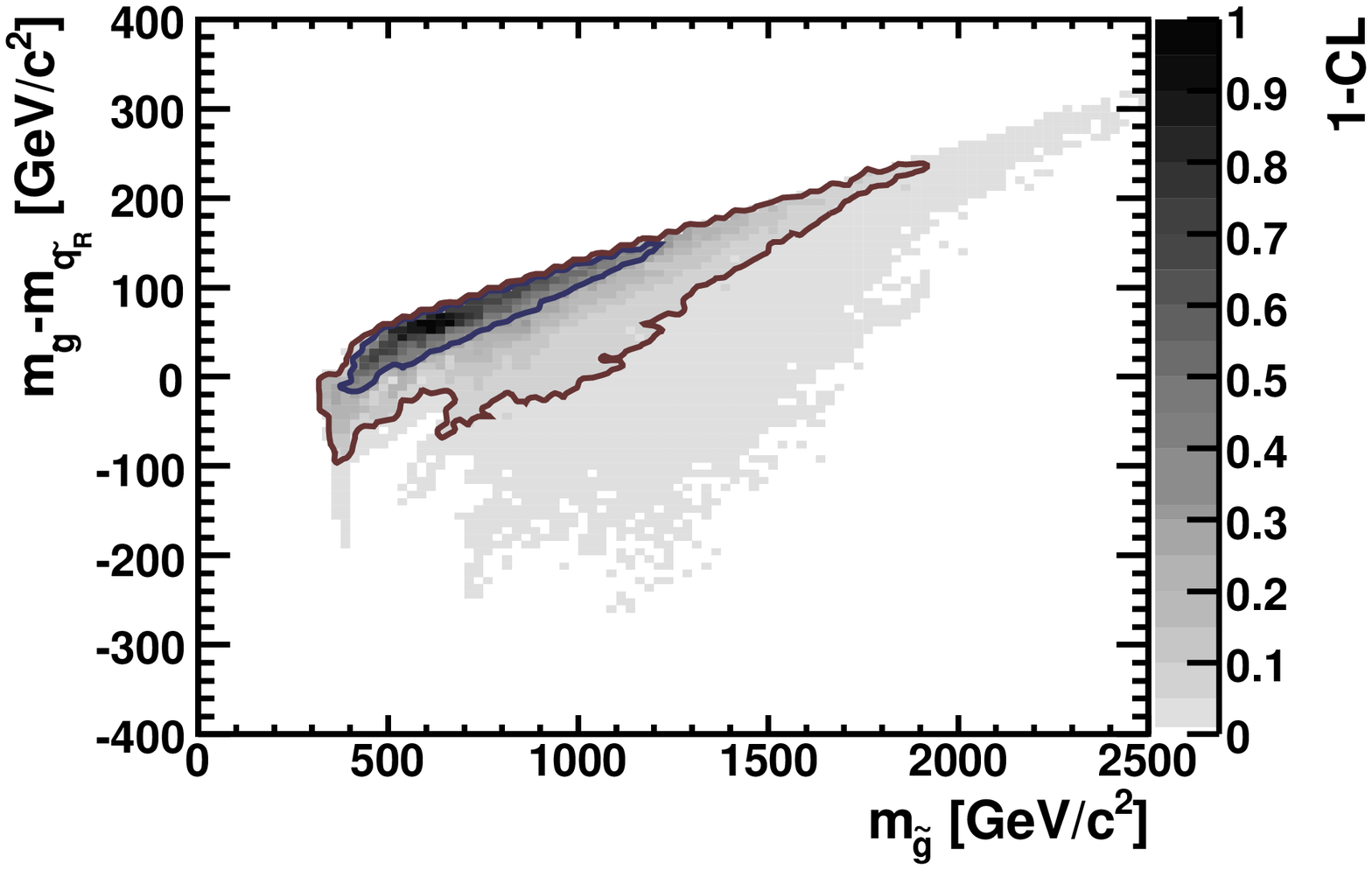}}
%%%%%%%%%%%%%%%%%%%%%%%%%%%%%%%
\vspace{-1em}
\caption{\it The differences between he gluino mass, $\mgl$,
and the masses of the the left- and right-handed partners
of the five light squark flavors, $m_{\tilde q_{L,R}}$ (upper and
lower panels, respectively)  
and in the CMSSM (left panels) and in the NUHM1
(right panels). 
}
\label{fig:mgmsqmassdiff}
\end{figure*}

These effects are more visible in Fig.~\ref{fig:mgmsqmassdiff},
where we plot the differences between the gluino and squark masses
in the CMSSM (left plots) and in the NUHM1 (right plots).
In the CMSSM, in the the cases of both the $\tilde q_{L}$ (upper left
panel) and $\tilde q_{R}$ (lower left panel), we see that the squarks
are always lighter than the gluino if $\mgl$ is itself light. However,
if $\mgl \gsim 1 \tev$, although $\mgl > \tilde q_{L,R}$ is still
favoured, this is not necessarily the case, and $\mgl < \tilde q_{L,R}$
becomes a possibility, because of the larger values of $m_0$ that occur
in the rapid-annihilation funnel that appears
 as $m_{1/2}$ and increases. In the case of 
the NUHM1 (right panels), $\mgl < \tilde q_{L,R}$ is a possibility also
at low $\mgl$, thanks to the possible appearance of a rapid-annihilation
funnel also at low $m_{1/2}$. 

%%%%%%%%%%%%%%%%%%%%%% F I G U R E %%%%%%%%%%%%%%%%%%%%%%%%%%%%%%%%%%%
%%%%%%%%%%%%%%%%%%%%%% F I G U R E %%%%%%%%%%%%%%%%%%%%%%%%%%%%%%%%%%%

\begin{figure*}[htb!]
%%%%%%%%%%%%%%%%%%%%%%%%%%%%%%%
\resizebox{8cm}{!}{\includegraphics{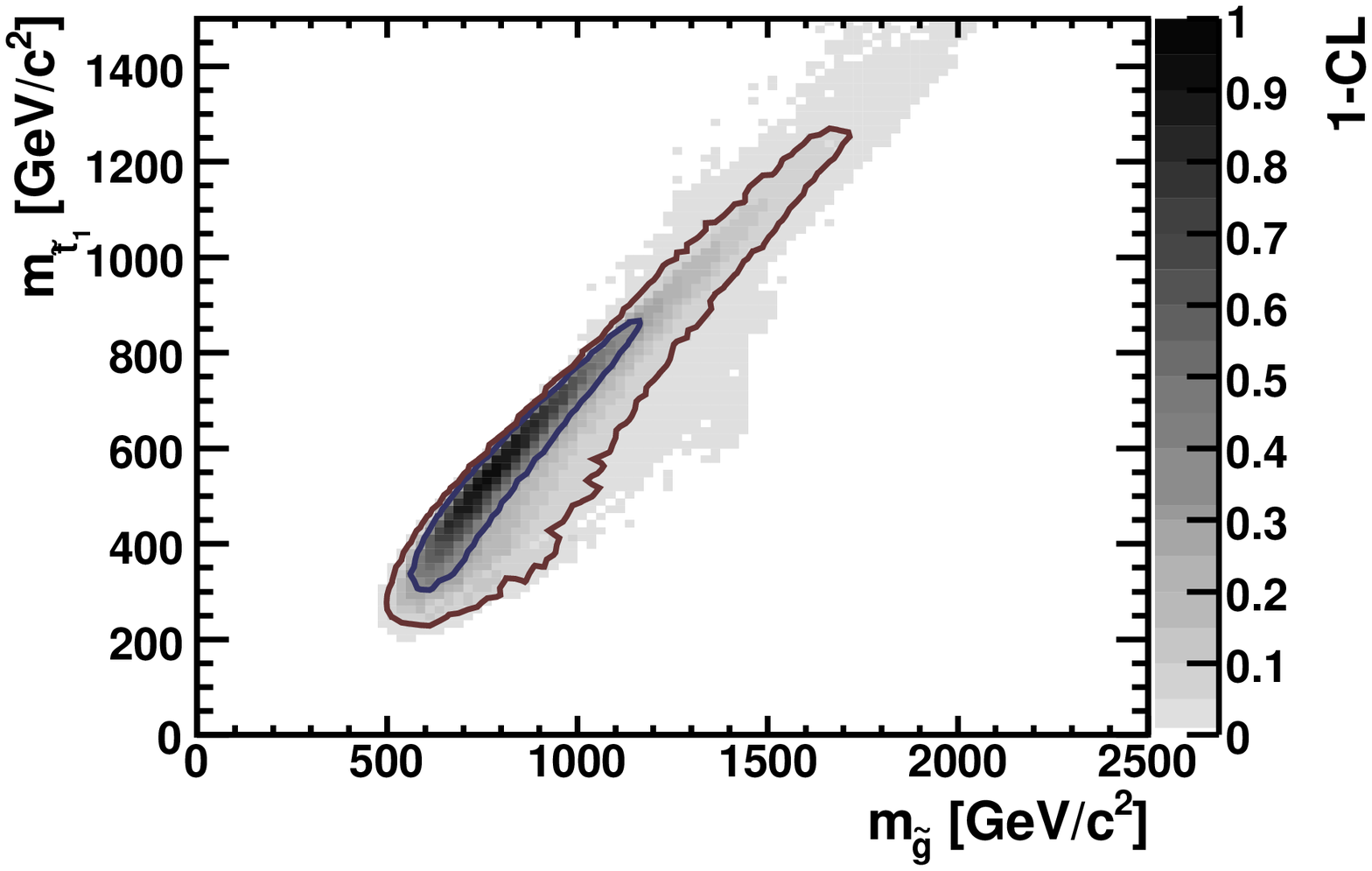}}
\resizebox{8cm}{!}{\includegraphics{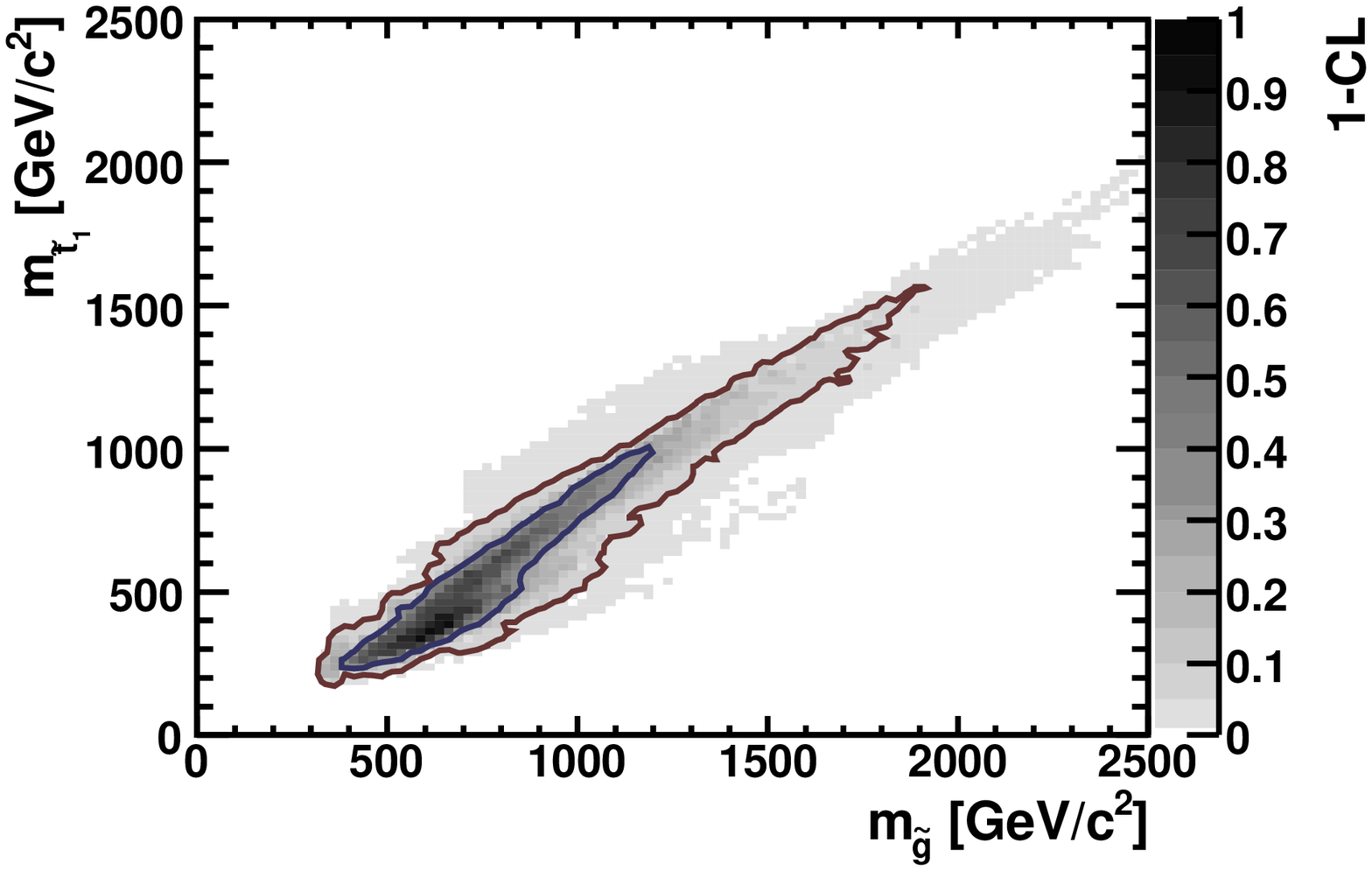}}
%%%%%%%%%%%%%%%%%%%%%%%%%%%%%%%
\vspace{-2em}
\caption{\it The correlation between $\mste$ and the gluino mass, $\mgl$, 
in the CMSSM (left panel) and in the NUHM1 (right panel).
}
\label{fig:mgmst}
\vspace{3em}
\end{figure*}

Fig.~\ref{fig:mgmst} displays the correlation between $\mgl$ and
$\mste$, which is somewhat weaker than the correlation between
$\mgl$ and the other squark masses. This is because, in addition to 
sharing the dependence on $m_0$ with the other squarks,
$\mste$ is sensitive, as commented
earlier, to the value of $\mu$ as well as
$m_{1/2}$ and $m_0$. We recall further that the preferred range of $\mu$
is broader in the NUHM1 than in the CMSSM, which explains why in this model the
preferred range of $\mste$ is broader for intermediate 
values of $\mgl$.

%%%%%%%%%%%%%%%%%%%%%% F I G U R E %%%%%%%%%%%%%%%%%%%%%%%%%%%%%%%%%%%
%%%%%%%%%%%%%%%%%%%%%% F I G U R E %%%%%%%%%%%%%%%%%%%%%%%%%%%%%%%%%%%

\begin{figure*}[htb!]
%%%%%%%%%%%%%%%%%%%%%%%%%%%%%%%
\resizebox{8cm}{!}{\includegraphics{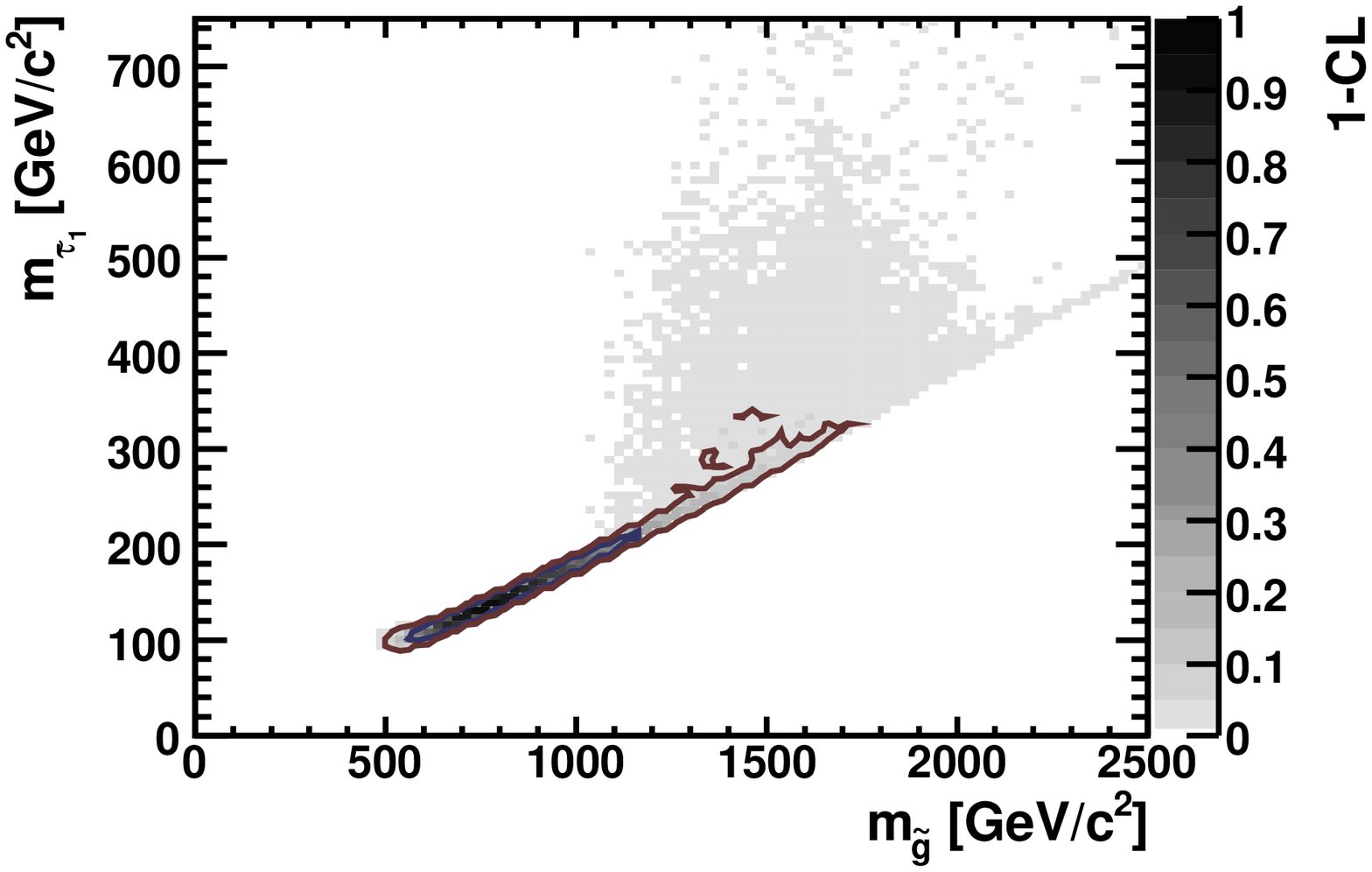}}
\resizebox{8cm}{!}{\includegraphics{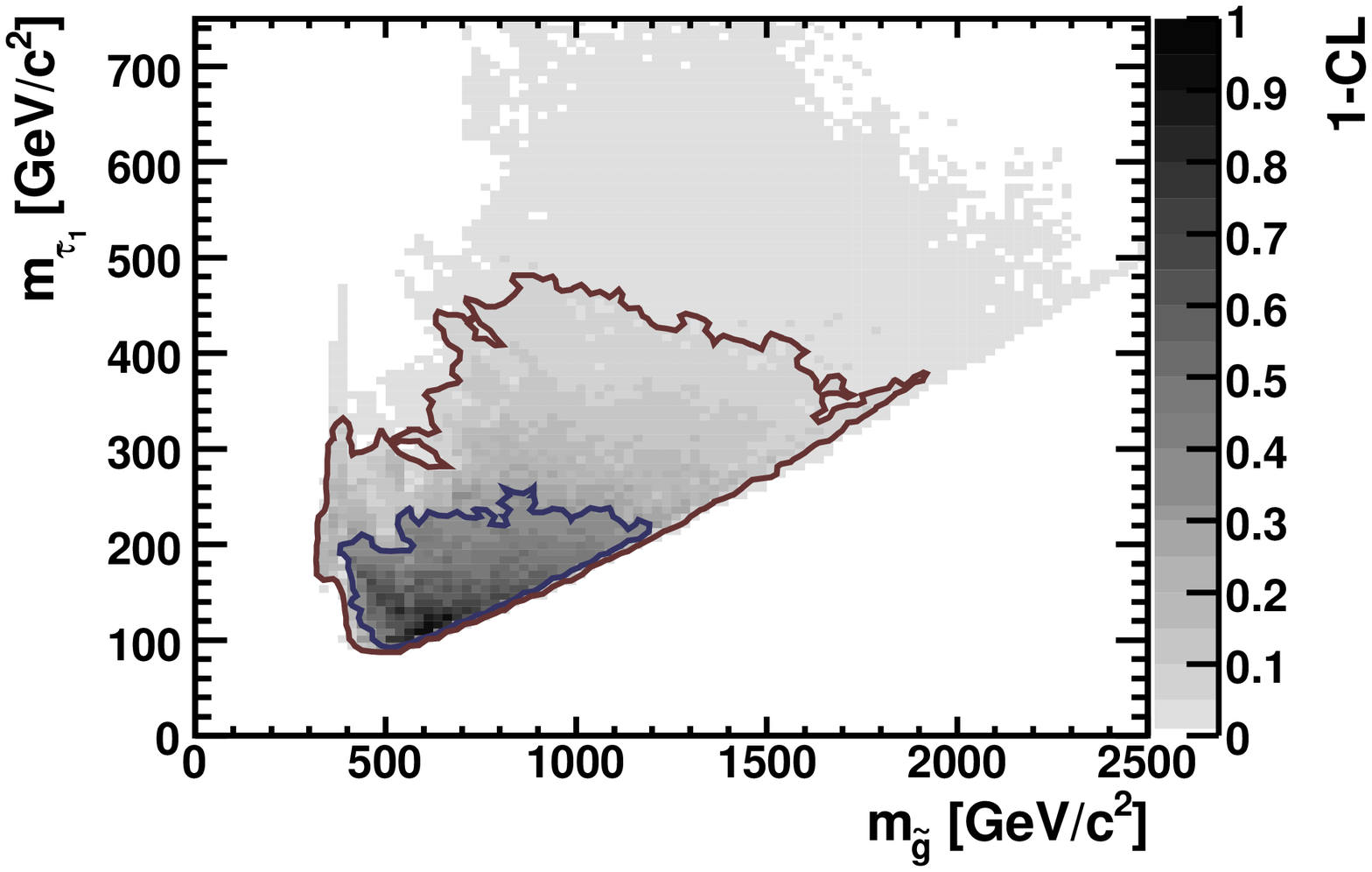}}
%%%%%%%%%%%%%%%%%%%%%%%%%%%%%%%
\vspace{-2em}
\caption{\it The correlation between $\mstaue$ and the gluino mass, $\mgl$, 
in the CMSSM (left panel) and in the NUHM1 (right panel).
}
\label{fig:mgmstau}
\vspace{1em}
\end{figure*}

Fig.~\ref{fig:mgmstau} displays the correlation between $\mstaue$ and
$\mgl$, which is generally proportional to the LSP mass, as discussed
earlier. The $\mstaue$ - $\mgl$ correlation is strikingly different in
the CMSSM (left 
panel) and the NUHM1 (right panel). The tight correlation in the CMSSM
reflects the fact that the favoured part of the parameter space is in
the $\neu{1}$-$\staue$ coannihilation region, where the $\neu{1}$-$\staue$
mass difference is very small. On the other hand, in
the NUHM1, as already commented, there are favoured regions away from
the coannihilation region, where rapid annihilation through
direct-channel $H, A$ poles keeps the relic density within the WMAP
range. 

%%%%%%%%%%%%%%%%%%%%%% F I G U R E %%%%%%%%%%%%%%%%%%%%%%%%%%%%%%%%%%%
%%%%%%%%%%%%%%%%%%%%%% F I G U R E %%%%%%%%%%%%%%%%%%%%%%%%%%%%%%%%%%%

\begin{figure*}[htb!]
%%%%%%%%%%%%%%%%%%%%%%%%%%%%%%%
\resizebox{8cm}{!}{\includegraphics{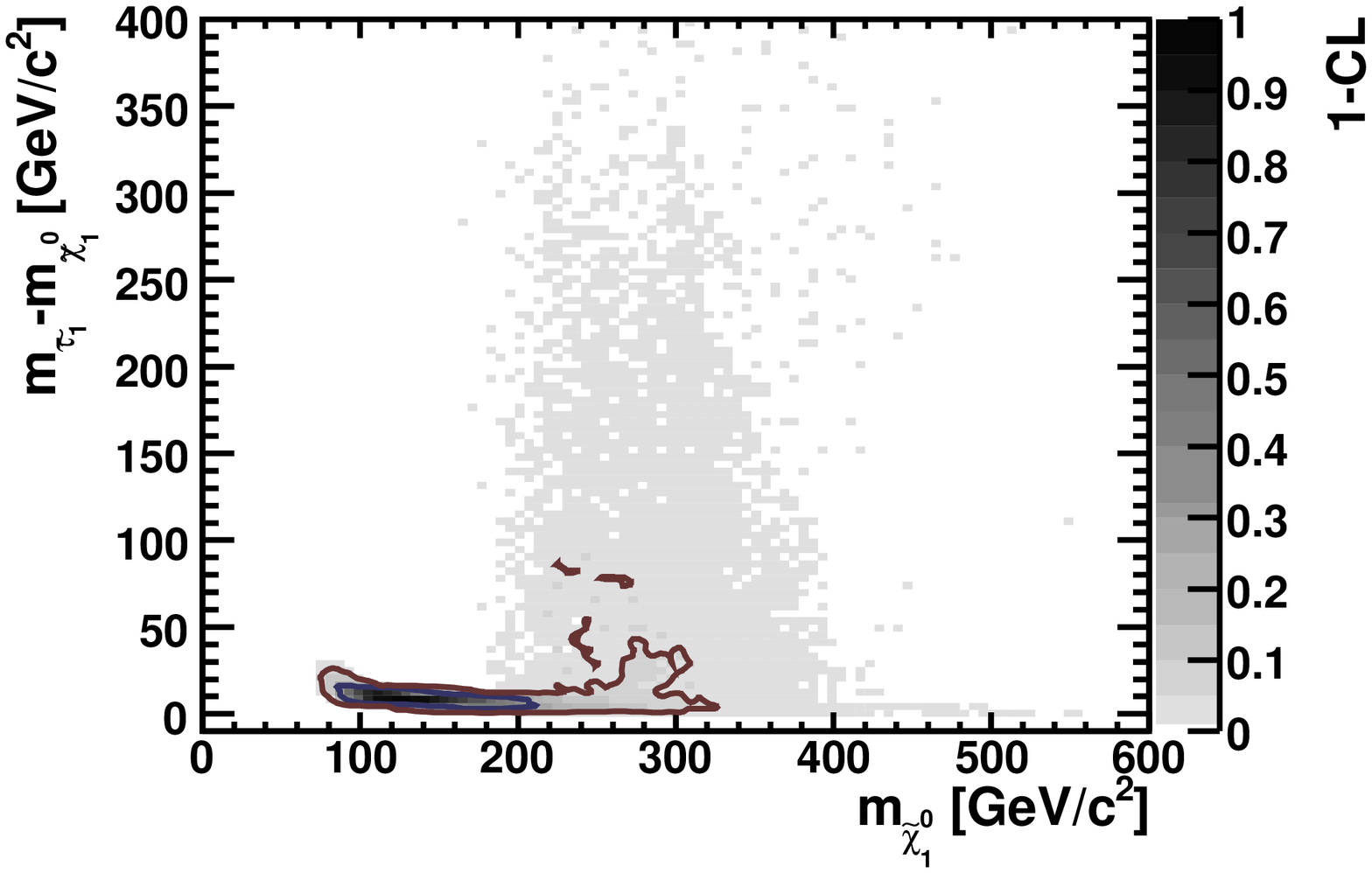}}
\resizebox{8cm}{!}{\includegraphics{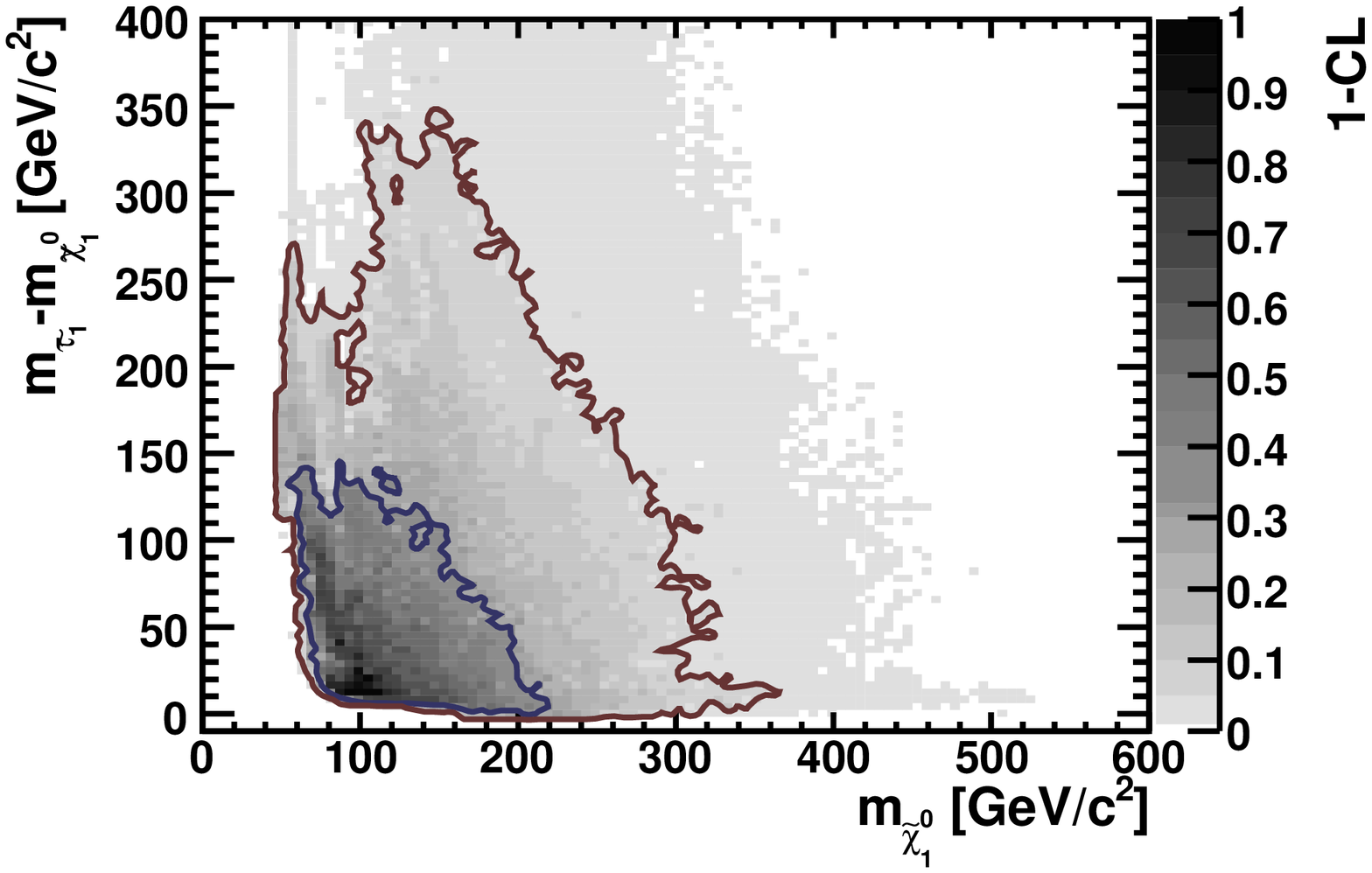}}
%%%%%%%%%%%%%%%%%%%%%%%%%%%%%%%
\vspace{-2em}
\caption{\it The correlation between the $\staue$ - $\neu{1}$ mass difference
as a function of the LSP mass, in the CMSSM (left panel) and in the
NUHM1 (right panel). 
}
\label{fig:mstau-mLSP}
\vspace{1em}
\end{figure*}

Fig.~\ref{fig:mstau-mLSP} demonstrates explicitly the big contrast
between the behaviours of the $\staue$ - $\neu{1}$ mass difference in
the CMSSM (left panel) and the NUHM1 (right panel). We see that in the
CMSSM small mass differences are always favoured, and are mandatory for
LSP masses $\lsim 200 \gev$, whereas larger mass differences are possible
for LSP masses $\gsim 200 \gev$, as the rapid-annihilation funnel opens
up. However, in the NUHM1 large mass differences are possible for all
LSP masses, {\it particularly} for LSP masses $\lsim 200 \gev$. This
means that, whereas in the CMSSM the `visible' $\staue$ pair-production
threshold at the ILC may be only slightly higher than the `invisible'
$\neu{1}$ pair-production threshold, it may be considerably higher in
the NUHM1, namely $\mstaue \lsim 400 \gev$ at the 95\% C.L..
This is a potentially crucial signature for distinguishing the NUHM1
from the CMSSM.

%%%%%%%%%%%%%%%%%%%%%% F I G U R E %%%%%%%%%%%%%%%%%%%%%%%%%%%%%%%%%%%
%%%%%%%%%%%%%%%%%%%%%% F I G U R E %%%%%%%%%%%%%%%%%%%%%%%%%%%%%%%%%%%

\begin{figure*}[htb!]
%%%%%%%%%%%%%%%%%%%%%%%%%%%%%%%
\resizebox{8cm}{!}{\includegraphics{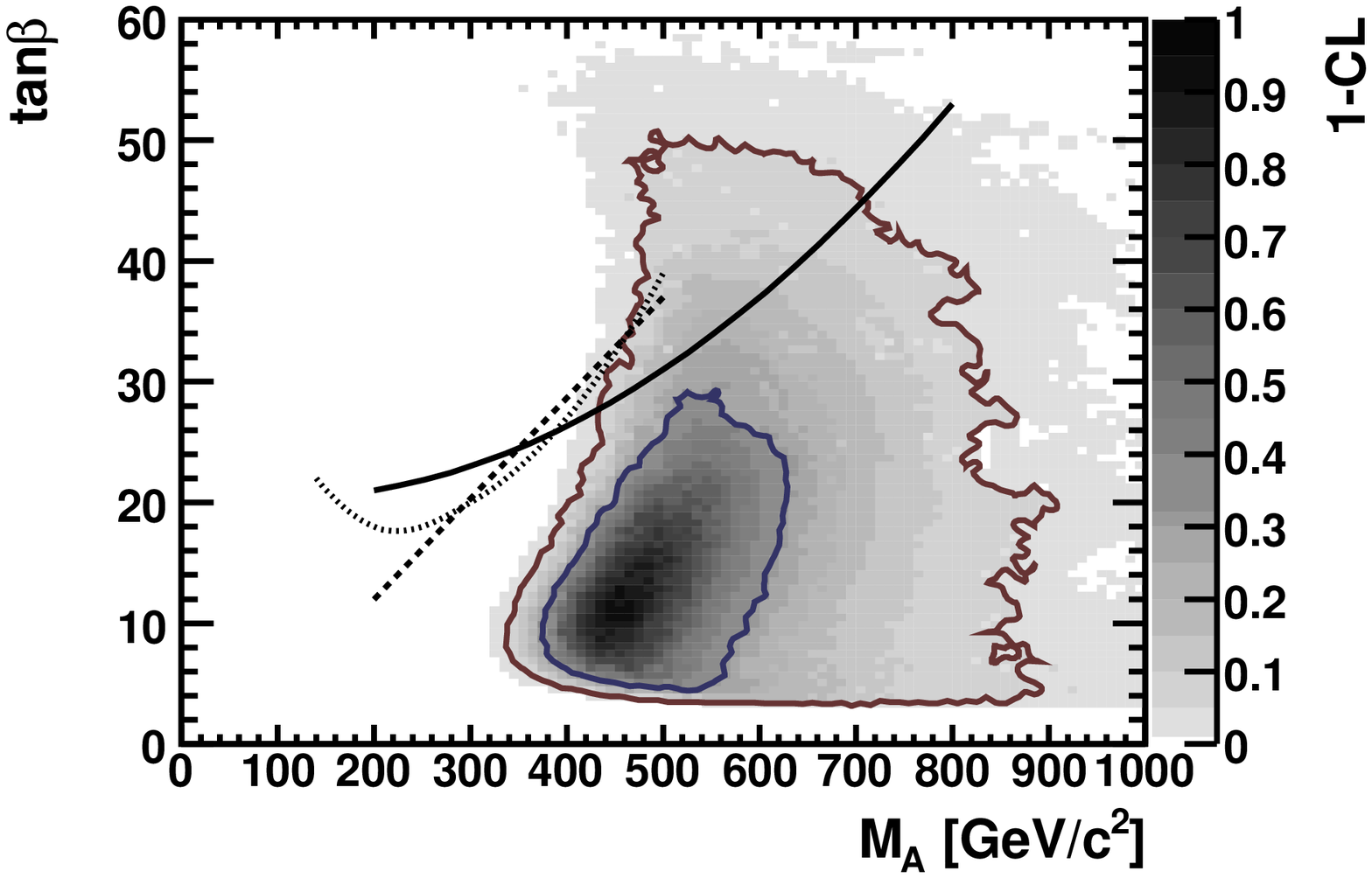}}
\resizebox{8cm}{!}{\includegraphics{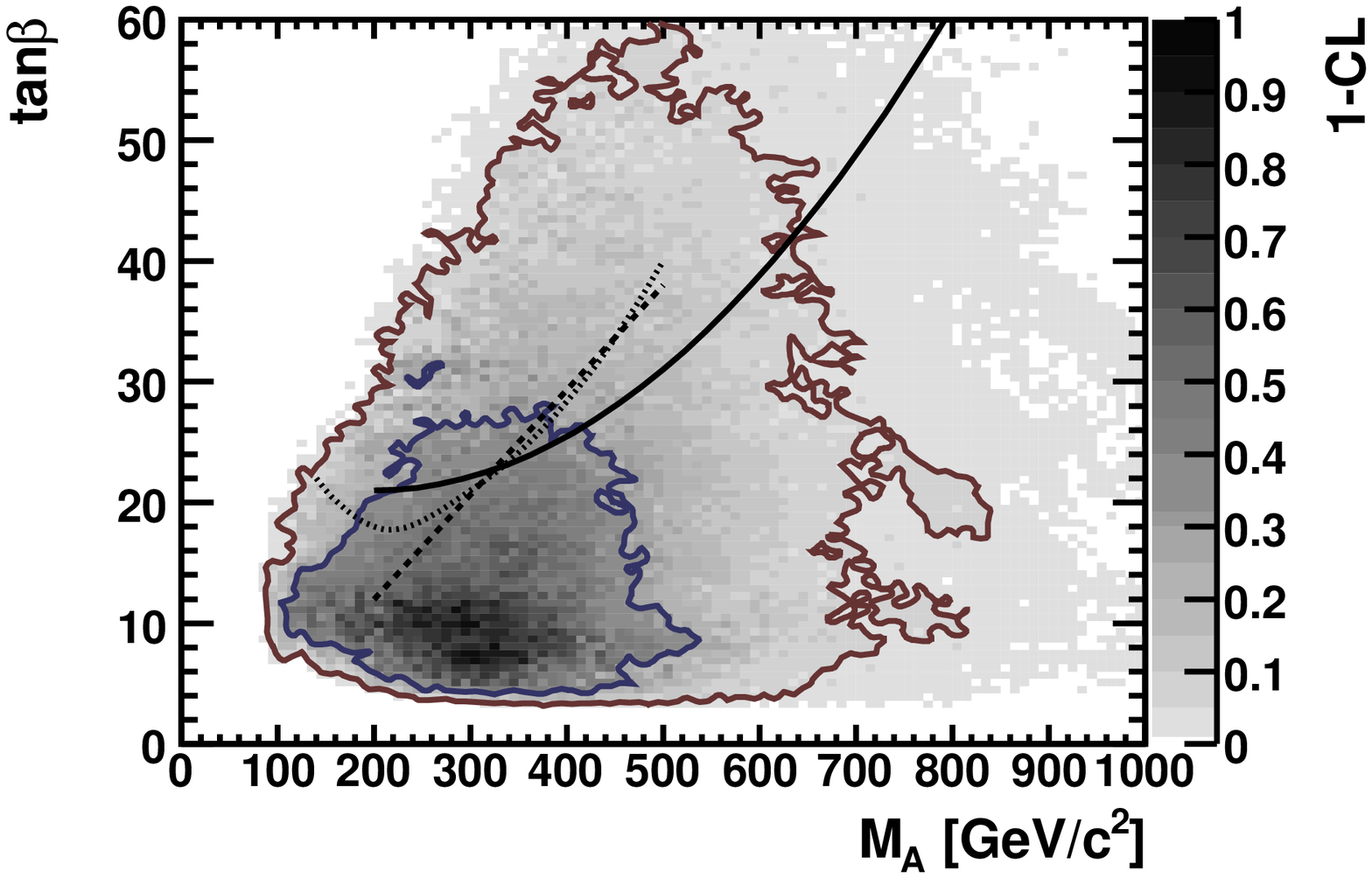}}
%%%%%%%%%%%%%%%%%%%%%%%%%%%%%%%
\vspace{-2em}
\caption{\it The correlations between $\MA$ and $\tb$
in the CMSSM (left panel) and in the NUHM1 (right panel).
Also shown are the 5-$\sigma$ discovery contours 
for observing the heavy MSSM Higgs bosons $H, A$ 
in the three decay channels
$H,A \to \tau^+\tau^- \to {\rm jets}$ (solid line), 
$\rm{jet}+\mu$ (dashed line), $\rm{jet}+e$ (dotted line)
at the LHC. The discovery contours have been obtained using an 
analysis that assumed 30 or 60~fb$^{-1}$ 
collected with the CMS detector~\cite{cmstdr,cmsHiggs}.
}
\label{fig:mAtb}
\vspace{3em}
\end{figure*}

Fig.~\ref{fig:mAtb} displays the favoured regions in the $(\MA, \tb)$ planes
for the CMSSM and NUHM1. We see that they are broadly similar, with little
correlation between the two parameters. 
$(\MA, \tb)$ planes in certain benchmark scenarios have often been 
used in the past to analyze the prospects for discovering heavy
Higgs bosons at the LHC~\cite{benchmark2,benchmark3}. Most of these 
analyses have been done in the context of scenarios that do not take the
relic-density constraint into account, for exceptions
see~\cite{Ellis:2007aa,Ellis:2007ka}.
The Higgs discovery contours determined in the various benchmark scenarios 
cannot directly be applied to the $(\MA, \tb)$ planes in
Fig.~\ref{fig:mAtb} displaying our fit 
results for the CMSSM and the NUHM1.
In order to assess the prospects for discovering heavy Higgs bosons at
the LHC in this context,
we follow the analysis in~\cite{cmsHiggs}, which assumed 30
or 60~fb$^{-1}$ collected with the CMS detector. For evaluating the
Higgs-sector observables including higher-order corrections we use 
the soft
SUSY-breaking parameters of the best-fit points in the CMSSM and the
NUHM1, respectively. We show in Fig.~\ref{fig:mAtb} the 5-$\sigma$
discovery contours for the three decay channels 
$H,A \to \tau^+\tau^- \to {\rm jets}$ (solid lines), $\rm{jet}+\mu$ (dashed
lines) and $\rm{jet}+e$ (dotted lines).
The parameter regions above and to the left of the curves are within reach 
of the LHC with about 30~fb$^{-1}$ of integrated luminosity.
We see that most of the highest-CL regions lie beyond this reach,
particularly in the CMSSM.
At the ILC(1000) masses up to $\MA \lsim 500 \gev$ can be probed. 
Within the CMSSM this includes the best-fit point, and within the NUHM1
nearly the whole 68\% C.L.\ area can be covered.

%%%%%%%%%%%%%%%%%%%%%% F I G U R E %%%%%%%%%%%%%%%%%%%%%%%%%%%%%%%%%%%
%%%%%%%%%%%%%%%%%%%%%% F I G U R E %%%%%%%%%%%%%%%%%%%%%%%%%%%%%%%%%%%

\begin{figure*}[htb!]
%%%%%%%%%%%%%%%%%%%%%%%%%%%%%%%
\resizebox{8cm}{!}{\includegraphics{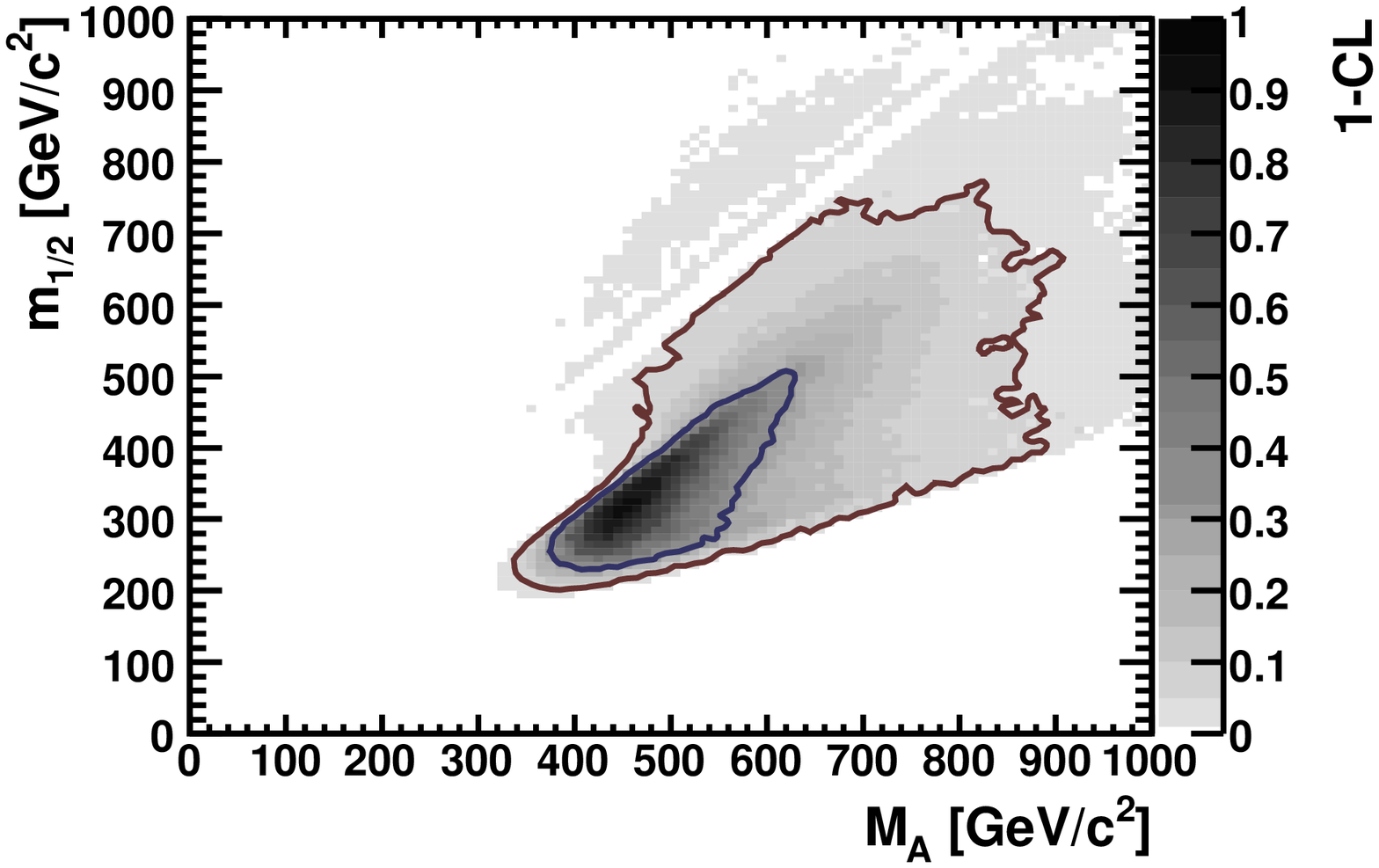}}
\resizebox{8cm}{!}{\includegraphics{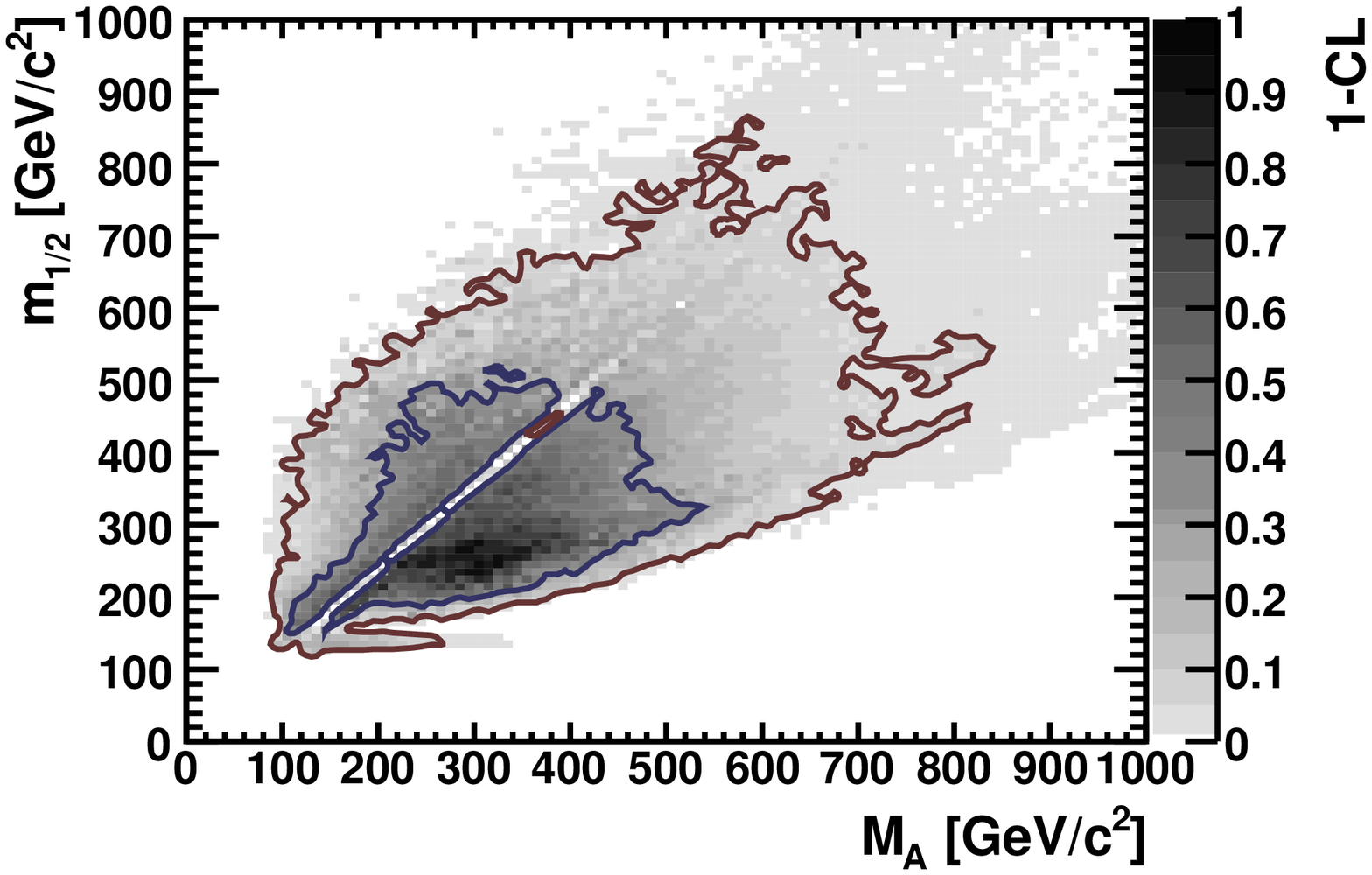}}
%%%%%%%%%%%%%%%%%%%%%%%%%%%%%%%
\vspace{-2em}
\caption{\it The correlations between $\MA$ and $m_{1/2}$
in the CMSSM (left panel) and in the NUHM1 (right panel).
}
\label{fig:mAm12}
\vspace{2em}
\end{figure*}

We display in Fig.~\ref{fig:mAm12} the correlations between $\MA$ and $m_{1/2}$
in the CMSSM and in the NUHM1. In the former case, the electroweak boundary
conditions fix $\MA$, and the effect is to force 
$\MA > 2 \mneu{1}$. However, $\MA$ 
becomes essentially a free parameter in the NUHM1, and values smaller
than $\mneu{1}$ 
become possible also. On the other hand, there is a narrow strip where
$\MA \sim 2 \mneu{1}$ which is disfavoured because there rapid direct-channel
annihilation suppresses the relic density below the range preferred by
astrophysics and cosmology. The points with $\MA < 2 \mneu{1}$ are a
qualitatively 
new possibility opened up within the NUHM1 as compared to the
CMSSM, and 
extend to relatively large values of $m_{1/2}$.

%%%%%%%%%%%%%%%%%%%%%% F I G U R E %%%%%%%%%%%%%%%%%%%%%%%%%%%%%%%%%%%
%%%%%%%%%%%%%%%%%%%%%% F I G U R E %%%%%%%%%%%%%%%%%%%%%%%%%%%%%%%%%%%

\begin{figure*}[htb!]
%%%%%%%%%%%%%%%%%%%%%%%%%%%%%%%
\resizebox{8cm}{!}{\includegraphics{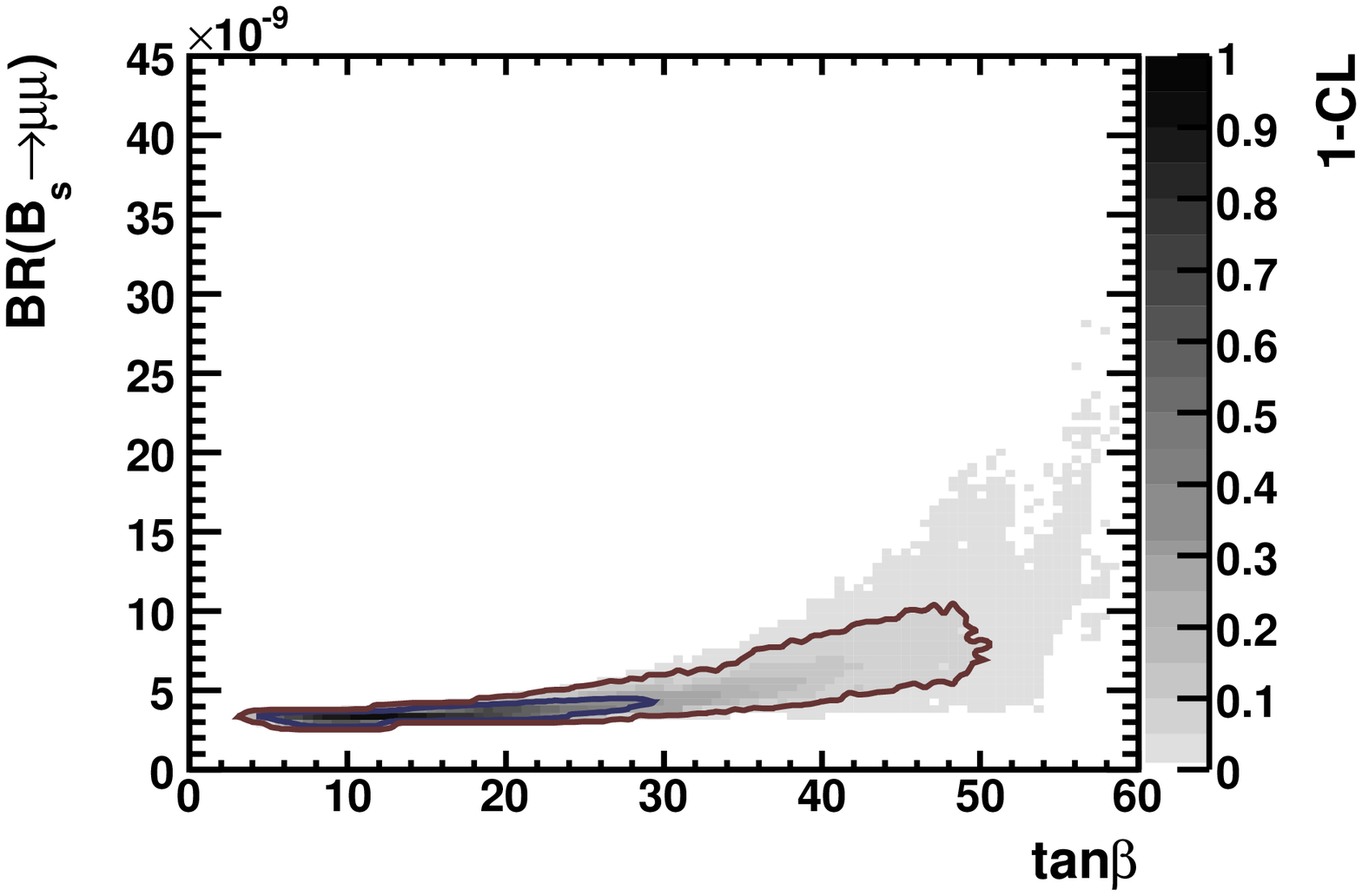}}
\resizebox{8cm}{!}{\includegraphics{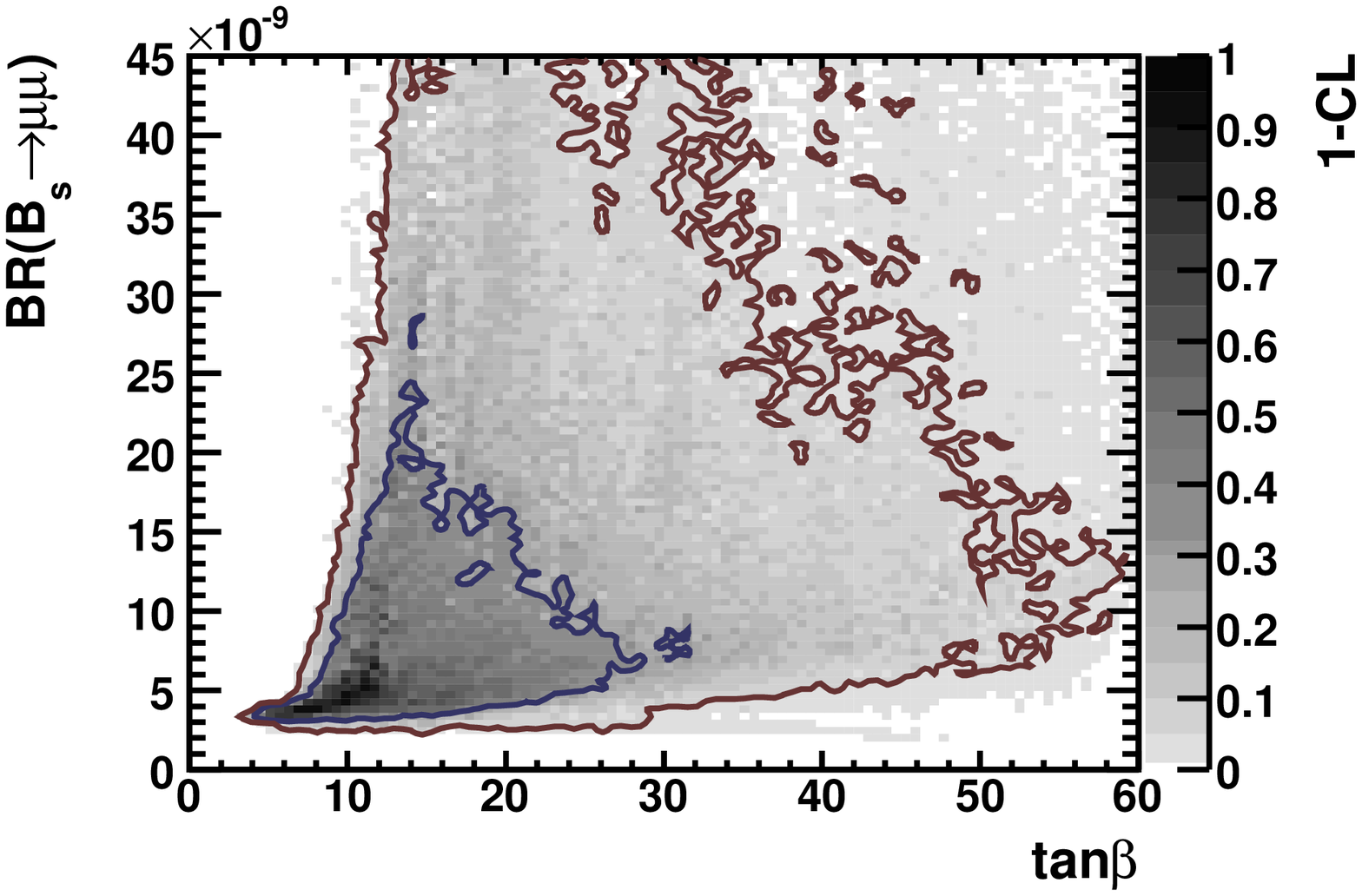}}
%%%%%%%%%%%%%%%%%%%%%%%%%%%%%%%
\vspace{-1cm}
\caption{\it The correlation between branching ratio for \bmm\ and $\tb$
in the CMSSM (left panel) and in the NUHM1 (right panel).
}
\label{fig:tbbsmumu}
\end{figure*}

Fig.~\ref{fig:tbbsmumu} displays the correlation between $\tb$
and the \bmm. As seen previously,
in the CMSSM the preferred values of the branching ratio are very close
to the value in the SM, though somewhat larger values
may occur at large $\tb$, which however have a lower likelihood.
The situation is completely different in the NUHM1, where much larger
values of the branching ratio for \bmm\ are possible,
even if $\tb \sim 10$. This increase reflects the possibility
that $\MA$ may be considerably smaller than in the CMSSM.
The upper right corner of the NUHM1 plot, i.e., simultaneous large $\tb$
and  large \bmm, is disfavoured because it would give rise to
values of \bsg\ that are too small.

%%%%%%%%%%%%%%%%%%%%%% F I G U R E %%%%%%%%%%%%%%%%%%%%%%%%%%%%%%%%%%%
%%%%%%%%%%%%%%%%%%%%%% F I G U R E %%%%%%%%%%%%%%%%%%%%%%%%%%%%%%%%%%%

\begin{figure*}[htb!]
%%%%%%%%%%%%%%%%%%%%%%%%%%%%%%%
\resizebox{8cm}{!}{\includegraphics{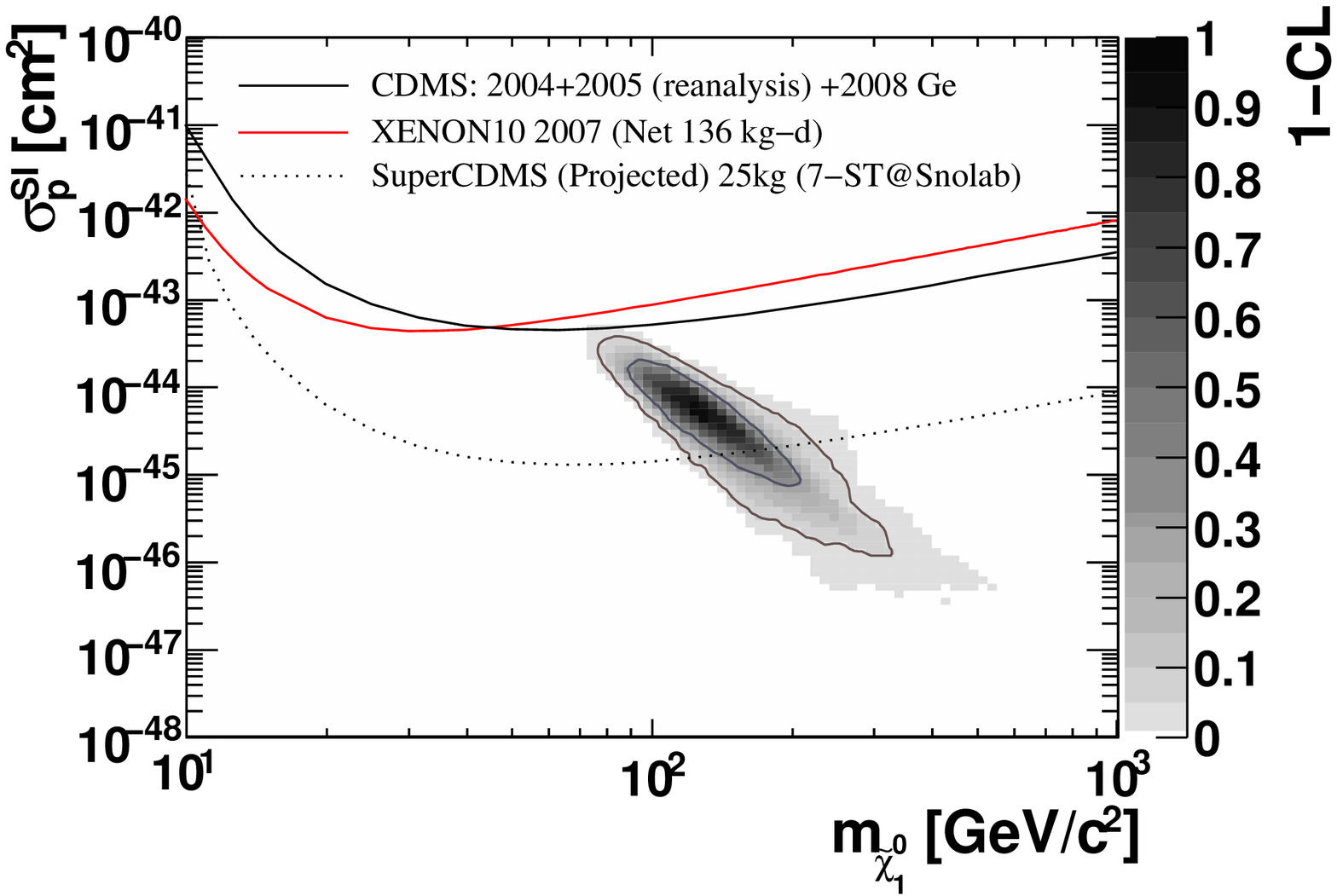}}
\resizebox{8cm}{!}{\includegraphics{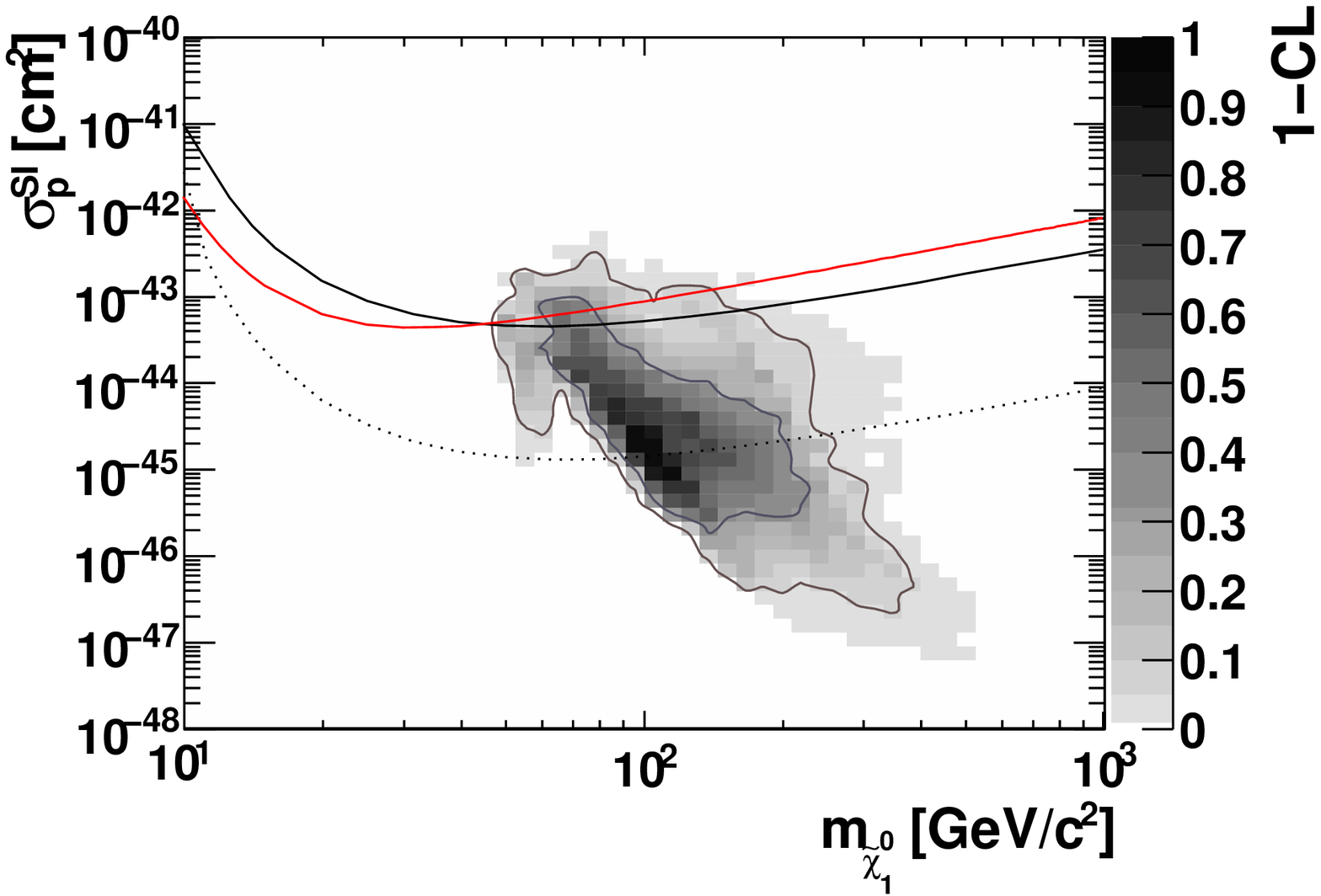}}
%%%%%%%%%%%%%%%%%%%%%%%%%%%%%%%
\vspace{-1cm}
\caption{\it The correlation between the spin-independent DM
scattering cross section $\ssi$ 
(calculated assuming a $\pi$-N scattering $\sigma$ term
$\Sigma_N = 64$~MeV) and $\mneu{1}$
in the CMSSM (left panel) and in the NUHM1 (right panel). The solid lines~\cite{DMtool}
are the present experimental upper limits from CDMS~\cite{CDMS} and XENON10\cite{Xe10}, 
and the dashed line~\cite{DMtool} indicates the
projected sensitivity of the SuperCDMS 
experiment~\cite{superCDMS}.
}
\label{fig:mchisig}
\end{figure*}

Fig.~\ref{fig:mchisig} displays the preferred range of the
spin-independent DM scattering cross section $\ssi$
(calculated assuming a $\pi$-N scattering $\sigma$ term
$\Sigma_N = 64$~MeV) as a function of $\mneu{1}$. 
In the case of the CMSSM, we see that
the expected range of $\ssi$ lies mainly between the present
experimental upper limits (solid lines) \cite{CDMS,Xe10}, 
which start to touch the preferred region,  
and the projected sensitivity of the SuperCDMS experiment (dashed line)~\cite{superCDMS},
which should cover the preferred region. As noted earlier, 
these experimental constraints were not applied in our analysis. 
The uncertainty in $\Sigma_N$ and the astrophysical uncertainties in the local dark matter density
(which are difficult to quantify), 
preclude including the value of $\ssi$ in the likelihood analysis presented here.
This region is in good agreement for neutralino masses between 100--300~GeV 
with that found in~\cite{Ellis:2009ai}, where a recent scan 
(without likelihood information) was performed.

As already commented, the range in the NUHM1
is larger than in the CMSSM. We see in Fig.~\ref{fig:mchisig} that the
larger cross-section values occur, as expected, for small $\mneu{1}$, in
particular in the small island of Higgsino-like DM that appears
close to the 95\% C.L. for $\mneu{1} < 100 \gev$. If $100~{\rm GeV} <
\mneu{1} < 200 \gev$, the allowed range of the cross section is larger
than in the CMSSM because of the wider range of possible values of
$\MA$ as found in \cite{Ellis:2009ai} for this neutralino mass range,
and the present experimental sensitivity is already below the values of $\ssi$
found for some favoured NUHM1 parameter values assuming the nominal values of
$\Sigma_N$ and the local LSP density. 
The smallest values of the cross section occur when $\mneu{1} >
200 \gev$, in models close to the 95\% C.L. limit for the NUHM1,
which have $\MA$ larger than in the CMSSM. In general,
we see that whereas the favoured values of $\ssi$ are close to the present
experimental upper limits~\cite{DMtool} in both the CMSSM and the
NUHM1, there is a greater possibility in the NUHM1 that the cross section may
lie beyond the projected sensitivity of SuperCDMS~\cite{superCDMS}.

%%%%%%%%%%%%%%%%%%%%%% F I G U R E %%%%%%%%%%%%%%%%%%%%%%%%%%%%%%%%%%%
%%%%%%%%%%%%%%%%%%%%%% F I G U R E %%%%%%%%%%%%%%%%%%%%%%%%%%%%%%%%%%%

%---------------------------------------------------------------------
\section{Dropping Constraints}
\label{sec:drop}

\subsection{Dropping the \boldmath{\gmt\ } Constraint}
\label{sec:dropgmt}
%---------------------------------------------------------------------
~\\
We have stressed above that the results in the previous section are strongly
dependent on the implementation of the \gmt\ constraint. In particular, we
have displayed in Fig.~\ref{fig:m0} above the likelihood functions for $m_0$
in the CMSSM (left) and the NUHM1 (right) both with the \gmt\ constraint
imposed (solid) and without it (dashed). We now discuss in more detail
the effect of dropping the \gmt\ constraint completely, calculating a new 
$\chi^2_{\rm loose}$ with no contribution from \gmt.

The $\chi^2$ function obtained for $m_0$ in the CMSSM without the
\gmt\ constraint, shown in the left panel of Fig.~\ref{fig:m0}, is much
flatter than the corresponding $\chi^2$ function obtained with the \gmt\
constraint. Nevertheless, we see non-trivial features in the $\chi^2$
function. One is that the location of the CMSSM global minimum is very similar
to the case with the \gmt\ constraint applied. We recall that the rise in
the $\chi^2$ function at small $m_0$ is determined essentially by the $\Mh$
and \bsg\ constraints, with \gmt\ not playing a role. However, it is perhaps
surprising that the other constraints cause $\chi^2$ to rise until $m_0
\sim 1000 \gev$, where $\Delta \chi^2 \sim 3$.
However, we see in Table~\ref{tab:chi2} that, in addition to \gmt,
several other constraints favour  
the best-fit CMSSM point over points with $m_0 > 1000 \gev$, including
\btn, $M_W$, $A_\ell$ and $R_\ell$. Continuing to larger $m_0$ in 
the left panel of Fig.~\ref{fig:m0},
we see that $\chi^2$ decreases again slightly, but that still 
$\Delta \chi^2 \gsim 2$.

Similar features are seen in the $\chi^2$ function obtained for $m_0$ in
the NUHM1 without the \gmt\ constraint, shown in the right panel of
Fig.~\ref{fig:m0}. Again, the value of $m_0$ at the best-fit NUHM1 point
is very similar, whether \gmt\ is included or not, and again $\Delta
\chi^2 \gsim 2$ at large $m_0$. However, there is no intermediate hump
at $m_0 \sim 1000 \gev$ analogous to that in the CMSSM, reflecting the
greater freedom in the NUHM1 to adjust parameters 
so as to obtain a lower value of $\chi^2$.

\begin{figure*}[htb!]
%%%%%%%%%%%%%%%%%%%%%%%%%%%%%%%
\resizebox{8cm}{!}{\includegraphics{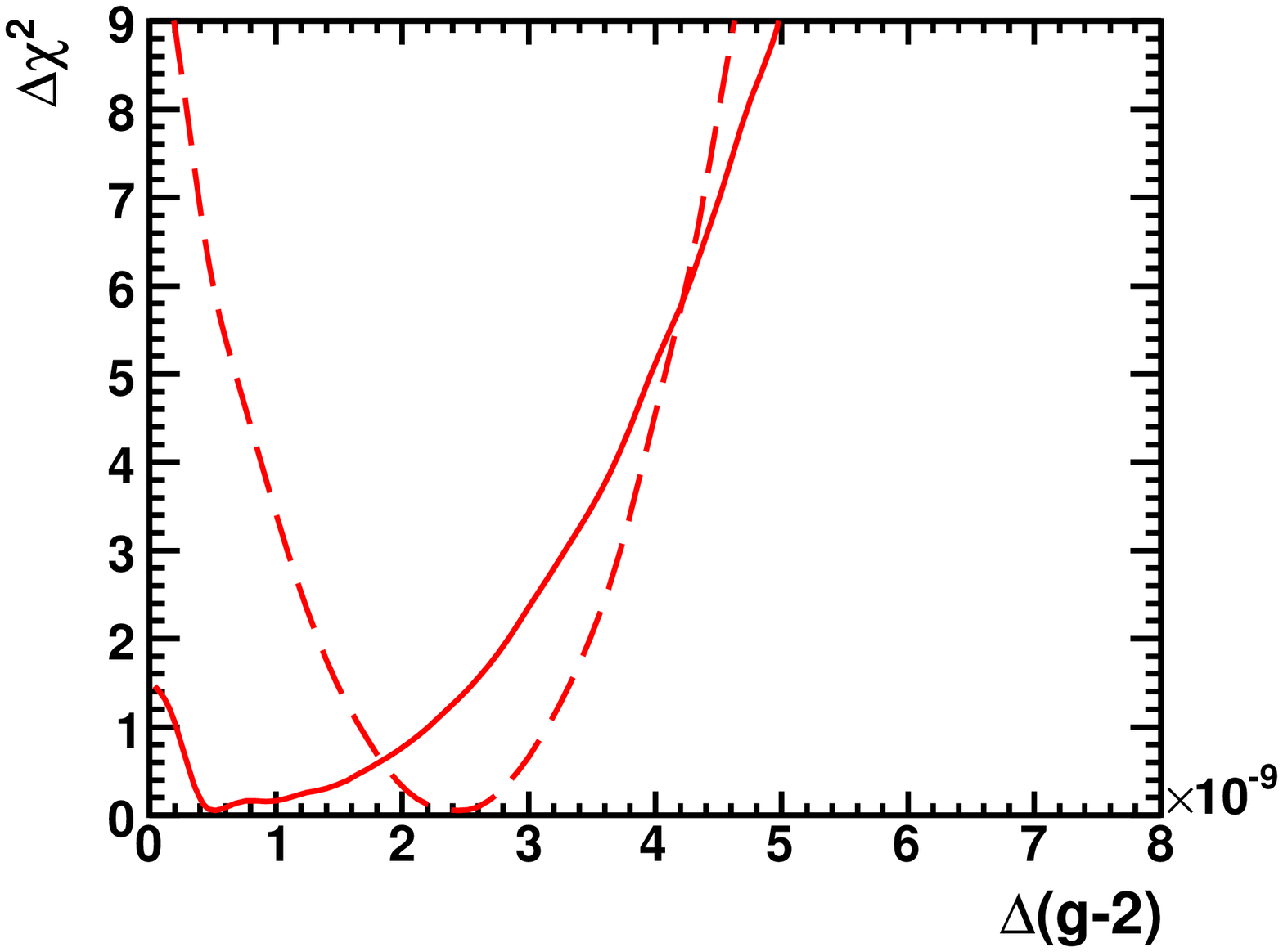}}
\resizebox{8cm}{!}{\includegraphics{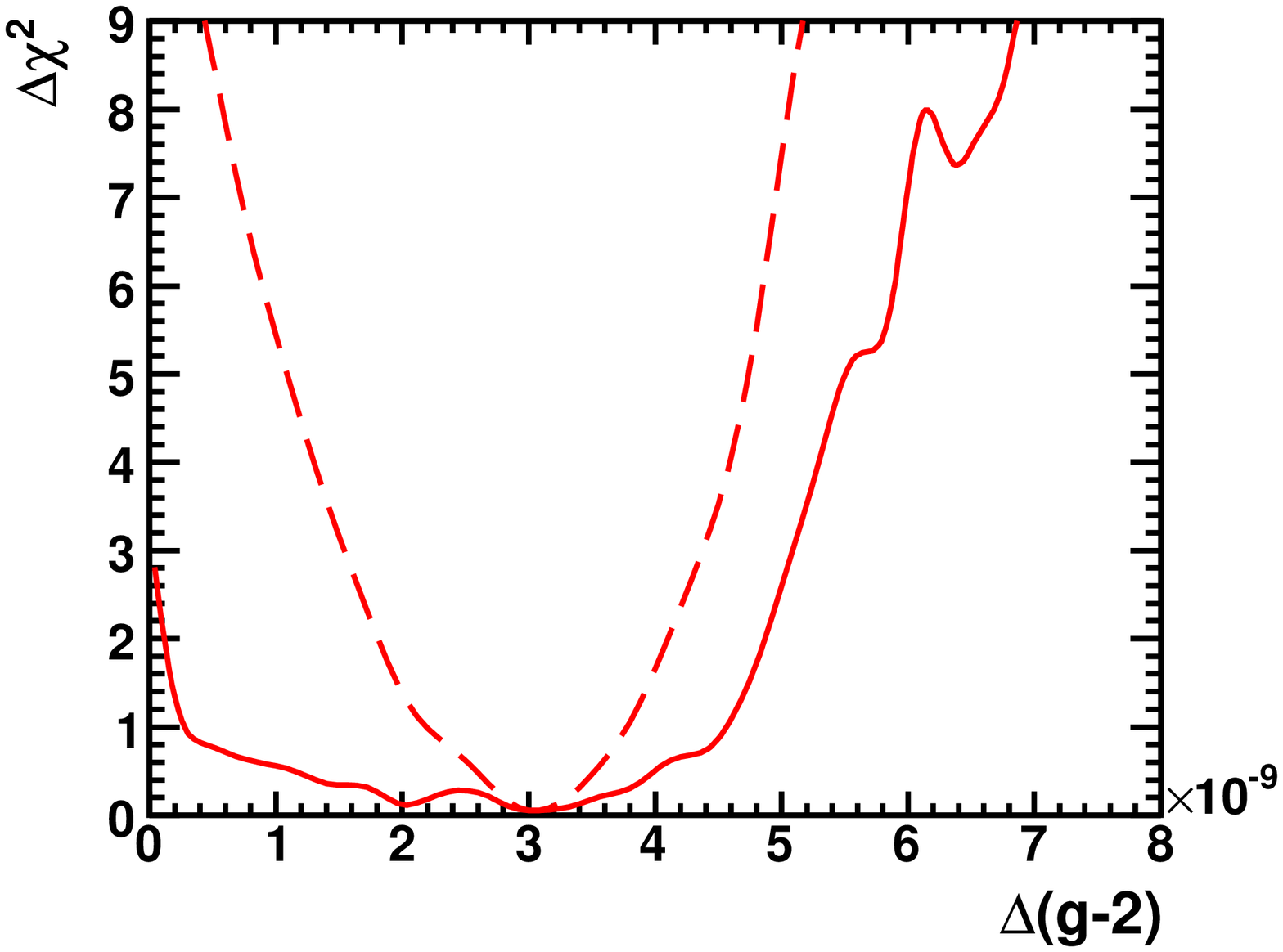}}
%%%%%%%%%%%%%%%%%%%%%%%%%%%%%%%
\vspace{-1cm}
\caption{\it The $\chi^2$ functions for the supersymmetric contributions
  to \gmt\ in the CMSSM (left) and the NUHM1 (right), as calculated
  using the other constraints except \gmt\ itself (solid line),
  and with all constraints included (dashed line).
}
\label{fig:gmt}
\end{figure*}

A corollary of the observations in the previous paragraphs is that, at
some level, the other constraints favour a non-zero supersymmetric
contribution to \gmt. This is indeed visible in Fig.~\ref{fig:gmt},
where we see the predicted values of the contributions of
supersymmetric particles to \gmt\ in the CMSSM (left) and the
NUHM1 (right). We show the $\chi^2$ functions only for positive
contributions to \gmt, since our points were all chosen 
to have $\mu > 0$. Nevertheless, the fact that the minima of the $\chi^2$
distributions are for $\Delta($\gmt$) \ne 0$ is non-trivial, because it
reflects the above observation that large values of the sparticle masses
are disfavoured, and the order of magnitude prediction for
$\Delta($\gmt$)$ agrees with estimates based on low-energy $e^+ e^-$ data.

%%%%%%%%%%%%%%%%%%%%%% F I G U R E %%%%%%%%%%%%%%%%%%%%%%%%%%%%%%%%%%%
%%%%%%%%%%%%%%%%%%%%%% F I G U R E %%%%%%%%%%%%%%%%%%%%%%%%%%%%%%%%%%%

\subsection{Dropping the \boldmath{\bsg\ } Constraint}

\begin{figure*}[htb!]
%%%%%%%%%%%%%%%%%%%%%%%%%%%%%%%
\resizebox{8cm}{!}{\includegraphics{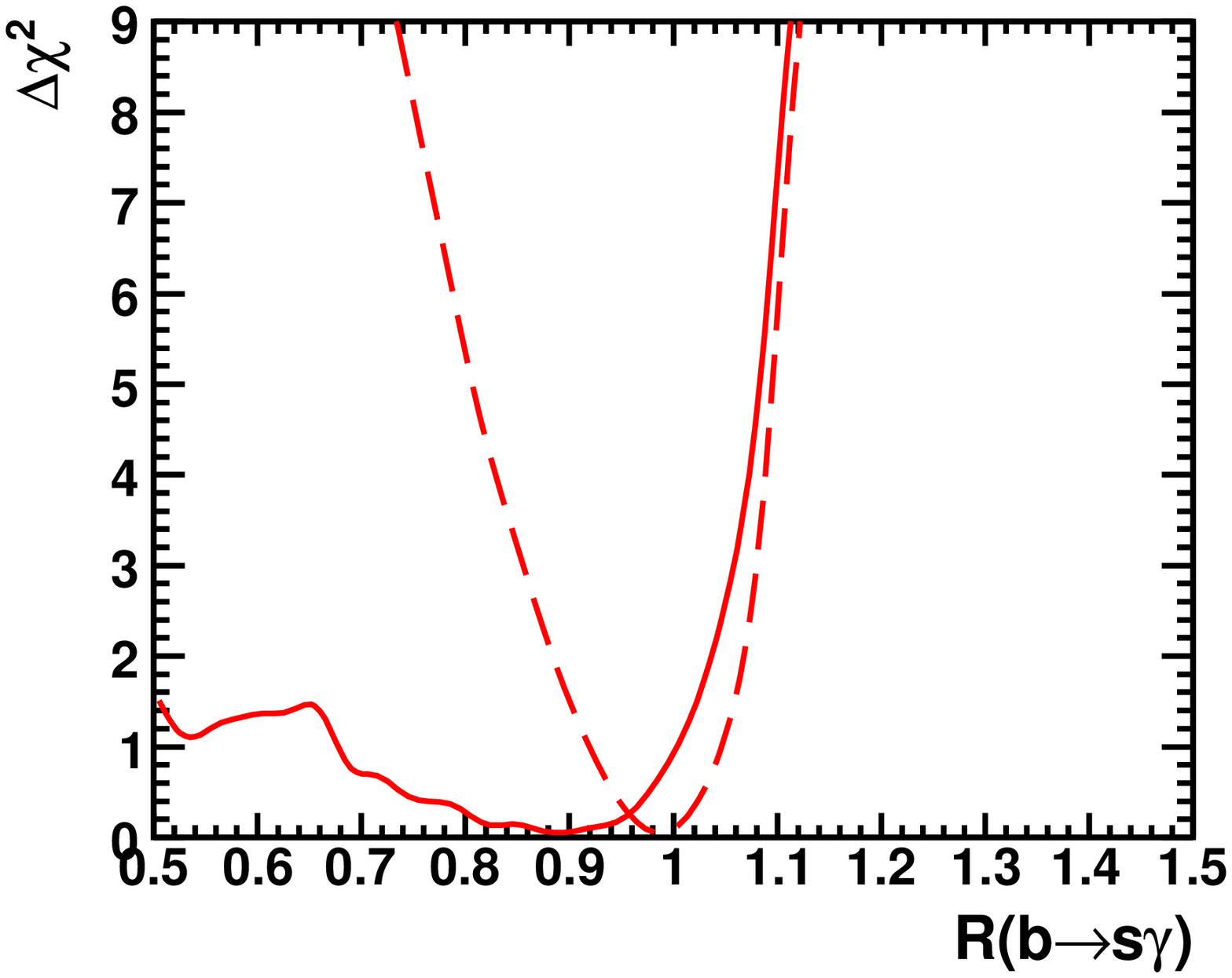}}
\resizebox{8cm}{!}{\includegraphics{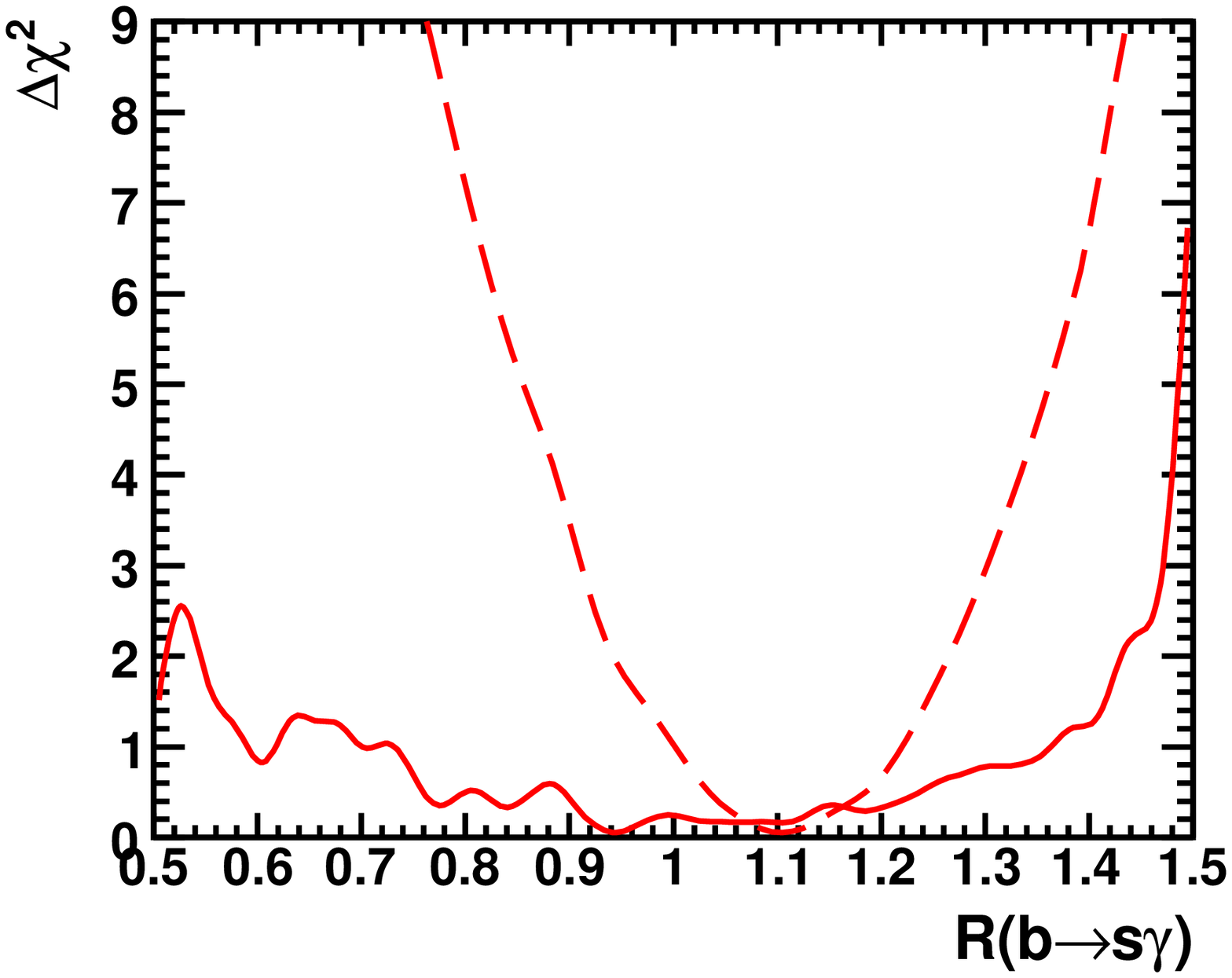}}
%%%%%%%%%%%%%%%%%%%%%%%%%%%%%%%
\vspace{-1cm}
\caption{\it The $\chi^2$ functions for the ratio 
of the MSSM prediction over the SM prediction to \bsg, 
$R(b \to s \gamma) \equiv {\rm BR}_{\rm b \to s \gamma}^{\rm SUSY}/ 
  {\rm BR}_{\rm b \to s \gamma}^{\rm SM}$,
in the CMSSM (left) and the NUHM1 (right), as calculated
  using the other constraints except \bsg\ itself (solid line),
  and with all constraints included (dashed line).
}
\label{fig:bsg}
\end{figure*}

~\\
We display in Fig.~\ref{fig:bsg} the effects on the CMSSM and NUHM1 fits
(left and right panels, respectively) of omitting the \bsg\ constraint from
the global fit, as obtained by calculating a new 
$\chi^2_{\rm loose}$ with no contribution from \bsg. 
In both models, we see that the predictions for \bsg\ based
on the other constraints (solid lines) are not very precise. The
best-fit values for \bsg\ 
are in both models quite close to the SM and hence the
experimental value, but the CMSSM permits much smaller values, and both
larger and smaller values are allowed in the NUHM1 with relatively
small increases in $\chi^2$.

The converse statement is that applying the \bsg\ constraint does not
impose a large $\chi^2$ price on the global minimum. 
This is apparent from Table~\ref{tab:chi2}, where we saw that 
\bsg\ contributes about $\Delta \chi^2 \sim 1$ to the total 
$\chi^2$ in the CMSSM and yields a negligible contribution in the NUHM1.
There is no tension between \bsg\
and the other constraints.

%%%%%%%%%%%%%%%%%%%%%% F I G U R E %%%%%%%%%%%%%%%%%%%%%%%%%%%%%%%%%%%
%%%%%%%%%%%%%%%%%%%%%% F I G U R E %%%%%%%%%%%%%%%%%%%%%%%%%%%%%%%%%%%

\subsection{Dropping the \boldmath{\Och\ } Constraint}

~\\
One of the most exciting predictions of the CMSSM and the NUHM1 is the
existence of a cold dark matter candidate in the form of the LSP,
which we assume here to be the lightest neutralino~\cite{EHNOS}. 
It is natural to
take the next step, and ask whether these models predict a relic LSP
density that is close to the experimental value of the cold dark
matter density. This density is determined
with an accuracy of a few percent, and 
an comparable accuracy in the prediction based on the other
available experimental constraints will be difficult. This will improve if
(when) the LHC discovers SUSY and its parameters are
measured more accurately at an $e^+e^-$ linear
collider~\cite{Battaglia:2003ab,Nojiri:2005ph,Baltz:2006fm,fittinoMC}. 

\begin{figure*}[htb!]
%%%%%%%%%%%%%%%%%%%%%%%%%%%%%%%
\resizebox{8cm}{!}{\includegraphics{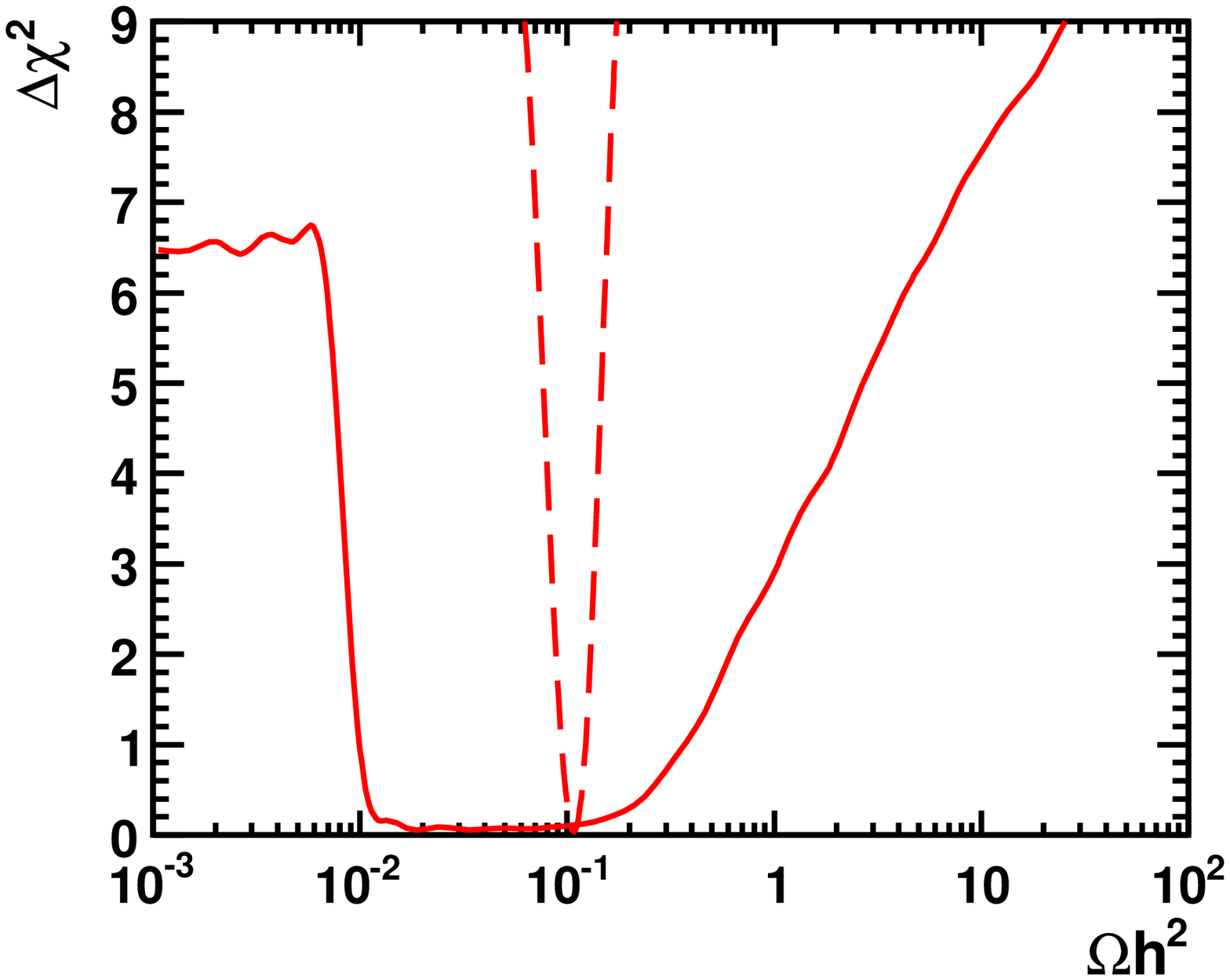}}
\resizebox{8cm}{!}{\includegraphics{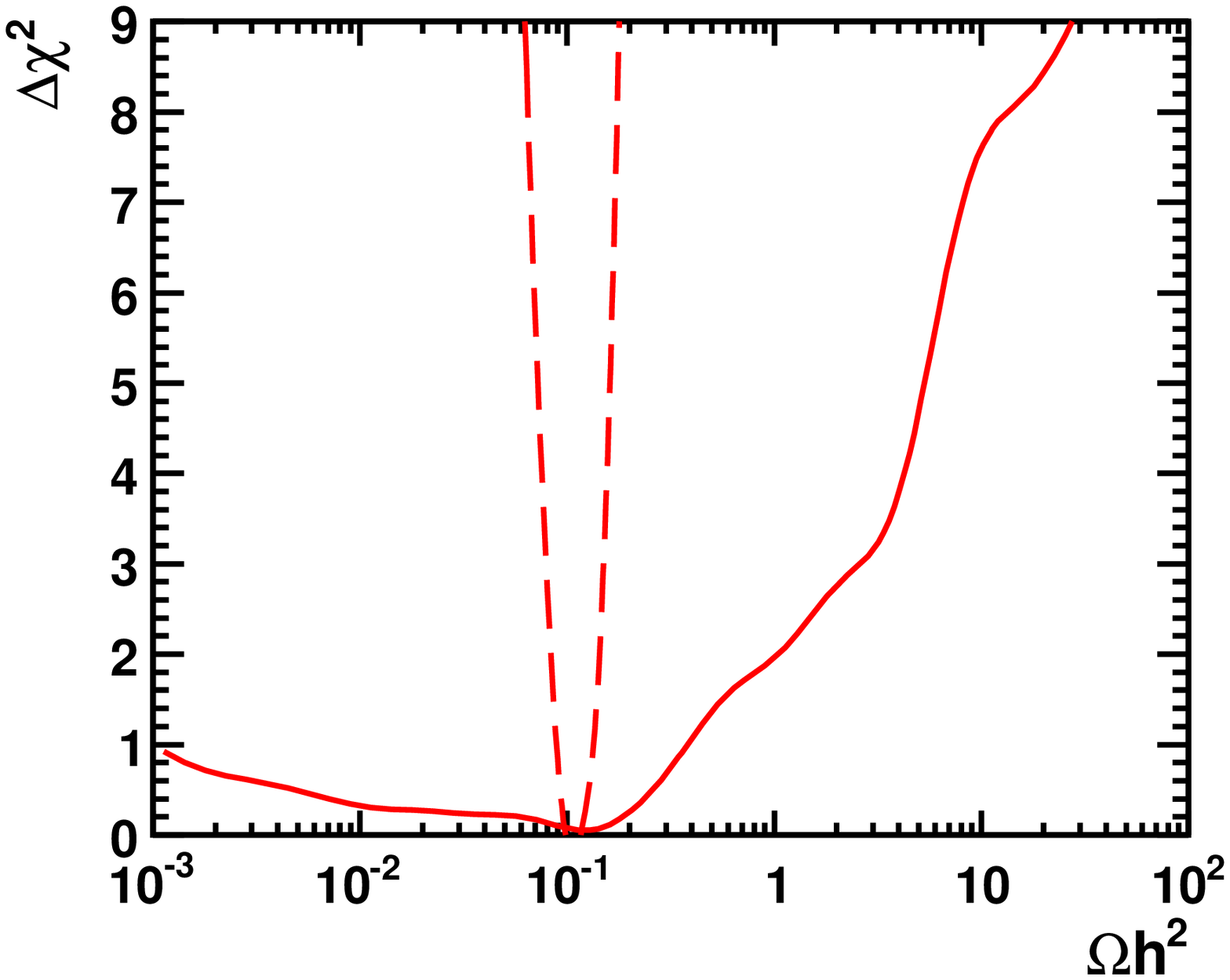}}
%%%%%%%%%%%%%%%%%%%%%%%%%%%%%%%
\vspace{-1cm}
\caption{\it The $\chi^2$ functions for the supersymmetric contributions
  to \Och\ in the CMSSM (left) and the NUHM1 (right), as calculated
  using the other constraints except \Och\ itself (solid line),
  and with all constraints included (dashed line).
}
\label{fig:och}
\end{figure*}

Nevertheless, calculating a new 
$\chi^2_{\rm loose}$ with no contribution from \Och,
it is interesting to see in Fig.~\ref{fig:och} that both
the CMSSM (left) and NUHM1 (right) favour ranges of \Och\ values
that include the measured values of the cold dark matter density.
In the case of the CMSSM, the prediction for \Och\ (solid line)
is within an order of magnitude above and below the measured value at the level
$\Delta \chi^2 < 4$. This is also true in the NUHM1 above the measured value,
but the relic LSP density could be two or more orders of magnitude below
the measured value with $\Delta \chi^2 < 1$. This is because there is a
possibility that the relic density may be suppressed by rapid
annihilation through direct-channel Higgs poles in the region of
relatively low $m_{1/2}$ and $\tb$ in the NUHM1 that is favoured by the
other constraints, notably \gmt. 

%%%%%%%%%%%%%%%%%%%%%% F I G U R E %%%%%%%%%%%%%%%%%%%%%%%%%%%%%%%%%%%
%%%%%%%%%%%%%%%%%%%%%% F I G U R E %%%%%%%%%%%%%%%%%%%%%%%%%%%%%%%%%%%

The converse statement is that applying the \Och\ constraint also does not
impose a large $\chi^2$ price on the global minimum. In fact, 
$\Delta \chi^2 \ll 1$ in both the CMSSM and the NUHM1, and therefore 
has not been listed in Table~\ref{tab:chi2}.
As in the case of \bsg, there is no tension between \Och\
and the other constraints.

\subsection{Dropping the \boldmath{$\Mh$} Constraint}

\label{sec:dropMh}

\begin{figure*}[htb!]
%%%%%%%%%%%%%%%%%%%%%%%%%%%%%%%
\resizebox{8cm}{!}{\includegraphics{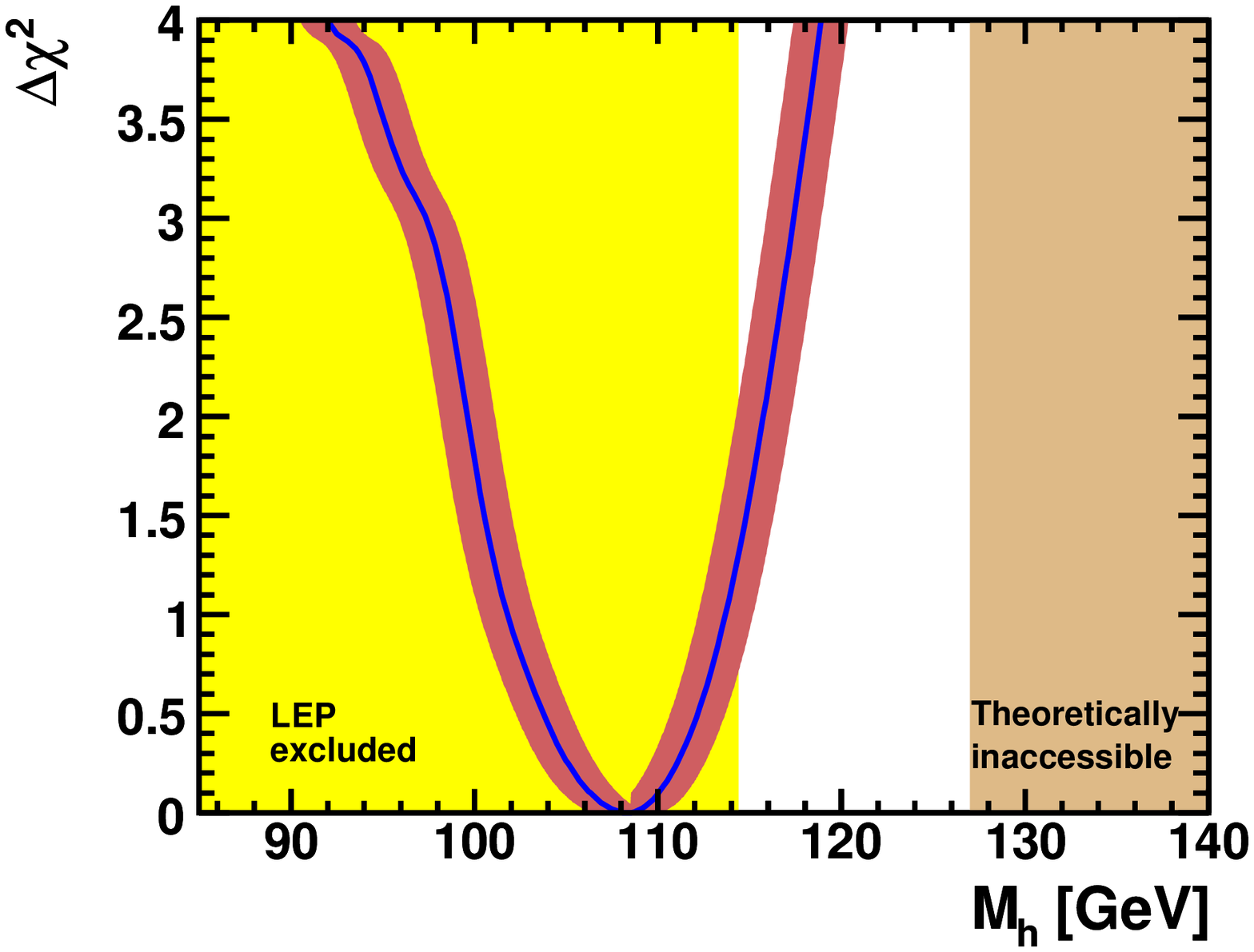}}
\resizebox{8cm}{!}{\includegraphics{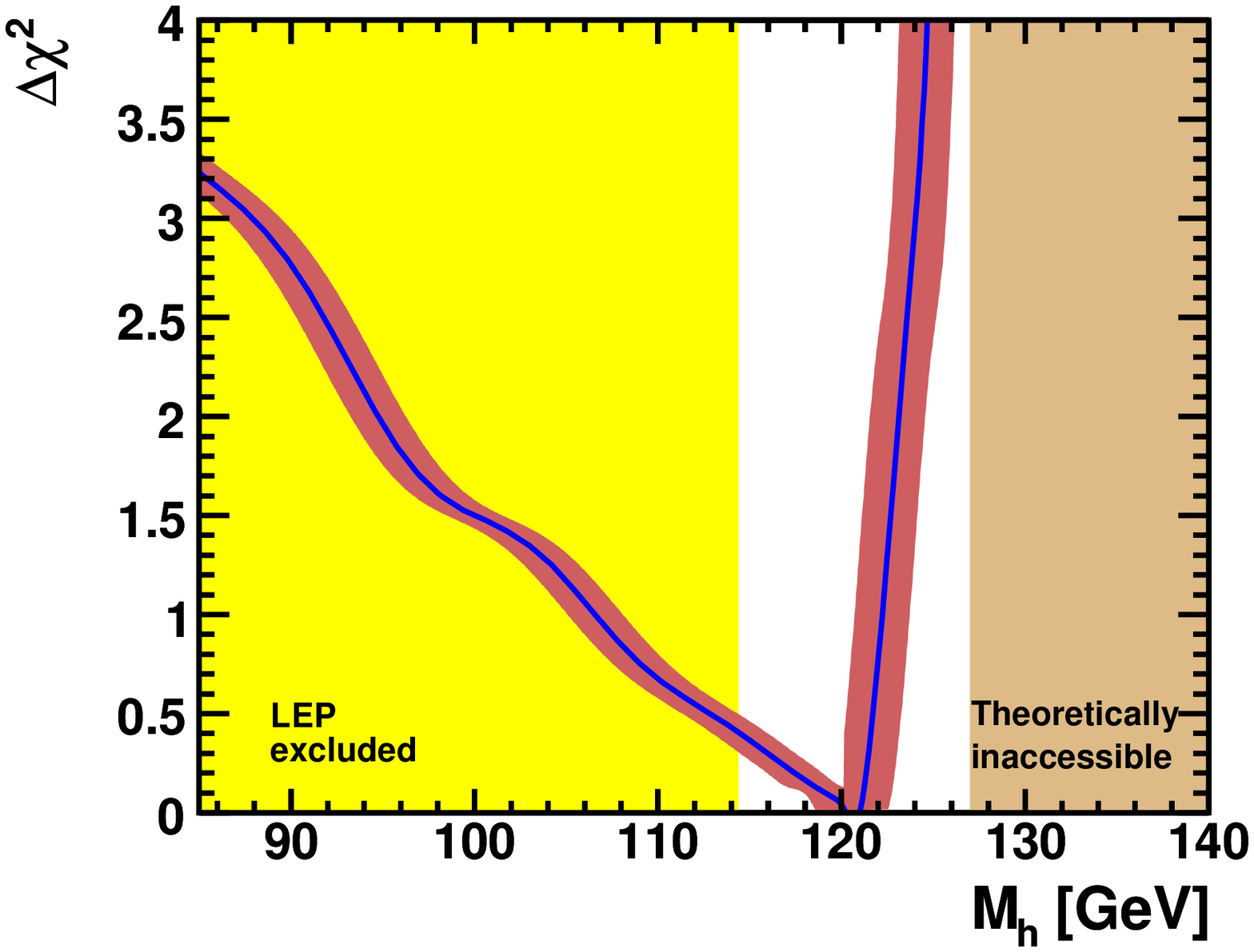}}
%%%%%%%%%%%%%%%%%%%%%%%%%%%%%%%
\vspace{-1cm}
\caption{\it The $\chi^2$ functions for $\Mh$ in the CMSSM (left) and
  the NUHM1 (right), 
including the theoretical uncertainties (red bands). Also shown is the mass
range excluded for a SM-like Higgs boson (yellow shading),
and the ranges theoretically inaccessible in the supersymmetric models
studied. 
}
\label{fig:redband}
\end{figure*}

~\\
We have already commented on the effect on the 
likelihood function for $\mneu{1}$ of dropping the LEP $\Mh$
constraint, see Fig.~\ref{fig:mchi}, and on the prediction for $\Mh$
itself, see Fig.~\ref{fig:mh}. We now discuss in more detail
the likelihood functions for $\Mh$ within the CMSSM and NUHM1 
frameworks obtained when dropping the
contribution to $\chi^2$ from the direct Higgs searches at LEP,
shown in the left and right panels of Fig.~\ref{fig:redband}, respectively. 
The left plot updates that for the CMSSM given in~\cite{Master1}. 

It is well
known that the central value of the Higgs mass in a SM
fit to the precision electroweak data lies below
100~GeV~\cite{ewpoMoriond2009,lepewwg}, 
but the theoretical (blue band) and experimental uncertainties 
in the SM fit are such that they are still compatible at the 
95\% C.L.\ 
with the direct lower limit of
114.4~GeV~\cite{Barate:2003sz} derived 
from searches at LEP. In the case of the CMSSM and NUHM1,
one may predict $\Mh$ on the basis of the underlying model
parameters, with a 1-$\sigma$ uncertainty of 1.5~GeV~\cite{Degrassi:2002fi},
shown as a red band in Fig.~\ref{fig:redband}. Also shown in
Fig.~\ref{fig:redband} are the LEP exclusion on a SM Higgs
(yellow shading)
and the ranges that are theoretically inaccessible in the
supersymmetric models studied (beige shading)~\footnote{It
is apparent that the current Tevatron exclusion~\cite{Tevatron} of a range
between 160 and 170~GeV does not impact supersymmetric
models.}. The LEP exclusion is directly applicable to the CMSSM,
since the $h$ couplings are essentially indistinguishable from
those of the SM Higgs boson~\cite{Ellis:2001qv,Ambrosanio:2001xb}, but
this is not 
necessarily the case in the NUHM1, as discussed earlier in this
paper. 

%%%%%%%%%%%%%%%%%%%%%% F I G U R E %%%%%%%%%%%%%%%%%%%%%%%%%%%%%%%%%%%
%%%%%%%%%%%%%%%%%%%%%% F I G U R E %%%%%%%%%%%%%%%%%%%%%%%%%%%%%%%%%%%

In the case of the CMSSM, we see in the left panel of
Fig.~\ref{fig:redband} that the minimum of the $\chi^2$
function occurs below the LEP exclusion limit. 
While the tension between the $\chi^2$ function for $\Mh$ 
arising from the CMSSM
fit and the LEP exclusion limit has slightly increased compared to the
earlier analysis performed in \cite{Master1}, the fit result is still
compatible at the 95\% C.L.\ with the search limit, 
similarly to the SM case. As we found in the analysis performed above, 
a global fit including the LEP constraint has acceptable $\chi^2$.
In the case of the NUHM1, shown in the right panel of
Fig.~\ref{fig:redband}, we see that the minimum of the $\chi^2$
function occurs {\it above} the LEP lower limit on the mass of a SM 
Higgs. Thus, 
within the NUHM1 the combination of all other experimental
constraints {\em naturally} evades the LEP Higgs constraints, and no
tension between $\Mh$ and the experimental bounds exists.

%---------------------------------------------------------------------
\section{Conclusions}
\label{sec:conx}
%---------------------------------------------------------------------

We have presented in this paper detailed results from
global fits to available experimental and cosmological data within
the CMSSM and NUHM1, using a frequentist approach. As already
reported in~\cite{Master2}, we find relatively small values of the key input
SUSY-breaking parameters $m_{1/2}$ and $m_0$ in both
models. Moreover, the values for these parameters
are quite similar in the two models,
indicating that the predictions are relatively robust and do not depend
strongly on the details of the Higgs sector.

We have presented details of the likelihood functions for
individual sparticle masses and the correlations between them.
As noted in~\cite{Master2}, the particle spectra are similar in the
two models, the most prominent differences being in the masses
of the heavier Higgs bosons $A, H$ and $H^\pm$, which are lighter
in the NUHM1, and the heavier neutralinos and chargino, which are
lighter in the CMSSM. These differences reflect the greater freedom
in varying the parameters of the Higgs sector in the NUHM1.
The favoured values of the particle masses in both models are such
that there are good prospects for detecting supersymmetric particles
even in the early phase of the LHC running with reduced centre-of-mass
energy and limited luminosity and for observing supersymmetric particles
and possibly the whole Higgs boson spectrum at a 1~TeV $e^+e^-$
collider (the latter refers in particular to the case of the NUHM1).

We find striking correlations between the different sparticle
masses in both the CMSSM and the NUHM1. This reflects the
fact that the dominant contributions to most of the sfermion masses
are those due to $m_{1/2}$, implying that most sparticle masses
are tightly correlated with those of the gluino and the LSP.
These correlations imply that, if the gluino is discovered at the LHC
and its mass determined by a combination of kinematic and
cross-section measurements, the predictions for the other
sparticle masses within the CMSSM and the NUHM1
could be refined considerably. In particular,
the masses for colour-neutral sparticles such as the
neutralino LSP and sleptons could be estimated more accurately, 
and hence also the energies of the corresponding thresholds in
$e^+ e^-$ annihilation within these models.
For some of the correlations, most notably the difference between the
LSP mass and the mass of the lighter stau, the pattern of the fit 
results in the NUHM1 drastically differs from the one in the CMSSM. 
Mass correlations of observed supersymmetric particles could therefore
provide very valuable information for distinguishing between different
models.

In addition to the sparticle masses, there are several other
observables that could serve to constrain (or provide evidence for)
the CMSSM or the NUHM1. One observable that could discriminate
between the CMSSM and the NUHM1, and might lead to 
%a `quick win'
an early discovery at
the LHCb experiment, is \bmm. In the CMSSM the rate
for \bmm\  obtained from the fit is expected to be close to the 
SM value, whereas the value may be considerably larger in the NUHM1 
without reducing the goodness of the fit.

A very exciting measurement would be that of the direct scattering
of astrophysical cold dark matter particles. We find in both the CMSSM
and the NUHM1 that the favoured rate for spin-independent dark matter
scattering lies quite close to the present experimental
upper limit, though with larger uncertainties in the NUHM1. In
view of the prospective improvements in the sensitivities of direct
dark matter search experiments in the near future, they may be able
to actually find the first indication of a supersymmetric particle
before the LHC, 
though a combination of astrophysical and collider measurements
would be needed to pin down its SUSY nature.

We have emphasized throughout this paper the sensitivity of our
conclusions to the imposition of the \gmt\ constraint. This plays
the dominant role in disfavouring large values of $m_{1/2}$ and
$m_0$ and hence, in particular, the focus-point region of the
CMSSM. In particular, \bsg\ plays no role in disfavouring the
focus-point region. Intriguingly, however, some other observables
seem slightly to prefer independently the coannihilation region,
such as $\MW$ and \btn. The net result is that the focus-point
region is disfavoured by $\Delta \chi^2 \sim 2$, even if the \gmt\
constraint is dropped. Conversely, the other data provide a hint 
that the supersymmetric contribution to \gmt\ might be of
comparable magnitude to the range required to reconcile the
experimental measurement of \gmt\ with the SM
calculation.

We have also explored the effect of dropping from the global fit
the experimental measurement of \bsg, and have shown that
there is no conflict between this observable and the other
constraints. We have shown as well that if \Och\ is dropped from
the global fit, the other constraints favour --- quite remarkably ---
a range within an
order of magnitude of the astrophysical cold dark matter density,
particularly within the CMSSM. These studies reveal no latent 
tensions between the data and either the CMSSM or NUHM1 fit.
Finally, we have discussed the impact of dropping the LEP
Higgs constraint from the global fits. While in the CMSSM there is
a slight tension between the fit result and the direct search limit, 
similarly to the SM case, 
the NUHM1 actually {\it favours} a
value for $\Mh$ significantly above the LEP limit.  The
discovery at the LHC of a Higgs boson weighing more than
120~GeV would favour the NUHM1 over the CMSSM.

Indirect constraints on supersymmetric model parameters are
fine in their own way, and it is encouraging that there are no
significant tensions in either the CMSSM or NUHM1 fits.
However, we hope that soon it will be possible to include in
such fits some experimental measurements of physics
beyond the SM. The fit results seem to indicate that there
are good prospects for sparticle and Higgs-boson production at the LHC, 
but that there may also be good chances at a similar time scale 
to obtain 
evidence for cold dark matter scattering or for a discrepancy with the
SM prediction for some other observable besides
\gmt, such as \bmm\ or \btn.

%---------------------------------------------------------------------
\section*{Acknowledgements}
%---------------------------------------------------------------------

We thank P.~Paradisi for earlier collaboration underlying this work,
as well as B.~Cousins and P.~Sandick for helpful discussions.
This work was supported in part by the European Community's Marie-Curie
Research Training Network under contracts MRTN-CT-2006-035505
`Tools and Precision Calculations for Physics Discoveries at Colliders'
and MRTN-CT-2006-035482 `FLAVIAnet', and by the Spanish MEC and FEDER under 
grant FPA2005-01678. The work of S.H. was supported 
in part by CICYT (grant FPA~2007--66387), and
the work of K.A.O. was supported in part
by DOE grant DE--FG02--94ER--40823 at the University of Minnesota.

%---------------------------------------------------------------------

\section*{Note added}

While we were finalizing this paper, the analysis~\cite{fittinoMC}
has appeared. This uses a previous version of the {\tt
MasterCode}~\cite{Master2,Master1} to fit available data within the
CMSSM and also adopts a frequentist Markov Chain Monte Carlo approach:
the results of the analysis are very similar to ours. 
Our paper also compares current CMSSM and NUHM1 fits,
whereas~\cite{fittinoMC} discusses  
the perspectives for fits using future data.

%---------------------------------------------------------------------

\newpage
\pagebreak

%---------------------------------------------------------------------

%---------------------------------------------------------------------

\end{document}